\documentclass[aps,pra,nobalancelastpage,superscriptaddress, 10pt, twocolumn]{revtex4-2}

\usepackage{color,ifthen,amsthm,amsmath,amsxtra,amsfonts,dsfont,graphicx,bm,tikz,scalerel,wasysym,bbm,graphicx,amsthm,physics,empheq,comment, tikz, tikz-network}
\usepackage[colorlinks=true,linkcolor=blue, citecolor=blue, urlcolor=blue, bookmarks]{hyperref}
\usepackage[normalem]{ulem}
\usetikzlibrary{patterns,decorations.pathreplacing}

\newcommand{\be}{\begin{equation}}
\newcommand{\ee}{\end{equation}}
\newcommand{\bea}{\begin{eqnarray}}
\newcommand{\eea}{\end{eqnarray}}
\newcommand{\bse}{\begin{subequations}}
\newcommand{\ese}{\end{subequations}}

\theoremstyle{plain}
\newcommand{\1}{\mathbbm{1}}

\theoremstyle{plain}

\theoremstyle{plain}
\newtheorem{assumption}{Assumption}
\theoremstyle{plain}

\newcommand{\Wgategreen}[2]{
\draw[very thick] (#1-0.75, #2 +0.75) -- (#1+0.75,#2-0.75);
\draw[very thick] (#1-0.75,#2-0.75) -- (#1+0.75,#2+0.75);
\draw[ thick, fill=mygreen, rounded corners=2pt] (#1-0.25,#2+0.25) rectangle (#1+0.25,#2-0.25);
\draw[thick] (#1,#2+0.15) -- (#1+0.15,#2+0.15) -- (#1+0.15,#2);
}

\newcommand{\Wgateolivegreen}[2]{
\draw[very thick] (#1-0.75, #2 +0.75) -- (#1+0.75,#2-0.75);
\draw[very thick] (#1-0.75,#2-0.75) -- (#1+0.75,#2+0.75);
\draw[ thick, fill=OliveGreen, rounded corners=2pt] (#1-0.25,#2+0.25) rectangle (#1+0.25,#2-0.25);
\draw[thick] (#1,#2+0.15) -- (#1+0.15,#2+0.15) -- (#1+0.15,#2);
}

\newcommand{\CircularGate}[2]{
\draw[very thick] (#1-0.75, #2 +0.75) -- (#1+0.75,#2-0.75);
\draw[very thick] (#1-0.75,#2-0.75) -- (#1+0.75,#2+0.75);
\draw[ thick, fill=OliveGreen, rounded corners=2pt] (#1,#2) circle (0.25);
}

\usepackage{color}
\definecolor{myred}{RGB}{232,102,102}
\definecolor{myblue}{RGB}{187,187,255}
\definecolor{myorange}{RGB}{202,52,51}
\definecolor{mygrey}{RGB}{105,105,105}
\definecolor{OliveGreen}{RGB}{85,107,47}
\definecolor{NavyBlue}{RGB}{0,0,128}
\definecolor{mygreen}{RGB}{34,139,34}
\definecolor{myY}{RGB}{220,255,203}
\definecolor{myYO}{RGB}{255, 220, 151}
\definecolor{myblue1}{RGB}{176,223,229}
\definecolor{myblue2}{RGB}{0,0,128}
\definecolor{myblue3}{RGB}{0,108,255}
\definecolor{myblue4}{RGB}{101,147,245}
\definecolor{myblue5}{RGB}{115,194,251}
\definecolor{myblue6}{RGB}{87,160,211}
\definecolor{myblue7}{RGB}{137,207,240}
\definecolor{myblue8}{RGB}{29,41,81}
\definecolor{myblue9}{RGB}{14,77,146}
\definecolor{myblue10}{RGB}{15,82,186}
\definecolor{myred1}{RGB}{255,36,0}
\definecolor{myred2}{RGB}{205,92,92}
\definecolor{myred3}{RGB}{178,34,34}
\definecolor{myred4}{RGB}{164,90,82}
\definecolor{myred5}{RGB}{255,8,0}
\definecolor{myred6}{RGB}{202,52,51}
\definecolor{myred7}{RGB}{226,13,9}
\definecolor{myred8}{RGB}{141,2,31}
\definecolor{myred9}{RGB}{250,128,114}
\definecolor{myred10}{RGB}{237,41,57}
\definecolor{myyellow1}{RGB}{254,220,86}
\definecolor{myyellow2}{RGB}{255,229,180}
\definecolor{myyellow3}{RGB}{238,220,130}
\definecolor{myyellow4}{RGB}{253,165,15}
\definecolor{myyellow5}{RGB}{255,195,11}
\definecolor{myyellow6}{RGB}{218,165,32}
\definecolor{myyellow7}{RGB}{255,211,0}
\definecolor{myyellow8}{RGB}{248,222,126}
\definecolor{myyellow9}{RGB}{245,245,220}
\definecolor{myyellow10}{RGB}{248,228,115}

\definecolor{mygray1}{RGB}{246,246,246}
\definecolor{mygray2}{RGB}{32,32,32}
\definecolor{mygray3}{RGB}{64,64,64}
\definecolor{mygray4}{RGB}{96,96,96}
\definecolor{mygray5}{RGB}{128,128,128}
\definecolor{mygray6}{RGB}{160,160,160}
\definecolor{mygray7}{RGB}{224,224,224}
\definecolor{mygray8}{RGB}{180,180,180}
\newcommand{\mcirc}{\mathbin{\scalerel*{\fullmoon}{G}}}

\begin{document}

\title{
Structural Stability Hypothesis of Dual Unitary Quantum Chaos
}

\author{Jonathon Riddell}
\affiliation{School of Physics and Astronomy, University of Nottingham, Nottingham, NG7 2RD, UK}
\affiliation{Centre for the Mathematics and Theoretical Physics of Quantum Non-Equilibrium Systems, University of Nottingham, Nottingham, NG7 2RD, UK}

\author{Curt von Keyserlingk}
\affiliation{Department of Physics, King’s College London, Strand WC2R 2LS, UK}

\author{Toma{\v z} Prosen}
\affiliation{Faculty of Mathematics and Physics, University of Ljubljana, Jadranska 19, SI1000 Ljubljana, Slovenia}

\affiliation{Institute of Mathematics, Physics, and Mechanics, Jadranska 19, SI1000 Ljubljana, Slovenia}

\author{Bruno Bertini}
\affiliation{School of Physics and Astronomy, University of Nottingham, Nottingham, NG7 2RD, UK}

\affiliation{Centre for the Mathematics and Theoretical Physics of Quantum Non-Equilibrium Systems, University of Nottingham, Nottingham, NG7 2RD, UK}

\begin{abstract}
Having spectral correlations that, over small enough energy scales, are described by random matrix theory is regarded as the most general defining feature of quantum chaotic systems as it applies in the many-body setting and away from any semiclassical limit. Although this property is extremely difficult to prove analytically for generic many-body systems, a rigorous proof has been achieved for dual-unitary circuits --- a special class of local quantum circuits that remain unitary upon swapping space and time. Here we consider the fate of this property when moving from dual-unitary to generic quantum circuits focussing on the \emph{spectral form factor}, i.e., the Fourier transform of the two-point correlation. We begin with a numerical survey that, in agreement with previous studies, suggests that there exists a finite region in parameter space where dual-unitary physics is stable and spectral correlations are still described by random matrix theory, although up to a maximal quasienergy scale. To explain these findings, we develop a perturbative expansion: it recovers the random matrix theory predictions, provided the terms occurring in perturbation theory obey a relatively simple set of assumptions. We then provide numerical evidence and a heuristic analytical argument  supporting these assumptions.
\end{abstract}

\maketitle


\section{Introduction}

Exact solutions of interacting many body systems are not just monuments to human ingenuity, they are also key instruments in both statistical mechanics and many-body dynamics. In fact, it is often implicitly assumed that each (dynamical) universality class in statistical physics should be endowed with at least one exactly solvable model through which one obtains deeper understanding of the whole universality class.

Until recently exact solutions in interacting systems were limited to the Yang-Baxter paradigm which underlies the so-called integrable systems in two spatial (or $1+1$) dimensions~\cite{baxter,faddeev}. The existence of an extensive number of conservation laws in these systems, however, makes their dynamical behaviour non-generic~\cite{calabrese2016introduction, bastianello2022introduction, bertini2021finitetemperature} and one was left to wonder how to describe generic dynamics. Namely, the ones of the so-called ``chaotic" quantum many-body systems, which have only a finite number of conserved charges. While various types of random matrix theories, depending on the model's time-reversal and interaction-locality, turned out to be a successful tool for modelling quantum chaotic systems~\cite{mehta1991random, guhr, syk, randomcircuits, nahum2017quantum, nahum2018operator, chan2018solution, vonkeyserlingk2018operator, khemani2018operator, rakovszky2019sub, friedman2019spectral}, only very recently the first class of exactly solvable non-integrable systems has been discovered~\cite{bertini2018exact,bertini2019exact} and has led to exact solutions for many dynamical problems even in the absence of explicit randomness~\cite{bertini2019exact, bertini2018exact, bertini2021random, bertini2019entanglement, gopalakrishnan2019unitary, claeys2020maximum, piroli2020exact, bertini2020scrambling, kos2023scrambling, rampp2023from, foligno2022growth, giudice2021temporal, foligno2023temporal, bertini2020operator, bertini2020operator2, reid2021entanglement, ho2022exact, claeys2022emergentquantum, ippoliti2022solvable}.

These are the so-called dual unitary circuits, which are expressed in the form of locally interacting quantum systems in discrete space time, and whose defining feature is that they generate a unitary evolution not only in time but also in space. This fundamental property is most clearly expressed in terms of the so-called space-time duality~\cite{akila2016particle}, a space-time swap symmetry of the tensor network diagram representing the physical observable of interest. Among other useful features, dual-unitary circuits allow for exact analytical computation of dynamical correlation functions of local operators~\cite{bertini2019exact}, as well as long-range two-point spectral correlations as expressed in the form of the spectral form factor (SFF)~\cite{bertini2018exact,bertini2021random}.

Given this novel class of exactly solvable chaotic systems a fundamental question concerns the stability or robustness of their dynamical features. One may trace the basic motivation for such a question to the notion of structural stability of hyperbolic flows in classical chaotic dynamical systems~\cite{robbin,robinson}. Contrary to integrable systems, which are structurally unstable and where perturbative expansions around them generically diverge (cf. Kolmogorov-Arnold-Moser theory in classical dynamical systems, or divergent Feynman diagram expansions of quantum field theories around their free/non-coupled limits), chaotic systems are expected to be robust against typical perturbations. In this work we formulate the hypothesis of structural stability of Floquet dual-unitary circuits (therefore focusing on discrete time-dynamics with explicit time translation symmetry), and streamline a simple strategy for its verification. 
Specifically, we conjecture that chaotic/ergodic dual-unitary circuits remain chaotic/ergodic under small, time-translation invariant perturbations. If we assume that chaos is equivalent to a linear-in-$t$ growth of the SFF in the thermodynamic limit, this hypothesis can be reformulated in terms of the so-called spectral Lyapunov exponents (SLE)~\cite{Chan20}. In that language, the hypothesis states that the leading SLE in each of the $t$ time-translation symmetry sectors decays to 1 exponentially fast in time $t$.
We note that a somewhat simpler approach to structural stability hypothesis has been addressed by two of us and P. Kos in Ref.~\cite{kos2021correlations}, which studied the robustness of local dynamical correlators. Importantly, however, the decay of local correlators is not sufficient for establishing quantum chaos and ergodicity, hence our current study of SFF, a global (non-local) dynamical observable, provides a more stringent characterisation. We also stress that our study goes substantially beyond the one presented in Ref.~\cite{braun2020transition}, where the stability of the SFF was investigated for a specific dual unitary circuit at first order in perturbation theory.

Clearly, this question is very challenging, and extremely difficult to address in full mathematical rigour. The purpose of our paper is to establish the minimal set of assumptions --- which turn out to be two --- needed for a rigorous proof of the stability of the chaotic SFF. We do this by formulating a perturbation theory for the SLE, and identifying sufficient conditions for the expansion to behave in a way that is consistent with ergodicity. We furthermore verify these two assumptions numerically.

We now summarise our assumptions and results in more detail. We study the SFF averaged over an ensemble of locally constructed unitaries $\mathcal{E}$ of an extended 1D system with $L$ sites $K(t,L)=\mathbb{E}_{\mathbb{U} \in \mathcal{E}} \left[|\mathrm{tr} \mathbb{U}^t|^2\right]$. The ensemble is parameterised by $\epsilon$, where the unperturbed point $\epsilon=0$ corresponds to an ensemble of dual unitary Floquet evolutions. Using the space-time duality noted above, the SFF may be recast as  ${K(t,L) = \tr \mathcal{T}^L}$ for a SFF \emph{transfer matrix} $\mathcal{T}$ acting on $2t$ sites~\cite{bertini2018exact,bertini2021random}. The linear-in-$t$ ramp in the SFF characteristic of ergodic systems (see Eq.~\eqref{eq:CUEform}) is obtained if the leading $t$ eigenvalues of $\mathcal{T}$
are are nearly degenerate, i.e., exponentially close to unity,  $\lambda_{j=1,\ldots,t}=1+e^{-O(t)}$. 

The unperturbed dual-unitary model provably has this eigenvalue structure; we investigate the circumstances under which it remains true in eigenvalue perturbation theory in $\epsilon$. We show that it follows from two assumptions.  The first of these stipulates a non-crossing of the leading eigenvalue of $\mathcal{T}$ as a function of $\epsilon$. Recalling the von Neumann-Wigner theorem~\cite{landau1981quantum}, this corresponds to a genericity assumption on the perturbation. The second, more substantial assumption can be expressed in terms of an exponential bound on certain multi-point correlation functions involving the unperturbed SFF transfer matrix. We numerically demonstrate the validity of these assumptions in a particular family of perturbed Floquet dual unitary circuits, and corroborate those results with a heuristic analytical argument involving the spectral decomposition of the resolvent of the unperturbed SFF transfer matrix.

The rest of the paper is structured as follows. In Sec.~\ref{sec:setting} we recall the precise setting considered, introduce a simple minimal model that we use for the numerical tests, and briefly describe the numerical methods used in our computations. In Sec.~\ref{sec:independentnumerics} we present a numerical survey suggesting that the dual unitary behaviour is indeed structurally stable. In Sec.~\ref{sec:pertubationtheory} we present our perturbative argument. In Sec.~\ref{sec:discussion} we discuss the validity of our second assumption, presenting numerical tests and an heuristic analytical argument. Finally, in Sec.~\ref{sec:conclusions} we report our conclusions and discuss the outlook of our research. Some technical details and proofs are reported in the three appendices.

\section{Setting}
\label{sec:setting}

\subsection{Physical System}

\begin{figure}
\begin{tikzpicture}[baseline=(current  bounding  box.center), scale=0.65]

\foreach \i in {0,...,4}{
\draw[ thick] (-2*\i+2-1.5,4) arc (135:-0:0.15);
\draw[ thick] (-2*\i+2-2.5,4) arc (-325:-180:0.15);
\draw[ thick] (-2*\i+2-1.5,-2) arc (-45:180:-0.15);
\draw[ thick] (-2*\i+2-2.5,-2) arc (45:-180:0.15);
}

\foreach \i in {0,...,4}{
\draw[thick, opacity=0.7, dashed] (-2*\i+0.75,4) -- (-2*\i+.75,-2);
\draw[thick, opacity=0.7, dashed] (-2*\i+0.75-1.5,4) -- (-2*\i+.75-1.5,-2);}

\foreach \i in {0,1,2}{
\draw[thick, opacity=0.7, dashed] (-9.5,2*\i-1.7) -- (0.4,2*\i-1.7);
\draw[thick, opacity=0.7, dashed] (-9.5,2*\i-1.3) -- (0.4,2*\i-1.3);
}

\foreach \i in {3,...,13}{
\draw[opacity=0.4] (-12.5+\i,-2.5) -- (-12.5+\i,4.5);
}
\foreach \i in {-1,...,5}{
\draw[opacity=0.4] (-10,3-\i) -- (1,3-\i);
}
\foreach \i in{1.5,2.5,3.5}{
\draw[thick] (0.5,2*\i-0.5-3.5) arc (45:-90:0.17);
\draw[thick] (-10+0.5+0,2*\i-0.5-3.5) arc (90:270:0.15);
}
\foreach \i in{0.5,1.5,2.5}
{
\draw[ thick] (0.5,1+2*\i-0.5-3.5) arc (-45:90:0.17);
\draw[ thick] (-10+0.5,1+2*\i-0.5-3.5) arc (270:90:0.15);
}
\foreach \i in {1,2}{
\Text[x=1.25,y=-2+2*\i]{\scriptsize$\i$}
}
\foreach \i in {1,3}{
\Text[x=1.25,y=-2+\i]{\small$\frac{\i}{2}$}
}
\foreach \i in {1,3,5}{
\Text[x=-7.5+\i-2,y=-2.9]{\small$\frac{\i}{2}$}
}
\foreach \i in {1,2,3}{
\Text[x=-7.5+2*\i-2,y=-2.9]{\scriptsize${\i}$
}
}
\foreach \jj[evaluate=\jj as \j using -2*(ceil(\jj/2)-\jj/2)] in {-1,-3,-5}{
\foreach \i in {1}
{
\draw[thick] (.5-2*\i-1*\j,-2-1*\jj) -- (1-2*\i-1*\j,-1.5-\jj);
\draw[thick] (1-2*\i-1*\j,-1.5-1*\jj) -- (1.5-2*\i-1*\j,-2-\jj);
\draw[thick] (.5-2*\i-1*\j,-1-1*\jj) -- (1-2*\i-1*\j,-1.5-\jj);
\draw[thick] (1-2*\i-1*\j,-1.5-1*\jj) -- (1.5-2*\i-1*\j,-1-\jj);
\draw[thick, fill=myred10, rounded corners=2pt] (0.75-2*\i-1*\j,-1.75-\jj) rectangle (1.25-2*\i-1*\j,-1.25-\jj);
\draw[thick] (-2*\i+2,-1.35-\jj) -- (-2*\i+2.15,-1.35-\jj) -- (-2*\i+2.15,-1.5-\jj);%
}
\foreach \i in {2}
{
\draw[thick] (.5-2*\i-1*\j,-2-1*\jj) -- (1-2*\i-1*\j,-1.5-\jj);
\draw[thick] (1-2*\i-1*\j,-1.5-1*\jj) -- (1.5-2*\i-1*\j,-2-\jj);
\draw[thick] (.5-2*\i-1*\j,-1-1*\jj) -- (1-2*\i-1*\j,-1.5-\jj);
\draw[thick] (1-2*\i-1*\j,-1.5-1*\jj) -- (1.5-2*\i-1*\j,-1-\jj);
\draw[thick, fill=myred9, rounded corners=2pt] (0.75-2*\i-1*\j,-1.75-\jj) rectangle (1.25-2*\i-1*\j,-1.25-\jj);
\draw[thick] (-2*\i+2,-1.35-\jj) -- (-2*\i+2.15,-1.35-\jj) -- (-2*\i+2.15,-1.5-\jj);%
}
\foreach \i in {3}
{
\draw[thick] (.5-2*\i-1*\j,-2-1*\jj) -- (1-2*\i-1*\j,-1.5-\jj);
\draw[thick] (1-2*\i-1*\j,-1.5-1*\jj) -- (1.5-2*\i-1*\j,-2-\jj);
\draw[thick] (.5-2*\i-1*\j,-1-1*\jj) -- (1-2*\i-1*\j,-1.5-\jj);
\draw[thick] (1-2*\i-1*\j,-1.5-1*\jj) -- (1.5-2*\i-1*\j,-1-\jj);
\draw[thick, fill=myred8, rounded corners=2pt] (0.75-2*\i-1*\j,-1.75-\jj) rectangle (1.25-2*\i-1*\j,-1.25-\jj);
\draw[thick] (-2*\i+2,-1.35-\jj) -- (-2*\i+2.15,-1.35-\jj) -- (-2*\i+2.15,-1.5-\jj);%
}
\foreach \i in {4}
{
\draw[thick] (.5-2*\i-1*\j,-2-1*\jj) -- (1-2*\i-1*\j,-1.5-\jj);
\draw[thick] (1-2*\i-1*\j,-1.5-1*\jj) -- (1.5-2*\i-1*\j,-2-\jj);
\draw[thick] (.5-2*\i-1*\j,-1-1*\jj) -- (1-2*\i-1*\j,-1.5-\jj);
\draw[thick] (1-2*\i-1*\j,-1.5-1*\jj) -- (1.5-2*\i-1*\j,-1-\jj);
\draw[thick, fill=myred7, rounded corners=2pt] (0.75-2*\i-1*\j,-1.75-\jj) rectangle (1.25-2*\i-1*\j,-1.25-\jj);
\draw[thick] (-2*\i+2,-1.35-\jj) -- (-2*\i+2.15,-1.35-\jj) -- (-2*\i+2.15,-1.5-\jj);%
}
\foreach \i in {5}
{
\draw[thick] (.5-2*\i-1*\j,-2-1*\jj) -- (1-2*\i-1*\j,-1.5-\jj);
\draw[thick] (1-2*\i-1*\j,-1.5-1*\jj) -- (1.5-2*\i-1*\j,-2-\jj);
\draw[thick] (.5-2*\i-1*\j,-1-1*\jj) -- (1-2*\i-1*\j,-1.5-\jj);
\draw[thick] (1-2*\i-1*\j,-1.5-1*\jj) -- (1.5-2*\i-1*\j,-1-\jj);
\draw[thick, fill=myred6, rounded corners=2pt] (0.75-2*\i-1*\j,-1.75-\jj) rectangle (1.25-2*\i-1*\j,-1.25-\jj);
\draw[thick] (-2*\i+2,-1.35-\jj) -- (-2*\i+2.15,-1.35-\jj) -- (-2*\i+2.15,-1.5-\jj);%
}
}
\foreach \jj[evaluate=\jj as \j using -2*(ceil(\jj/2)-\jj/2)] in {-4,-2,0}{
\foreach \i in {1}
{
\draw[thick] (.5-2*\i-1*\j,-2-1*\jj) -- (1-2*\i-1*\j,-1.5-\jj);
\draw[thick] (1-2*\i-1*\j,-1.5-1*\jj) -- (1.5-2*\i-1*\j,-2-\jj);
\draw[thick] (.5-2*\i-1*\j,-1-1*\jj) -- (1-2*\i-1*\j,-1.5-\jj);
\draw[thick] (1-2*\i-1*\j,-1.5-1*\jj) -- (1.5-2*\i-1*\j,-1-\jj);
\draw[thick, fill=myred5, rounded corners=2pt] (0.75-2*\i-1*\j,-1.75-\jj) rectangle (1.25-2*\i-1*\j,-1.25-\jj);
\draw[thick] (-2*\i+1,-1.35-\jj) -- (-2*\i+1.15,-1.35-\jj) -- (-2*\i+1.15,-1.5-\jj);%
}
\foreach \i in {2}
{
\draw[thick] (.5-2*\i-1*\j,-2-1*\jj) -- (1-2*\i-1*\j,-1.5-\jj);
\draw[thick] (1-2*\i-1*\j,-1.5-1*\jj) -- (1.5-2*\i-1*\j,-2-\jj);
\draw[thick] (.5-2*\i-1*\j,-1-1*\jj) -- (1-2*\i-1*\j,-1.5-\jj);
\draw[thick] (1-2*\i-1*\j,-1.5-1*\jj) -- (1.5-2*\i-1*\j,-1-\jj);
\draw[thick, fill=myred4, rounded corners=2pt] (0.75-2*\i-1*\j,-1.75-\jj) rectangle (1.25-2*\i-1*\j,-1.25-\jj);
\draw[thick] (-2*\i+1,-1.35-\jj) -- (-2*\i+1.15,-1.35-\jj) -- (-2*\i+1.15,-1.5-\jj);%
}
\foreach \i in {3}
{
\draw[thick] (.5-2*\i-1*\j,-2-1*\jj) -- (1-2*\i-1*\j,-1.5-\jj);
\draw[thick] (1-2*\i-1*\j,-1.5-1*\jj) -- (1.5-2*\i-1*\j,-2-\jj);
\draw[thick] (.5-2*\i-1*\j,-1-1*\jj) -- (1-2*\i-1*\j,-1.5-\jj);
\draw[thick] (1-2*\i-1*\j,-1.5-1*\jj) -- (1.5-2*\i-1*\j,-1-\jj);
\draw[thick, fill=myred3, rounded corners=2pt] (0.75-2*\i-1*\j,-1.75-\jj) rectangle (1.25-2*\i-1*\j,-1.25-\jj);
\draw[thick] (-2*\i+1,-1.35-\jj) -- (-2*\i+1.15,-1.35-\jj) -- (-2*\i+1.15,-1.5-\jj);%
}
\foreach \i in {4}
{
\draw[thick] (.5-2*\i-1*\j,-2-1*\jj) -- (1-2*\i-1*\j,-1.5-\jj);
\draw[thick] (1-2*\i-1*\j,-1.5-1*\jj) -- (1.5-2*\i-1*\j,-2-\jj);
\draw[thick] (.5-2*\i-1*\j,-1-1*\jj) -- (1-2*\i-1*\j,-1.5-\jj);
\draw[thick] (1-2*\i-1*\j,-1.5-1*\jj) -- (1.5-2*\i-1*\j,-1-\jj);
\draw[thick, fill=myred2, rounded corners=2pt] (0.75-2*\i-1*\j,-1.75-\jj) rectangle (1.25-2*\i-1*\j,-1.25-\jj);
\draw[thick] (-2*\i+1,-1.35-\jj) -- (-2*\i+1.15,-1.35-\jj) -- (-2*\i+1.15,-1.5-\jj);%
}
\foreach \i in {5}
{
\draw[thick] (.5-2*\i-1*\j,-2-1*\jj) -- (1-2*\i-1*\j,-1.5-\jj);
\draw[thick] (1-2*\i-1*\j,-1.5-1*\jj) -- (1.5-2*\i-1*\j,-2-\jj);
\draw[thick] (.5-2*\i-1*\j,-1-1*\jj) -- (1-2*\i-1*\j,-1.5-\jj);
\draw[thick] (1-2*\i-1*\j,-1.5-1*\jj) -- (1.5-2*\i-1*\j,-1-\jj);
\draw[thick, fill=myred1, rounded corners=2pt] (0.75-2*\i-1*\j,-1.75-\jj) rectangle (1.25-2*\i-1*\j,-1.25-\jj);
\draw[thick] (-2*\i+1,-1.35-\jj) -- (-2*\i+1.15,-1.35-\jj) -- (-2*\i+1.15,-1.5-\jj);%
}
}
\Text[x=-2,y=-2.9]{$\cdots$}
\Text[x=0.47,y=-2.9]{\small $L\equiv 0$}
\Text[x=-4,y=-3.5]{\small $x$}
\Text[x=1.25,y=4]{\small $t$}
\Text[x=2,y=1]{\small$\tau$}
\Text[x=1.25,y=3.2]{$\vdots$}
\end{tikzpicture}
\caption{Diagrammatic representation of $\tr[\mathbb{U}^t]$. The boxes represent local gates and different legs act on different spatial sites. Matrix product is represented by joining legs and goes from bottom to top. The lines at the left and right edges are joined because we consider periodic boundary conditions ($L\equiv 0$), while those at the top and bottom are joined because of the trace. The background grid specifies the space-time lattice. The gates in the same vertical column are identical.}
\label{fig:quantumcircuit}
\end{figure}
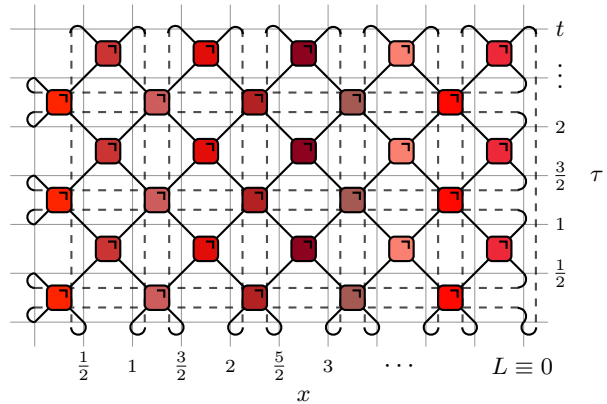

We consider a unitary quantum circuit acting on a chain of $2L$ qubits (the local Hilbert space has dimension $d=2$) at half-integer positions that are evolved by discrete applications of the Floquet operator $\mathbb{U} =  \mathbb{U}_o \mathbb{U}_e$ such that 
\begin{equation}
\mathbb{U}_o  = U_0 \otimes\cdots\otimes U_{L-1}, \quad \mathbb{U}_e =U_{1/2} \otimes\cdots\otimes U_{L-1/2}.
    \label{eq:evolutionoperator}
\end{equation}
Here $\{U_{x}\}_{x=0, 1/2,\ldots, L-1/2}$ are the local gates, i.e., unitary matrices acting on two adjacent qubits, at positions $x$ and $x+1/2$. Matrices acting at different positions are generically different and we denote by the subscript $x$ the leftmost site where the matrix acts non-trivially. The local gates can be parameterised as 
\be
 U_x =  V_x \cdot (u_x \otimes v_x),
    \label{eq:localgates}
 \ee
with 
\begin{equation}
\begin{aligned}
 &V_x \equiv e^{i \sum_{k=1}^3 J_{k, x} \sigma^{(k)}_x \sigma^{(k)}_{x+1/2}},\,\,\\
 & u_x \equiv e^{i \boldsymbol{\theta}_x  \cdot  \boldsymbol{\sigma}_x},\,\, v_x \equiv e^{i \boldsymbol{\phi}_x \cdot \boldsymbol{\sigma}_{x+1/2}},
 \end{aligned}
 \label{paramVuv}
\end{equation}
where $\boldsymbol{\sigma}=(\sigma^{(1)},\sigma^{(2)},\sigma^{(3)})$ is a vector of Pauli matrices, and $\boldsymbol{\sigma}_x$ is the corresponding local embedding in $(\mathbb C^2)^{\otimes 2L}$, while $(u_x \otimes v_x)$ is a tensor product of two one-site unitaries $u,v$ positioned at sites $x$ and $x+1/2$ respectively. Since the operator $\mathbb{U}$ is a special kind of matrix-product-operator, it can be depicted using the standard diagrammatic representation of tensor networks~\cite{cirac2021matrix}.  In this rendition tensors with $n$ indices are represented as two dimensional shapes with $n$ protruding legs. For instance, the local gate is represented as
\be
\label{eq:Ugate}
U_{x}=\begin{tikzpicture}[baseline=(current  bounding  box.center), scale=.7]
\draw[ thick] (-4.25,0.5) -- (-3.25,-0.5);
\draw[ thick] (-4.25,-0.5) -- (-3.25,0.5);
\draw[ thick, fill=myred, rounded corners=2pt] (-4,0.25) rectangle (-3.5,-0.25);
\draw[thick] (-3.75,0.15) -- (-3.75+0.15,0.15) -- (-3.75+0.15,0);
\Text[x=-4.25,y=-0.75]{}
\end{tikzpicture}.
\ee
The sum over indices is represented by connecting the corresponding legs and matrix multiplication acts bottom to top. For example, see Fig.~\ref{fig:quantumcircuit} for a portray of $\tr[\mathbb{U}^t]$ where each local gate is represented by
\be
\label{eq:Ugate}
U_{x}=\begin{tikzpicture}[baseline=(current  bounding  box.center), scale=.7]
\draw[ thick] (-4.25,0.5) -- (-3.25,-0.5);
\draw[ thick] (-4.25,-0.5) -- (-3.25,0.5);
\draw[ thick, fill=myred, rounded corners=2pt] (-4,0.25) rectangle (-3.5,-0.25);
\draw[thick] (-3.75,0.15) -- (-3.75+0.15,0.15) -- (-3.75+0.15,0);
\Text[x=-4.25,y=-0.75]{}
\end{tikzpicture},
\ee
and different shades denote in principle different matrices.

Note that, choosing $J_{1,x}  = J_{2,x}  = {\pi}/{4}$ the quantum circuit becomes dual-unitary~\cite{bertini2019exact}, and also the left-to-right contraction of the diagram can be thought of as the trace of a power of unitary operator~\cite{bertini2018exact, bertini2021random}.   

For simplicity from now on we focus on the case where the interaction term is the same at each half step. Namely, we consider
\be
J^{\phantom{\prime}}_{j,x}=J^{\phantom{\prime}}_{j},\quad J_{j,x+1/2}'=J_{j}', \quad j=1,2,3,\,\, x \in \mathbb Z_L,
\label{eq:JTI}
\ee
and set
\be
V = e^{i \sum_{k=1}^3 J_k\sigma^{(k)} \otimes \sigma^{(k)}}, \quad W = e^{i \sum_{k=1}^3 J_k'\sigma^{(k)} \otimes \sigma^{(k)}},
\label{eq:gatesVW}
\ee
while the one-site gates $u_x$ and $v_x$ are position dependent, i.e., 
$\boldsymbol{\theta}_x,
\boldsymbol{\phi}_x$ in (\ref{paramVuv}) are explicitly $x-$dependent.
 
\subsection{Spectral Form Factor and Space Transfer Matrix}
\label{sec:SFF}

Our aim is to characterise the spectral statistics of the Floquet operator~\eqref{eq:evolutionoperator} for generic choices of the local gates. Namely, we want to understand the general features of the distribution of the eigenvalues of $\mathbb U$, i.e., 
\be
{\rm spect}[\mathbb U]= \{e^{i \varphi_j};\,j=1,2\ldots,2^{2L}\}\,,
\ee
where the  \emph{quasienergies} $\varphi_j$ can be taken in $[0,2\pi)$. To this end we compute the spectral form factor (SFF)
\begin{equation} 
\label{eq:SFF}
    K(t,L) = \mathbb E\left[\sum_{j,j'=1}^{2^{2L}} e^{i (\varphi_j-\varphi_{j'}) t}\right],
\end{equation}
which measures spectral correlations over arbitrary distance and, over the last few years, has emerged as the standard spectral-correlation measure in extended systems, see, e.g., Refs.~\cite{kos2018many, chan2018spectral, bertini2018exact, friedman2019spectral, suntajs2020quantum, bertini2021random, moldgalya2021spectral, suntajs2021spectral, garratt2021local, garratt2021manybody, bertini2022exactspectral, shivam2023many, chan2022many, suntajs2023localization}. Here $\mathbb{E}[\cdot]$ denotes an expectation value over an ensemble of similar systems, which we conveniently generate by taking the local gates $v_x, u_x$ in Eq.~\eqref{eq:localgates} to be independent and identically-distributed random matrices (equivalently one can take the angles $\{\boldsymbol \theta_x, \boldsymbol \eta_x\}$ to be i.i.d.\ random). The specific distribution of $\{v_x, u_x\}$ is irrelevant for the discussion below, and we will make a concrete choice when discussing the parameterisation used in our numerical studies (cf. Sec.~\ref{sec:numerics}). Introducing the average allows us to filter out the system-specific details and obtain a universal result which is expected to only depend on gross properties of the system. Note that the SFF is not self averaging~\cite{prange1997the}, therefore without disorder averaging one should not expect universal results (see, e.g., Refs.~\cite{pineda2007universal}).
We expect our results to apply for any finite --- no matter how small --- magnitude of disorder in the thermodynamic limit.

Specifically, in ergodic systems  the SFF is expected to take a universal form that coincides with that observed in random matrices of the same size. The specific random-matrix ensemble to compare with depends on the anti-unitary symmetries of the Floquet operator (e.g.\ time reversal symmetry). In the generic case of no anti-unitary symmetries, which is the one of interest here, the relevant prediction is that of Dyson's {\em Circular Unitary Ensemble} (CUE), which reads as~\cite{mehta1991random,haake2001quantum} 
\begin{align}
K_{\rm CUE}(t,L) &= {\rm min}(t, 2^{2L})\,,
\label{eq:CUEform}
\end{align}
showing a characteristic ramp-like shape. On the other hand, whenever the system is strongly non-ergodic (e.g.\ integrable, or localised) the energy levels are expected to be statistically independent. This means that the SFF should reproduce the Poissonian-distribution result
\be
K_{\rm Poisson}(t,L) = 2^{2L}\,.
\ee

\begin{figure}
\begin{tikzpicture}[baseline=(current  bounding  box.center), scale=0.65]
\def\shiftX{10}
\def\shiftXX{20}
\foreach \i in {1,...,5}{
\draw[thick,  opacity=0.7, dashed] (2*\i+2-12.5+0.255,-1.75-0.1) -- (2*\i+2-12.5+0.255,4.25-0.1);
\draw[thick,  opacity=0.7, dashed] (2*\i+2-11.5-0.255,-1.75-0.1) -- (2*\i+2-11.5-0.255,4.25-0.1);}

\foreach \i in {1,...,5}{
\draw[thick] (2*\i+2-11.5,4) arc (-45:175:0.15);
\draw[thick] (2*\i+2-11.5,-2) arc (315:180:0.15);
\draw[thick] (2*\i+2-0.5-12,-2) arc (-135:0:0.15);
}
\foreach \i in {2,...,6}{
\draw[thick] (2*\i+2-2.5-12,4) arc (225:0:0.15);
}
\foreach \i in {0,1,2}{
\draw[thick,  opacity=0.7, dashed] (-9.5,2*\i-1.745) -- (0.4,2*\i-1.745);
\draw[thick,  opacity=0.7, dashed] (-9.5,2*\i-1.255) -- (0.4,2*\i-1.255);
}
\foreach \i in{1.5,2.5,3.5}{
\draw[thick] (0.5,2*\i-0.5-3.5) arc (45:-90:0.15);
\draw[thick] (-10+0.5+0,2*\i-0.5-3.5) arc (45:270:0.15);
}
\foreach \i in{0.5,1.5,2.5}
{
\draw[ thick] (0.5,1+2*\i-0.5-3.5) arc (-45:90:0.15);
\draw[ thick] (-10+0.5,1+2*\i-0.5-3.5) arc (315:90:0.15);
}
\foreach \jj[evaluate=\jj as \j using -2*(ceil(\jj/2)-\jj/2)] in {-1,-3,-5}{
\foreach \i in {1,...,5}
{
\draw[thick] (.5-2*\i-1*\j,-2-1*\jj) -- (1-2*\i-1*\j,-1.5-\jj);
\draw[thick] (1-2*\i-1*\j,-1.5-1*\jj) -- (1.5-2*\i-1*\j,-2-\jj);
\draw[thick] (.5-2*\i-1*\j,-1-1*\jj) -- (1-2*\i-1*\j,-1.5-\jj);
\draw[thick] (1-2*\i-1*\j,-1.5-1*\jj) -- (1.5-2*\i-1*\j,-1-\jj);
\draw[thick, fill=myorange, rounded corners=2pt] (0.75-2*\i-1*\j,-1.75-\jj) rectangle (1.25-2*\i-1*\j,-1.25-\jj);
\draw[thick] (-2*\i+2,-1.35-\jj) -- (-2*\i+2.15,-1.35-\jj) -- (-2*\i+2.15,-1.5-\jj);%
}
}
\foreach \jj[evaluate=\jj as \j using -2*(ceil(\jj/2)-\jj/2)] in {-4,-2,0}{
\foreach \i in {1,...,5}
{
\draw[thick] (.5-2*\i-1*\j,-2-1*\jj) -- (1-2*\i-1*\j,-1.5-\jj);
\draw[thick] (1-2*\i-1*\j,-1.5-1*\jj) -- (1.5-2*\i-1*\j,-2-\jj);
\draw[thick] (.5-2*\i-1*\j,-1-1*\jj) -- (1-2*\i-1*\j,-1.5-\jj);
\draw[thick] (1-2*\i-1*\j,-1.5-1*\jj) -- (1.5-2*\i-1*\j,-1-\jj);
\draw[thick, fill=myred, rounded corners=2pt] (0.75-2*\i-1*\j,-1.75-\jj) rectangle (1.25-2*\i-1*\j,-1.25-\jj);
\draw[thick] (-2*\i+1,-1.35-\jj) -- (-2*\i+1.15,-1.35-\jj) -- (-2*\i+1.15,-1.5-\jj);%
}
}
\foreach \jj in {0,2,4}{
\draw[ thick, fill=myblue1, rounded corners=2pt] (0.5,-1+\jj) circle (.15);
\draw[ thick, fill=myblue2, rounded corners=2pt] (-0.5,-1+\jj) circle (.15);
\draw[ thick, fill=myblue3, rounded corners=2pt] (-1.5,-1+\jj) circle (.15);
\draw[ thick, fill=myblue4, rounded corners=2pt] (-2.5,-1+\jj) circle (.15);
\draw[ thick, fill=myblue5, rounded corners=2pt] (-3.5,-1+\jj) circle (.15);
\draw[ thick, fill=myblue6, rounded corners=2pt] (-4.5,-1+\jj) circle (.15);
\draw[ thick, fill=myblue7, rounded corners=2pt] (-5.5,-1+\jj) circle (.15);
\draw[ thick, fill=myblue8, rounded corners=2pt] (-6.5,-1+\jj) circle (.15);
\draw[ thick, fill=myblue9, rounded corners=2pt] (-7.5,-1+\jj) circle (.15);
\draw[ thick, fill=myblue10, rounded corners=2pt] (-8.5,-1+\jj) circle (.15);
\draw[ thick, fill=myblue8, rounded corners=2pt] (0.5,\jj) circle (.15);
\draw[ thick, fill=myblue9, rounded corners=2pt] (-0.5,\jj) circle (.15);
\draw[ thick, fill=myblue1, rounded corners=2pt] (-1.5,\jj) circle (.15);
\draw[ thick, fill=myblue3, rounded corners=2pt] (-2.5,\jj) circle (.15);
\draw[ thick, fill=myblue2, rounded corners=2pt] (-3.5,\jj) circle (.15);
\draw[ thick, fill=myblue4, rounded corners=2pt] (-4.5,\jj) circle (.15);
\draw[ thick, fill=myblue10, rounded corners=2pt] (-5.5,\jj) circle (.15);
\draw[ thick, fill=myblue9, rounded corners=2pt] (-6.5,\jj) circle (.15);
\draw[ thick, fill=myblue8, rounded corners=2pt] (-7.5,\jj) circle (.15);
\draw[ thick, fill=myblue7, rounded corners=2pt] (-8.5,\jj) circle (.15);
}
\foreach \jj in {0,-2,-4}{
\foreach \i in {1,...,5}{
\draw[thick] (-2*\i+1.4,-.95-\jj) -- (-2*\i+1.55,-.95-\jj) -- (-2*\i+1.55,-1.1-\jj);}
\foreach \i in {1,...,5}{
\draw[thick] (-2*\i+2.4,0.05-\jj) -- (-2*\i+2.55,0.05-\jj) -- (-2*\i+2.55,-0.1-\jj);}
\foreach \i in {1,...,5}{
\draw[thick]  (-2*\i+1.45,-0.1-\jj) -- (-2*\i+1.45,0.05-\jj) -- (-2*\i+1.6,0.05-\jj);}
\foreach \i in {0,...,4}{
\draw[thick]  (-2*\i+.45,-1.1-\jj) -- (-2*\i+.45,-0.95-\jj) -- (-2*\i+.6,-0.95-\jj);}
}
\def\shiftx{0}
\def\shifty{-9}
\foreach \i in {1,...,5}{
\draw[ thick,  opacity=0.7, dashed] (2*\i+2-1.485+0.25+\shiftx-11,-2.5-0.1+\shifty+2.5) -- (2*\i+2-1.485+0.25+\shiftx-11,3.5-0.1+\shifty+2.5);
\draw[ thick,  opacity=0.7, dashed] (2*\i+2-0.525-0.25+\shiftx-11,-2.5-0.1+\shifty+2.5) -- (2*\i+2-0.525-0.25+\shiftx-11,3.5-0.1+\shifty+2.5);
}
\foreach \i in {0,1,2}{
\draw[ thick,  opacity=0.7, dashed] (1.75+\shiftx-11,2*\i-1.25+\shifty+2.5) -- (11.5+\shiftx-11,2*\i-1.25+\shifty+2.5);
\draw[ thick,  opacity=0.7, dashed] (1.5+\shiftx-11,2*\i-.76+\shifty+2.5) -- (11.5+\shiftx-11,2*\i-.76+\shifty+2.5);
}
\foreach \i in {1,...,5}
{
\draw[ thick] (2*\i+2-1.5-11+\shiftx,3.5+\shifty+2.5) arc (135:-0:0.15);
\draw[ thick] (2*\i+2-.5-11+\shiftx,3.5+\shifty+2.5) arc (-325:-180:0.15);
\draw[ thick] (2*\i+2-1.5-11+\shiftx,-2.5+\shifty+2.5) arc (-45:180:-0.15);
\draw[ thick] (2*\i+2-0.5-11+\shiftx,-2.5+\shifty+2.5) arc (45:-180:0.15);
}
\foreach \i in {3,...,5}
{
\draw[ thick] (\shiftx+.5,2*\i-0.5-3.5+\shifty) arc (45:-90:0.15);
\draw[ thick] (\shiftx-10+0.5+0,2*\i-0.5-3.5+\shifty) arc (45:280:0.15);
}
\foreach \i in{2,...,4}
{
\draw[ thick] (\shiftx+.5,1+2*\i-0.5-3.5+\shifty) arc (-45:90:0.15);
\draw[ thick] (\shiftx-10+0.5,1+2*\i-0.5-3.5+\shifty) arc (-45:-280:0.15);
}
\foreach \i in {1,...,5}
{
\draw[ thick] (\shiftx+.5-2*\i,6+\shifty) -- (\shiftx+1-2*\i,5.5+\shifty);
\draw[ thick] (\shiftx+1.5-2*\i,6+\shifty) -- (\shiftx+1-2*\i,5.5+\shifty);
}
\foreach \jj[evaluate=\jj as \j using -2*(ceil(\jj/2)-\jj/2)] in {0,...,3}
\foreach \i in {1,...,5}
{
\draw[ thick] (\shiftx+.5-2*\i-1*\j,2+1*\jj+\shifty) -- (\shiftx+1-2*\i-1*\j,1.5+\jj+\shifty);
\draw[ thick] (\shiftx+1-2*\i-1*\j,1.5+1*\jj+\shifty) -- (\shiftx+1.5-2*\i-1*\j,2+\jj+\shifty);
}
\foreach \i in {0,...,4}
{
\draw[ thick] (\shiftx-.5-2*\i,1+\shifty) -- (\shiftx+0.5-2*\i,0+\shifty);
\draw[ thick] (\shiftx-0.5-2*\i,0+\shifty) -- (\shiftx+0.5-2*\i,1+\shifty);
\draw[ thick, fill=mygreen, rounded corners=2pt] (\shiftx-0.25-2*\i,0.25+\shifty) rectangle (\shiftx+.25-2*\i,0.75+\shifty);
\draw[thick] (\shiftx-2*\i,0.65+\shifty) -- (\shiftx+.15-2*\i,.65+\shifty) -- (\shiftx+.15-2*\i,0.5+\shifty);
}
\foreach \jj[evaluate=\jj as \j using -2*(ceil(\jj/2)-\jj/2)] in {-1,1,3}
\foreach \i in {1,...,5}
{
\draw[ thick] (\shiftx+.5-2*\i-1*\j,1+1*\jj+\shifty) -- (\shiftx+1-2*\i-1*\j,1.5+\jj+\shifty);
\draw[ thick] (\shiftx+1-2*\i-1*\j,1.5+1*\jj+\shifty) -- (\shiftx+1.5-2*\i-1*\j,1+\jj+\shifty);
\draw[ thick, fill=myblue4, rounded corners=2pt] (\shiftx+0.75-2*\i-1*\j,1.75+\jj+\shifty) rectangle (\shiftx+1.25-2*\i-1*\j,1.25+\jj+\shifty);
\draw[thick] (\shiftx+1-2*\i-1*\j,1.65+1*\jj+\shifty) -- (\shiftx+1.15-2*\i-1*\j,1.65+1*\jj+\shifty) -- (\shiftx+1.15-2*\i-1*\j,1.5+1*\jj+\shifty);
}
\foreach \jj[evaluate=\jj as \j using -2*(ceil(\jj/2)-\jj/2)] in {0,2,4}{
\foreach \i in {1,...,5}
{
\draw[ thick] (\shiftx+.5-2*\i-1*\j,1+1*\jj+\shifty) -- (\shiftx+1-2*\i-1*\j,1.5+\jj+\shifty);
\draw[ thick] (\shiftx+1-2*\i-1*\j,1.5+1*\jj+\shifty) -- (\shiftx+1.5-2*\i-1*\j,1+\jj+\shifty);
\draw[ thick, fill=myblue, rounded corners=2pt] (\shiftx+0.75-2*\i-1*\j,1.75+\jj+\shifty) rectangle (\shiftx+1.25-2*\i-1*\j,1.25+\jj+\shifty);
\draw[thick] (\shiftx+1-2*\i-1*\j,1.65+1*\jj+\shifty) -- (\shiftx+1.15-2*\i-1*\j,1.65+1*\jj+\shifty) -- (\shiftx+1.15-2*\i-1*\j,1.5+1*\jj+\shifty);
}}
\foreach \jj in {0.5,2.5,4.5}{
\draw[ thick, fill=mygray1, rounded corners=2pt] (.5+\shiftx,-.5+\jj+\shifty) circle (.15);
\draw[ thick, fill=mygray4, rounded corners=2pt] (-.5+\shiftx,-.5+\jj+\shifty) circle (.15);
\draw[ thick, fill=mygray3, rounded corners=2pt] (-1.5+\shiftx,-.5+\jj+\shifty) circle (.15);
\draw[ thick, fill=mygray2, rounded corners=2pt] (-2.5+\shiftx,-.5+\jj+\shifty) circle (.15);
\draw[ thick, fill=mygray1, rounded corners=2pt] (-3.5+\shiftx,-.5+\jj+\shifty) circle (.15);
\draw[ thick, fill=mygray5, rounded corners=2pt] (-4.5+\shiftx,-.5+\jj+\shifty) circle (.15);
\draw[ thick, fill=mygray4, rounded corners=2pt] (-5.5+\shiftx,-.5+\jj+\shifty) circle (.15);
\draw[ thick, fill=mygray8, rounded corners=2pt] (-6.5+\shiftx,-.5+\jj+\shifty) circle (.15);
\draw[ thick, fill=mygray7, rounded corners=2pt] (-7.5+\shiftx,-.5+\jj+\shifty) circle (.15);
\draw[ thick, fill=mygray6, rounded corners=2pt] (-8.5+\shiftx,-.5+\jj+\shifty) circle (.15);
}
\foreach \jj in {1.5,3.5,5.5}
{
\draw[ thick, fill=mygray3, rounded corners=2pt] (.5+\shiftx,-.5+\jj+\shifty) circle (.15);
\draw[ thick, fill=mygray6, rounded corners=2pt] (-.5+\shiftx,-.5+\jj+\shifty) circle (.15);
\draw[ thick, fill=mygray1, rounded corners=2pt] (-1.5+\shiftx,-.5+\jj+\shifty) circle (.15);
\draw[ thick, fill=mygray2, rounded corners=2pt] (-2.5+\shiftx,-.5+\jj+\shifty) circle (.15);
\draw[ thick, fill=mygray3, rounded corners=2pt] (-3.5+\shiftx,-.5+\jj+\shifty) circle (.15);
\draw[ thick, fill=mygray4, rounded corners=2pt] (-4.5+\shiftx,-.5+\jj+\shifty) circle (.15);
\draw[ thick, fill=mygray5, rounded corners=2pt] (-5.5+\shiftx,-.5+\jj+\shifty) circle (.15);
\draw[ thick, fill=mygray6, rounded corners=2pt] (-6.5+\shiftx,-.5+\jj+\shifty) circle (.15);
\draw[ thick, fill=mygray7, rounded corners=2pt] (-7.5+\shiftx,-.5+\jj+\shifty) circle (.15);
\draw[ thick, fill=mygray8, rounded corners=2pt] (-8.5+\shiftx,-.5+\jj+\shifty) circle (.15);
}
\foreach \jj in {0,-2,-4}{
\foreach \i in {1,...,5}{
\draw[thick] (-2*\i+1.4+\shiftx,-.95-\jj-0.5+\shifty+1.5) -- (-2*\i+1.55+\shiftx,-.95-\jj-0.5+\shifty+1.5) -- (-2*\i+1.55+\shiftx,-1.1-\jj-0.5+\shifty+1.5);}
\foreach \i in {1,...,5}{
\draw[thick] (-2*\i+2.4+\shiftx,0.05-\jj-0.5+\shifty+1.5) -- (-2*\i+2.55+\shiftx,0.05-\jj-0.5+\shifty+1.5) -- (-2*\i+2.55+\shiftx,-0.1-\jj-0.5+\shifty+1.5);}
\foreach \i in {1,...,5}{
\draw[thick]  (-2*\i+1.45+\shiftx,-0.1-\jj-0.5+\shifty+1.5) -- (-2*\i+1.45+\shiftx,0.05-\jj-0.5+\shifty+1.5) -- (-2*\i+1.6+\shiftx,0.05-\jj-0.5+\shifty+1.5);}
\foreach \i in {0,...,4}{
\draw[thick]  (-2*\i+.45+\shiftx,-1.1-\jj-0.5+\shifty+1.5) -- (-2*\i+.45+\shiftx,-0.95-\jj-0.5+\shifty+1.5) -- (-2*\i+.6+\shiftx,-0.95-\jj-0.5+\shifty+1.5);}
}

\draw[thin, rounded corners=2pt, fill= gray, opacity=0.5] (1-4.3,4.4) rectangle (1+2-4.3,\shifty-0.4);

\draw (-2.3,5.4) node {$\mathcal T_{x}$};

\draw (-2.3,4.8) node[rotate=90] {$=$};

\end{tikzpicture}
\caption{Graphical representation of $K(t,L)$ (cf.\ Eq.~\eqref{eq:SFFU}). The symbols represent the local gates as per Eqs.~\eqref{eq:diaggates}--\eqref{eq:diaggatesdag}. Random one-site gates along the same column coincide, while those on different columns are uncorrelated.}
\label{fig:SFF}
\end{figure}
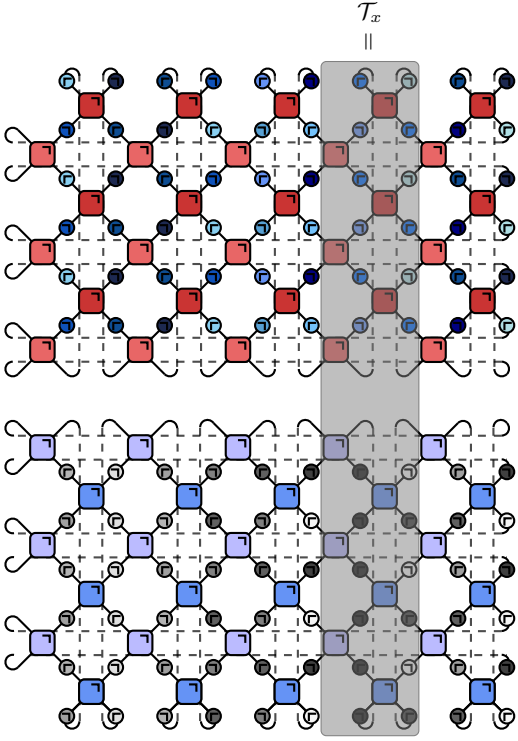

As discussed in Refs.~\cite{bertini2018exact, bertini2021random} (see also Refs.~\cite{garratt2021local, garratt2021manybody, shivam2023many} where this approach is applied to non-dual-unitary systems), the spectral form factor of a quantum circuit can be rewritten in terms of the trace of the $L$-th power of a transfer matrix acting along the space direction. The idea is to exploit the symmetry of the diagram in Fig.~\ref{fig:quantumcircuit} under a 90\textdegree\ degree rotation (space-time swap), and the fact that the disorder is uncorrelated in space. The main steps go as follows. First we observe that, for $t>0$, the SFF in Eq.~\eqref{eq:SFF} can be written as 
\be
\label{eq:SFFU}
K(t,L) = \mathbb{E} \left[ \left| \tr \left( \mathbb{U}^t \right) \right|^2 \right] =  \mathbb{E} \left[ \tr \left( \mathbb{U}\otimes \mathbb{U}^* \right)^t \right]. 
\ee
Using now the parameterisation (\ref{eq:JTI}, \ref{eq:gatesVW}) we can depict the quantity inside the average on the r.h.s.\ as in Fig.~\ref{fig:SFF} where we introduced the symbols 
\begin{align}
 V & = 
\begin{tikzpicture}[baseline=(current  bounding  box.center), scale=.7]
\draw[ thick] (-4.25,0.5) -- (-3.25,-0.5);
\draw[ thick] (-4.25,-0.5) -- (-3.25,0.5);
\draw[ thick, fill=myred, rounded corners=2pt] (-4,0.25) rectangle (-3.5,-0.25);
\draw[thick] (-3.75,0.15) -- (-3.75+0.15,0.15) -- (-3.75+0.15,0);
\Text[x=-4.25,y=-0.75]{}
\end{tikzpicture}\,,
& W &=\begin{tikzpicture}[baseline=(current  bounding  box.center), scale=.7]
\draw[ thick] (-4.25,0.5) -- (-3.25,-0.5);
\draw[ thick] (-4.25,-0.5) -- (-3.25,0.5);
\draw[ thick, fill=myorange, rounded corners=2pt] (-4,0.25) rectangle (-3.5,-0.25);
\draw[thick] (-3.75,0.15) -- (-3.75+0.15,0.15) -- (-3.75+0.15,0);
\Text[x=-4.25,y=-0.75]{}
\end{tikzpicture}\,,
&
u_{x}, w_{x}=
\begin{tikzpicture}[baseline=(current  bounding  box.center), scale=.7]
\draw[ thick] (-4.25,0.5) -- (-4.25,-0.5);
\draw[ thick, fill=myblue4, rounded corners=2pt] (-4.25,0) circle (.15);
\draw[thick, rotate around = {-45:(0.525-4.77,0.375-0.4)}]  (.45-4.77,0.3-0.4) -- (.45-4.77,0.45-0.4) -- (.6-4.77,0.45-0.4);
\Text[x=-4.25,y=-0.75]{}
\end{tikzpicture}\,,\label{eq:diaggates}\\
 V^\dag & = 
\begin{tikzpicture}[baseline=(current  bounding  box.center), scale=.7]
\draw[ thick] (-4.25,0.5) -- (-3.25,-0.5);
\draw[ thick] (-4.25,-0.5) -- (-3.25,0.5);
\draw[ thick, fill=myblue, rounded corners=2pt] (-4,0.25) rectangle (-3.5,-0.25);
\draw[thick] (-3.75,0.15) -- (-3.75+0.15,0.15) -- (-3.75+0.15,0);
\Text[x=-4.25,y=-0.75]{}
\end{tikzpicture}\,,
& W^\dag &=\begin{tikzpicture}[baseline=(current  bounding  box.center), scale=.7]
\draw[ thick] (-4.25,0.5) -- (-3.25,-0.5);
\draw[ thick] (-4.25,-0.5) -- (-3.25,0.5);
\draw[ thick, fill=myblue4, rounded corners=2pt] (-4,0.25) rectangle (-3.5,-0.25);
\draw[thick] (-3.75,0.15) -- (-3.75+0.15,0.15) -- (-3.75+0.15,0);
\Text[x=-4.25,y=-0.75]{}
\end{tikzpicture}\,,
&
u_{x}^\dag, w_{x}^\dag=
\begin{tikzpicture}[baseline=(current  bounding  box.center), scale=.7]
\draw[ thick] (-4.25,0.5) -- (-4.25,-0.5);
\draw[ thick, fill=mygray4, rounded corners=2pt] (-4.25,0) circle (.15);
\draw[thick, rotate around = {-45:(0.525-4.77,0.375-0.4)}]  (.45-4.77,0.3-0.4) -- (.45-4.77,0.45-0.4) -- (.6-4.77,0.45-0.4);
\Text[x=-4.25,y=-0.75]{}
\end{tikzpicture}\,,\label{eq:diaggatesdag}
\end{align}
and random one-site gates of the same shade of colour are the same. As it is clear from the figure, this quantity can be equivalently represented as the trace of a matrix acting on the vertical lattice (of $2t$ qubits). 

More precisely, we denote by $\tilde{\mathbb V} $ and $\tilde{\mathbb W}$  the many-body operators composed by vertical columns of gates acting on the time lattice, i.e.,
\begin{align}
\tilde{\mathbb V} &\equiv \tilde V^{\otimes t},&
\tilde{\mathbb W} &\equiv \tilde W^{\otimes t}\,,
\label{eq:Utildeo}
\end{align}
where $\tilde V$ and $\tilde W$ are obtained by reshuffling the indices of the local gates to propagate from left to right. Namely, the are obtained from $V$ and $W$ through the mapping $\,\tilde{(\cdot )}\,:{\rm End}(\mathbb C^{2}\otimes \mathbb C^{2}) \rightarrow {\rm End}(\mathbb C^{2}\otimes \mathbb C^{2})$ as 
\be
\tilde{O}_{ki;lj} := O_{ij,kl},\qquad
i,j,k,l\in\{0,1\}\,.
\label{eq:duallocalgate}
\ee
Using these objects we define the transfer matrix 
\begin{align}
\mathcal T_{x} = & (\tilde{\mathbb V} \!\otimes_r\! \tilde{\mathbb V}^*) ((u_x \otimes\1)^{\otimes t}\otimes_r (u^*_x \otimes\1)^{\otimes t})\notag\\
&((\1\otimes v^T_x )^{\otimes t}\otimes_r (\1\otimes v^\dag_x)^{\otimes t})  (\Pi_{2t}\!\otimes_r\! \Pi^*_{2t})(\tilde{\mathbb W} \!\otimes_r\!\tilde{\mathbb W}^*)  \notag\\
&(\Pi^{\dag}_{2t}\!\otimes_r\! \Pi^{T}_{2t}) ((u^T_{x+1/2} \!\otimes\!\1)^{\otimes t}\!\otimes_r\! (u^\dag_{x+1/2} \!\otimes\!\1)^{\otimes t})\label{eq:TMnonave}\\
&((\1\otimes v_{x+1/2} )^{\otimes t}\otimes_r (\1\otimes v^*_{x+1/2})^{\otimes t}),\notag
\end{align}
which is highlighted in the shaded box of Fig.~\ref{fig:SFF}. Here $(\cdot)^T$ denotes transposition, $\Pi_{n}$ is the one-site shift operator in a chain of $n$ qubits, and the tensor product $\otimes_r$ combines operators acting in the forward evolving time-sheet with those acting on the backward evolving one (cf.\ Eq.~\eqref{eq:SFF}): from now on we will refer to the lattices over which these operators act respectively as the \emph{forward} and \emph{backward} time lattice. 

Using the transfer matrix in Eq.~\eqref{eq:TMnonave} we can express the SFF as 
\be
\label{eq:SFFaveout}
K(t,L) =  \mathbb{E} \left[ \tr \left( \mathcal T_{1}\cdots \mathcal T_{L} \right)\right].
\ee
Recalling now that random gates at different positions (i.e.\ those along different columns in the figure) are uncorrelated, we can bring the average $\mathbb{E} \left[\cdot\right]$ inside the trace to obtain  
\begin{equation} 
\label{eq:transfermat}
    K(t,L) = \tr \mathcal{T}^L,\qquad \mathcal T \equiv  \mathbb{E} \left[\mathcal T_{x}\right]\,.
\end{equation}
Explicitly, we have 
\begin{equation}
\mathcal{T} \!=\!(\tilde{\mathbb V} \!\otimes_r\! \tilde{\mathbb V}^*) \mathcal{O}^\dag  (\Pi_{2t}\!\otimes_r\! \Pi^*_{2t}) (\tilde{\mathbb W} \!\otimes_r\!\tilde{\mathbb W}^*)  (\Pi^{-1}_{2t}\!\otimes_r\! \Pi^{T}_{2t}) \mathcal{O}^{\phantom{\dag}}\!\!\!,
\label{eq:SFFTM2}
\end{equation}
where $\mathcal O$ is a non-expanding map implementing the average over the local disorder~\footnote{The average of a set of unitary operators is non-expanding.}, i.e., 
\be
\begin{aligned}
\mathcal{O} & \equiv {\mathbb E}[ (u_x \otimes\1)^{\otimes t}\!\otimes_r\! (u^*_x \otimes\1)^{\otimes t}] \\
&\,\times  {\mathbb E}[ (\1\otimes v^T_x )^{\otimes t}\!\otimes_r\! (\1\otimes v^\dag_x)^{\otimes t}].
\end{aligned}
\ee

Note that, although we cannot generically prove that $\mathcal T$ is diagonalisable, Eq.~\eqref{eq:transfermat} implies that the SFF is solely determined by its spectrum and the size of its Jordan blocks. To see this we write the Jordan decomposition of $\mathcal T$ as follows 
\be
\label{eq:Jordandec}
\mathcal T = \mathcal{R} (\mathcal{D}+\mathcal{K}) \mathcal{R}^{-1},
\ee
where $\mathcal{R}$ is invertible, $\mathcal{D}$ is diagonal, and $\mathcal{K}$ is strictly upper triangular with zeros on the diagonal.
The eigenvalues of $\mathcal{D}$ coincide with those of $\mathcal{T}$ and the degeneracy of a given eigenvalue $\lambda$ is given by 
\be
d_\lambda = \sum_{j=1}^{N_\lambda} {\rm dim}(J_{j, \lambda}),
\ee 
where $j$ labels all the Jordan blocks corresponding to $\lambda$ (their total number is $N_\lambda$), designated as $J_{j, \lambda}$, while ${\rm dim}(A)$ denotes the dimension of the matrix $A$. Plugging the decomposition \eqref{eq:Jordandec} into \eqref{eq:transfermat} we find 
\be\label{eq:sff_eigen}
K(t,L) = \tr \mathcal{D}^L
\ee
where we used that products of $\mathcal D$ and at least one $\mathcal K$ are traceless (using fact $\mathcal K$ strictly upper diagonal).

A notable property of $\mathcal{T}$ is that it has a global $\mathbb{Z}_{t}\times \mathbb{Z}_{t}$ symmetry under independent two-site translations in the forward and backward lattices, i.e.
\begin{equation}
\label{eq:doubletranslation}
    [\Pi^{2\tau_1}_{2t} \otimes \Pi^{2\tau_2}_{2t} ,\mathcal{T}] = 0,\qquad \tau_1,\tau_2=0,\ldots,t-1.
\end{equation}
As a result, we can block-diagonalise it by considering a fixed double-momentum sector labelled by $(\nu,\nu')$, with $\nu,\nu'=0,\ldots,t-1$. Therefore, the eigenvalues of $\mathcal T$ (or $\mathcal{D}$) can be labelled as 
\be
\lambda_{a,{(\nu,\nu')}},\quad \nu,\nu'=0,\ldots,t-1, 
\ee 
where $a=0,\ldots, N_{{(\nu,\nu')}}-1$ with $N_{{(\nu,\nu')}}$ the size of the $(\nu,\nu')$ sector. Noting that the projector onto the sector $(\nu,\nu')$ can be written as $Y^{(\nu)}\otimes Y^{(\nu')}$, where we introduced  
\be
\label{eq:Yproj}
Y^{(\nu)} = \frac{1}{t} \sum_{\tau=0}^{t-1} \exp\left(\frac{2\pi i \tau \nu}{t}\right) \Pi_{2t}^{2\tau},
\ee 
such that $Y^{(\nu)}Y^{(\nu')} = Y^{(\nu')}Y^{(\nu)}= \delta_{\nu,\nu'}Y^{(\nu)}$, we find  
\be
N_{(\nu,\nu')}= \tr[\smash{Y^{(\nu)}}]\tr[\smash{Y^{(\nu')}}]\,.
\ee
Ref.~\cite{bertini2021random} proved that in the dual unitary limit of the models considered here and away from the trivial non-interacting point (specifically for $J'_{1,2}=J_{1,2}=\pi/4$ and $J'_{3},J_{3}\neq\pi/4$ in Eq.~\eqref{eq:gatesVW}), the transfer matrix $\mathcal{T}$ has exactly $t$ eigenvalues $\lambda = 1$, corresponding to one-dimensional Jordan blocks, while all other eigenvalues are contained in a disk of radius $r < 1$. In fact, there is exactly one maximal-magnitude eigenvalue in each diagonal sector $(\nu,\nu)$, and their corresponding eigenvectors read as 
\be
\ket{1_{{(\nu,\nu)}}}=\ket*{Y^{(\nu)}}, \qquad \nu=0,\ldots,t-1. 
\label{eq:eigs}
\ee
In the r.h.s.\ of Eq.~\eqref{eq:eigs} we represented the operator in Eq.~\eqref{eq:Yproj} as a state of a Hilbert space with doubled dimension using to the operator-to-state mapping $(\mathbb C^2)^{\otimes 2t} \otimes_r (\mathbb C^2)^{\otimes 2t} \ni \ket{A} \mapsto A \in {\rm End}((\mathbb C^2)^{\otimes 2t})$ such that 
\be
\label{eq:mapping}
\braket{i_1\cdots i_{2t} j_1\cdots j_{2t}}{A}= 
 \bra{i_1\cdots i_{2t}}A
\ket{j_1\cdots j_{2t}}\,.
\ee 
The presence of $t$ dominant eigenvalues with unit magnitude allows one to simply show that $K(t,L)$ is indeed described by the CUE form in Eq.~\eqref{eq:CUEform} for large enough $L$. As pointed out in Ref.~ \cite{garratt2021manybody}, this $t$-fold degeneracy of $\mathcal T$, and the fact the eigenvectors preserve only the diagonal part of its $\mathbb{Z}_{t}\times \mathbb{Z}_{t}$ symmetry, indicates that the ramp in the spectral form factor is a manifestation of a spontaneous breaking of symmetry $\mathbb{Z}_{t}\times \mathbb{Z}_{t}$ to $\mathbb{Z}_{t}$. The goal of this paper is to understand if and why this spontaneous symmetry breaking is stable as one moves away from the dual unitary point, from the perspective of perturbation theory.

\subsection{Minimal Example}
\label{sec:minexample}

\begin{figure}[t]
\centering
\includegraphics[width=0.9\linewidth]{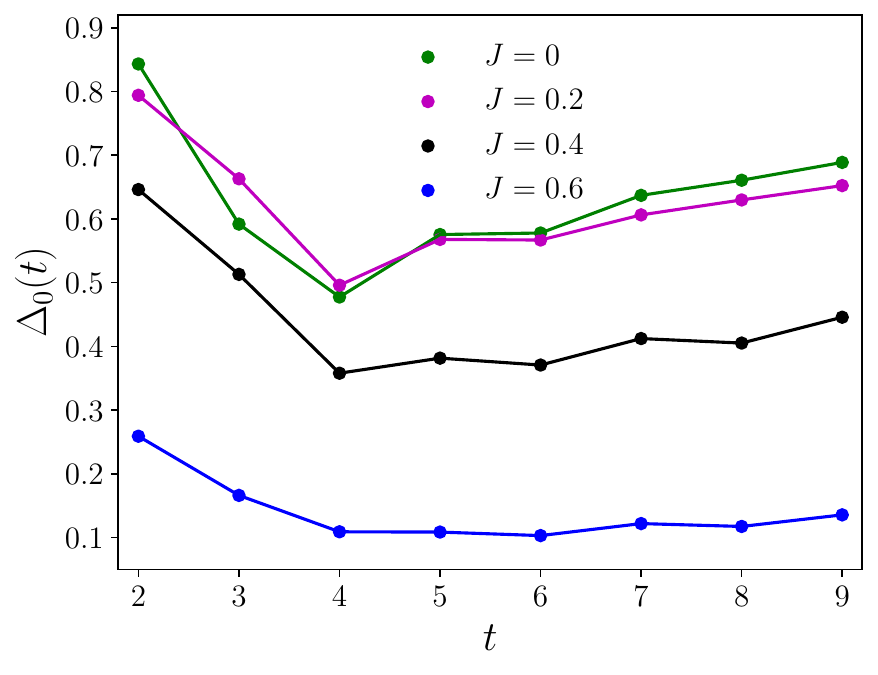}
\includegraphics[width=0.9\linewidth]{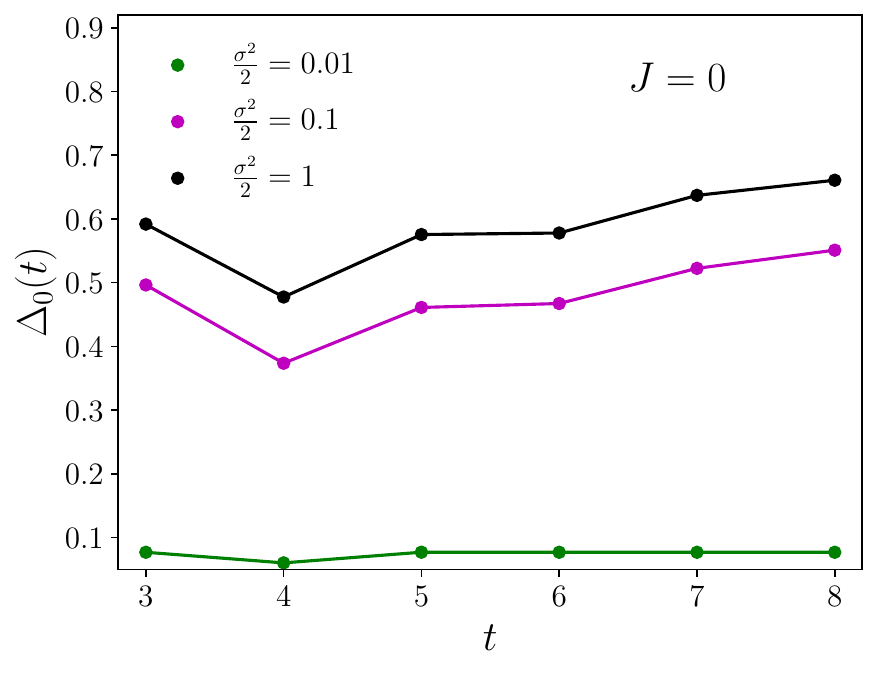}
\caption{The gap $\Delta_0(t) = 1-|\lambda_1|$ at the dual unitary point for various $t$ and $J$, all data is collected from $(\nu,\nu') = (0,0)$ symmetry sector. The top plot takes $\alpha \to \infty$ as in Eq. \eqref{eq:disorder} while the bottom plot sets $J = 0$ and takes $\sigma$ finite. The bottom plot is the only data set in this article where $\sigma$ is taken to be finite, all other datasets and figures take $\sigma\to \infty$.
The data showcased in the top plot with $J = 0$ is featured in Fig. \ref{fig:betalin}.}
\label{fig:dualgap}
\end{figure}

Even though our theoretical analysis can be carried out for the full family of circuits,
Eqs.~\eqref{eq:diaggates}--\eqref{eq:diaggatesdag},
 for our numerical investigations it is useful to fix some of the parameters and consider a minimal toy model example. Specifically, we consider
\begin{equation} 
\label{eq:model}
V=W \equiv U= e^{i\left( (\tfrac{\pi}{4}-\epsilon_1)\sigma^{(1)} \otimes \sigma^{(1)} + (\tfrac{\pi}{4}-\epsilon_2) \sigma^{(2)} \otimes\sigma^{(2)}\right)}, 
\end{equation} 
set $\theta^{(2)}_x = \phi^{(2)}_x = 0$, and average over $\theta^{(1)}_x, \theta^{(3)}_x, \phi^{(1)}_x, \phi^{(3)}_x$ with a Gaussian measure with zero mean and infinite variance, i.e., maximal disorder strength. This gives 
\begin{equation}
\label{eq:model2}
    \mathcal{O} = \mathcal{O}_{0}^{(1)} \mathcal{O}_{1/2}^{(1)}\mathcal{O}_{0}^{(3)}\mathcal{O}_{1/2}^{(3)}, 
\end{equation}
where we set 
\begin{equation} \label{eq:disorder}
\!\!\!\mathcal{O}_{s}^{(\alpha)} \!=\! \lim_{\sigma\to\infty} \!\!\exp\!\!\left[- \frac{\sigma^2}{2} \!\!\left(\sum_{\tau \in \mathbb Z_t} \sigma^{(\alpha)}_{\tau+s}\!\otimes_r\! \1 \!-\! \1 \!\otimes_r\! \sigma^{(\alpha)}_{\tau+s}\right)^{\!\!2}\right]\!\!, 
\end{equation}
$s=0,1/2$. These choices simplify drastically our numerical analysis as they reduce the number of parameters to only two, however, they still contain a rich phenomenology. The two extremal cases are found for $\epsilon_1 = \epsilon_2 =0$, when the model corresponds to an ensemble of ergodic dual unitary circuits, and for $\epsilon_1 = \epsilon_2 = {\pi}/{4}$ when the model is trivially localised as there is no coupling between different sites. We verified that including the term $J \sigma^{(3)} \otimes \sigma^{(3)}$ in the local gate ($J$ has been set to 0 in Eq.~\eqref{eq:model}), or a finite disorder variance, does not qualitatively modify the numerical results. For instance, in Fig.~\ref{fig:dualgap} we show the gap in the 0,0) momentum sector of $\mathcal T$ at the dual unitary point $\epsilon_1=\epsilon_2=0$, i.e., 
\be
\Delta_0 = 1 - |\lambda_1|,
\ee
as a function of $J$ and $t$. We note that the gap is largest when $J= 0$, and closes ($\Delta_0\to 0$) when $J\to {\pi}/{4}$, i.e., when the local unitary $U$ equals the swap matrix and the circuit becomes non-interacting. 

 Moreover, in Eq. \eqref{eq:disorder} we took $\sigma \to \infty$. To confirm that this choice does not introduce non-generic behaviour we investigated the gap of the space transfer matrix for various values of finite variance. A representative example of this is reported in the bottom panel of Fig.~\ref{fig:dualgap}, where we took $J=0$. We see that $\sigma^2/2 = 1$ we are already close to the $\sigma = \infty$ limit. For example at $t = 8, \enspace J = 0, \enspace \sigma^2/2 = 1$  the gap is found to be $\Delta_0(8) \approx 0.660709$. Instead taking $\sigma \to \infty$ one gets $\Delta_0(8) \approx 0.6607102$. That is, the correction to the gap is observed in the fifth decimal place. 
We also emphasise that the choice of an infinite variance $\sigma^2\to\infty$, i.e., a maximal disorder strength, should be the most challenging for the survival of dual unitary behaviour away from the dual unitary point. Indeed, this is the optimal regime in which one might expect Floquet many-body localisation (MBL).

The number of parameters can be reduced further by fixing the ratio between $\epsilon_1$ and $\epsilon_2$. The value of the ratio, however, affects rather drastically how dual-unitarity is broken and has to be chosen with care. Here we choose the two values of $\epsilon_2/\epsilon_1$ corresponding to weakest and strongest breaking of dual-unitarity. To find them we note that $\rho_{\tilde U}=\tilde U \tilde U^\dag/4$ is a positive matrix with unit trace~\footnote{The latter property follows from the fact that, since $\tilde{(\cdot)}$ is just an index reshuffling, we have $\tr\smash{[\tilde U \tilde U^\dag]}=\tr\smash{[U U^\dag]}=4$.} and can be interpreted as a quantum state. Therefore, the dual-unitarity breaking can be estimated by computing the fidelity between $\rho_{\tilde U}$ and the maximally mixed state $\rho_\infty = \1/4$. In particular, an explicit calculation gives 
\be
F(\rho_\infty, \rho_{\tilde U}) = \tr|\sqrt{\rho_\infty}\sqrt{\rho_{\tilde U}}|^2= \cos(\epsilon_1)^2\cos(\epsilon_2)^2, 
\ee
where $|A|= \sqrt{A A^\dag}$. This means that, considering without loss of generality 
\be
\label{eq:epsdef}
\epsilon_2\leq \epsilon_1 =: \epsilon,
\ee
the two cases corresponding respectively to the weakest and strongest breaking of dual unitarity are $\epsilon_2=0$ and $\epsilon_2=\epsilon$. In the following we consider both these cases referring to them, respectively, as Case I and Case II.

\subsection{Numerical Approaches}
\label{sec:numerics}

Besides standard exact diagonalisation of \eqref{eq:evolutionoperator} for small systems, in this paper we provide three numerical tests to characterise the space transfer matrix $\mathcal{T}$. The first is an Arnoldi exact diagonalisation method to converge several leading eigenvalues simultaneously. We use this to study the full spectrum of $\mathcal T$ without resolving the translation symmetry in Eq.~\eqref{eq:doubletranslation}. Due to the number of eigenvalues we want to compare and the behaviour of the spectrum this method is limited to $t\leq 6$ (our full forward-backward time lattice comprises of $4t$ qubits). 

Next, we isolate a specific double-momentum sector $(\nu,\nu)$ and characterise the action of the transfer matrix within the sector (a detailed discussion of how this is achieved is presented in Appendix ~\ref{sec:doublemoneta}). This removes degeneracy near the spectral edges of $\mathcal{T}$ and in most cases allows us to study the spectrum with the power iteration method, allowing us to access times $t\leq 9$ (for some data sets we had to use the Arnoldi method also within a given sector as the gap was too small). Moreover, we use this representation to evaluate the coefficients in our pertubative analysis of Sec.~\ref{sec:pertubationtheory}. 

Our third method is a Monte Carlo-based approach that approximates the maximal eigenvalue in a given double-momentum sector by
\emph{stochastically unraveling} the average in Eq.~\eqref{eq:SFFaveout} and allowing us to approach times $t\leq 12$. The idea is to observe that, before the average, the transfer matrices in Eq.~\eqref{eq:SFFaveout} are written as the tensor product of two operators (related by complex conjugation) acting only on the forward and backward lattices. Therefore, if we do not perform the average explicitly we can consider only one of the lattices, halving the number of qubits we need to simulate. More concretely we proceed as follows. Instead of considering directly Eq.~\eqref{eq:transfermat} we construct local gates on the time lattice as 
\begin{equation}
\Tilde{U}_{\Vec{\theta}_x} = \Tilde{U}\cdot  (u_x \otimes v_x), 
\end{equation}
where 
\begin{align}
    & u =  \exp \left(i \theta^{(1)} \sigma^{(1)}\right)  \exp \left( i \theta^{(3)} \sigma^{(3)}\right),\\
    &v =  \exp \left(i \phi^{(1)} \sigma^{(1)}\right)  \exp \left( i \phi^{(3)} \sigma^{(3)}\right). 
\end{align}
Therefore, the transfer matrix is characterised by four angles $\Vec{\theta} = (\theta^{(1)}_x,\theta^{(3)}_x,\phi^{(1)}_x,\phi^{(3)}_x)$, which are uniformly generated random numbers, and can be written as 
\be
\mathcal{T}_{\Vec{\theta}_x} = \tilde{\mathbb U}_{\Vec{\theta}_x} \otimes_r \tilde{\mathbb U}_{\Vec{\theta}_x}^{*}, \qquad   \tilde{\mathbb U}_{\Vec{\theta}_x} = \Tilde{U}_{\Vec{\theta}_x}^{\otimes t} \Pi_{2t}  \Tilde{U}_{\Vec{\theta}_x}^{\otimes t} \Pi_{2t}^{-1}\,.
\ee
 We then observe that for large enough $N$
\begin{align} 
\mathbb{E}\left[\prod_{n=1}^N\mathcal{T}_{\Vec{\theta}_n} |\psi \rangle\otimes_r |\psi\rangle^* \right]
\approx \lambda_0^N c_0|\lambda_0\rangle + \dots, \label{eq:monte} 
\end{align}
where $\ket{\psi}$ is a random state on the forward lattice, $c_0 = \langle \lambda_0| (|\psi\rangle\otimes_r| \psi\rangle^*)$, and the average is performed over all choices of $\Vec{\theta}_n$, $n=1\dots N$. We now estimate the the left hand side of Eq.~\eqref{eq:monte} by sampling over the choices of $\Vec{\theta}_n$. Letting 
\be
\ket{\psi_N}=\prod_{n=1}^N \tilde{\mathbb U}_{\Vec{\theta}_n} |\psi\rangle,
\ee
we have 
\begin{equation}
     \mathbb{E}\left[\prod_{n=1}^N\mathcal{T}_{\Vec{\theta}_n} |\psi \rangle |\psi\rangle \right]\approx \frac{1}{\Lambda} \sum_{m=1}^{\Lambda} \ket{\psi_N}_m \otimes_r \ket{\psi_N}_m,
\end{equation}
where $m$ labels the samples of $\Vec{\theta}_1, \ldots, \Vec{\theta}_n$ and $\Lambda$ denotes the sample size. Therefore, plugging back into \eqref{eq:monte} we obtain 
\begin{equation} \label{eq:monte2}
    \frac{1}{\Lambda} \sum_{m=1}^{\Lambda} |\langle \psi |\psi_N\rangle_m|^2 \approx  \lambda_0^N c_0 (\langle \psi | \otimes_r \langle\psi|^* )|\lambda_0\rangle + \dots.
\end{equation}
The left hand side of this equation can be easily computed for a large number of samples and moderately large $t$ because it is defined only on the forward lattice (the objects involved live in a vector space of half the size). A simple way to extract $\lambda_0$ from this relation is to take the logarithm of both sides and find the slope of the data as a function of $N$. In what follows we take $\Lambda=10^7$ for small $t$ and reduce our sample size to $\Lambda= 10^5$ for $t=11,12$. This method works best when $\lambda_0$ is significantly larger than one. 

Finally we mention that if one is only interested in characterising the evolution in time of the leading eigenvalues of $\mathcal T$ one can use the time-evolution-based method recently introduced in Ref.~\cite{garratt2021local}. This method assumes a certain structure for the spectrum of $\mathcal T$ and evaluates the leading eigenvalues by performing an evolution in time in a quantum circuit with finite length and judiciously selected twisted boundary conditions. Here, however, we are interested in an assumption-free characterisation of $\mathcal T$ which goes beyond its leading eigenvalues (cf.\ Sec.~\ref{sec:pertubationtheory}). Therefore, we do not use this approach.

\section{Stability of the ergodic phase: Numerical Survey}
\label{sec:independentnumerics}

\begin{figure*}[t]
\centering
\includegraphics[width=0.45\linewidth]{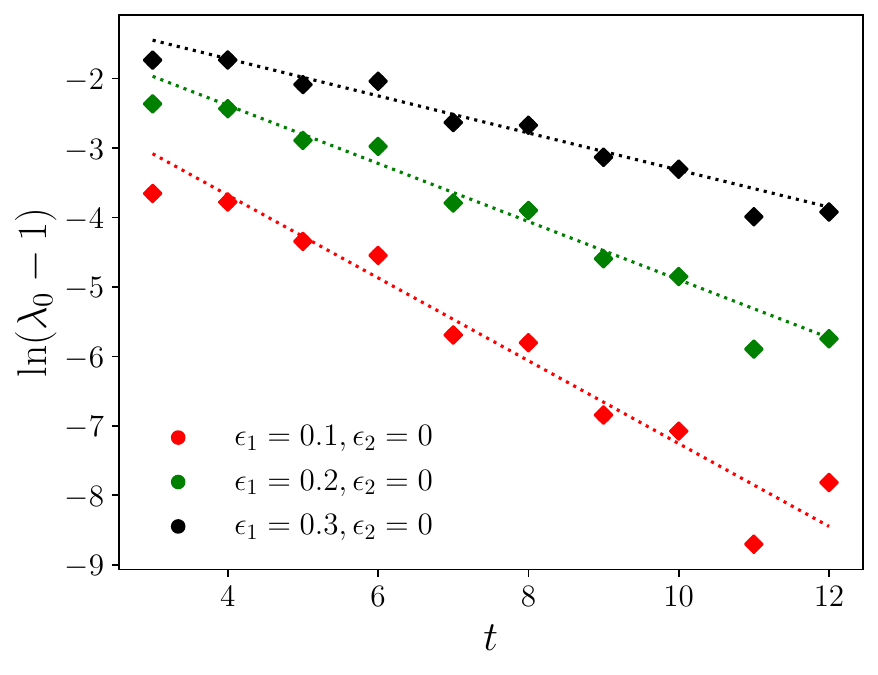}
\includegraphics[width=0.45\linewidth]{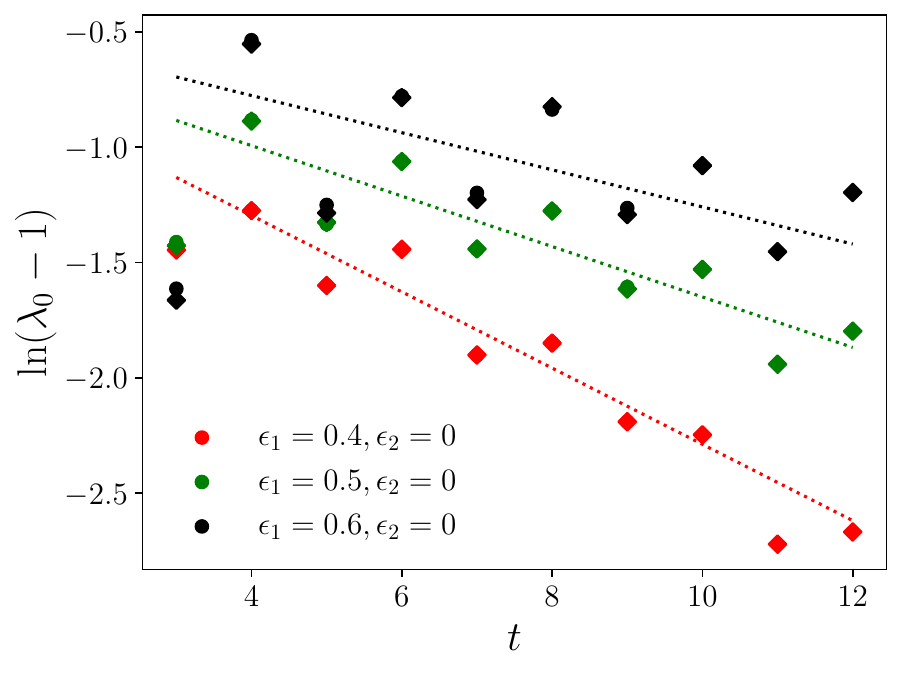}
\includegraphics[width=0.45\linewidth]{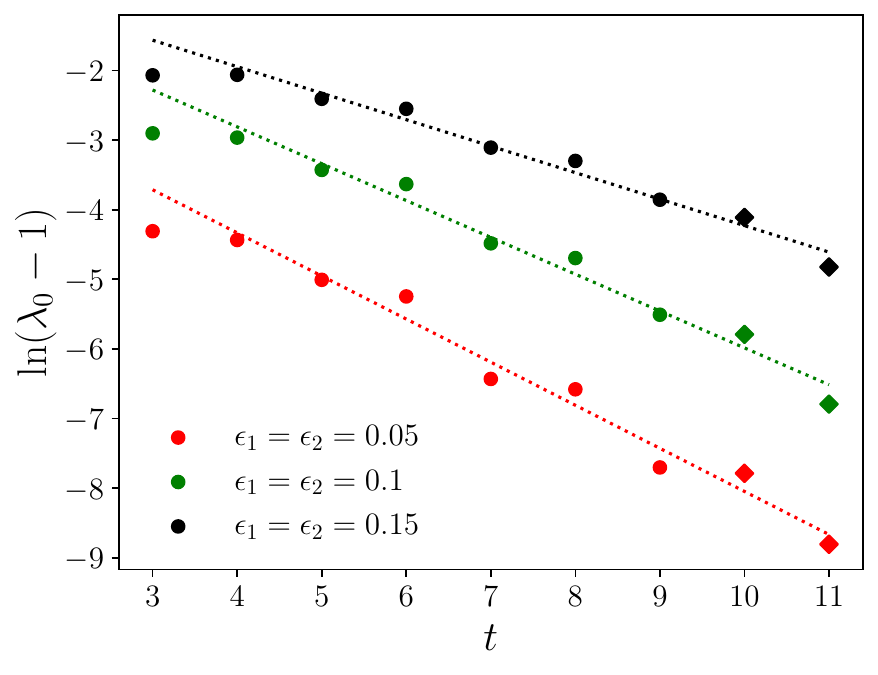}
\includegraphics[width=0.45\linewidth]{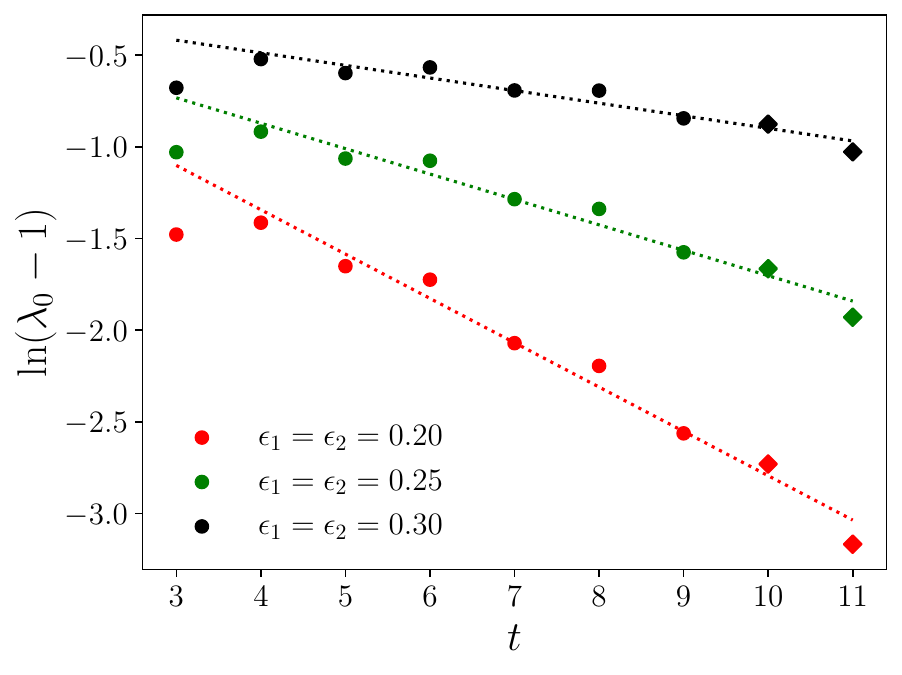}
\caption{  $\ln(\lambda_0(t)-1)$ as a function of time for various $\epsilon_1,\epsilon_2$. Dotted lines indicate numerical fits. Circle data points are retrieved with the power method isolated in the $(\nu,\nu') = (0,0)$ symmetry sector. Diamond data points are calculated using the Monte Carlo. Monte Carlo data consists of $10^7$ samples for $t\leq 10$ and $t >10 $ use $10^5$ data points. In the top two panels we supply exact values and Monte Carlo estimates for all times $t$, while in the bottom two panels we simply plot exact values for $t\leq 9$ and the remaining points are Monte Carlo estimations.}
\label{fig:lamexpdecay}
\end{figure*}

As recalled in Sec.~\ref{sec:SFF}, on the dual-unitary manifold the quantum circuit is chaotic in the sense that its spectral form factor exhibits the linear ramp characteristic of random matrix theory (cf. Eq.~\eqref{eq:CUEform}).  This property is found to correspond to the space transfer matrix $\mathcal{T}$ having $t$ eigenvalues equal to one. In this section we investigate how the spectrum of $\mathcal{T}$ behaves when we move away from the dual unitary point using the model in Eq.~\eqref{eq:model}.  

Our expectation is that the $\epsilon$-dependence of the spectrum of $\cal T$ should be smooth for small enough $\epsilon$. This means that there should exist a phase where the spectrum of $\cal T$ is qualitatively similar to the that of the dual-unitary point: The leading eigenvalue, or SLE, should have an approximate $t$-fold degeneracy near unity (which becomes increasingly exact as $t\rightarrow \infty$) and the corresponding eigenvectors should lie in the diagonal momentum sectors (i.e.\ $(\nu,\nu)$). This phenomenology would be consistent with the spontaneous symmetry breaking scenario proposed in Ref.~\cite{garratt2021manybody} (cf.~Sec.~\ref{sec:SFF}). More concretely, we expect that at late enough times Eq.~\eqref{eq:sff_eigen} is dominated by the leading eigenvalues so that
\be
\label{eq:K_approx}
K(t,L) \approx \sum^{t-1}_{
\nu=0} \lambda^L_{0,(\nu,\nu)}.
\ee
For $\epsilon=0$ this approximation is exact, the eigenvalues are all equal to $1$, so that $K(t,L)=t$ for large enough $L$. When $\epsilon \neq 0$, the requirement is that a linear ramp ensues on time scales in excess of the so called Thouless time (expected to be sub-polynomial in $L$ in the absence of conservation laws). Combined with Eq.~\eqref{eq:K_approx}, this suggests that the leading eigenvalues in each sector are split from $1$ by an amount that decays exponentially in time, i.e.,  
\be
\label{eq:eigenvalueapprox}
\lambda_{0,(\nu,\nu)}=1+O(e^{- \gamma t}),\qquad  \gamma >0.
\ee
On the other hand, an MBL phase should be characterised by a single leading eigenvalue going to $\lambda_{0,(0,0)} = 4$ for large $t$, while all remaining eigenvalues having magnitude less than unity. In the language of Ref.~\cite{garratt2021manybody} this corresponds to a symmetry-unbroken phase.

\begin{figure*}[t]
\centering
\includegraphics[width=0.45\linewidth]{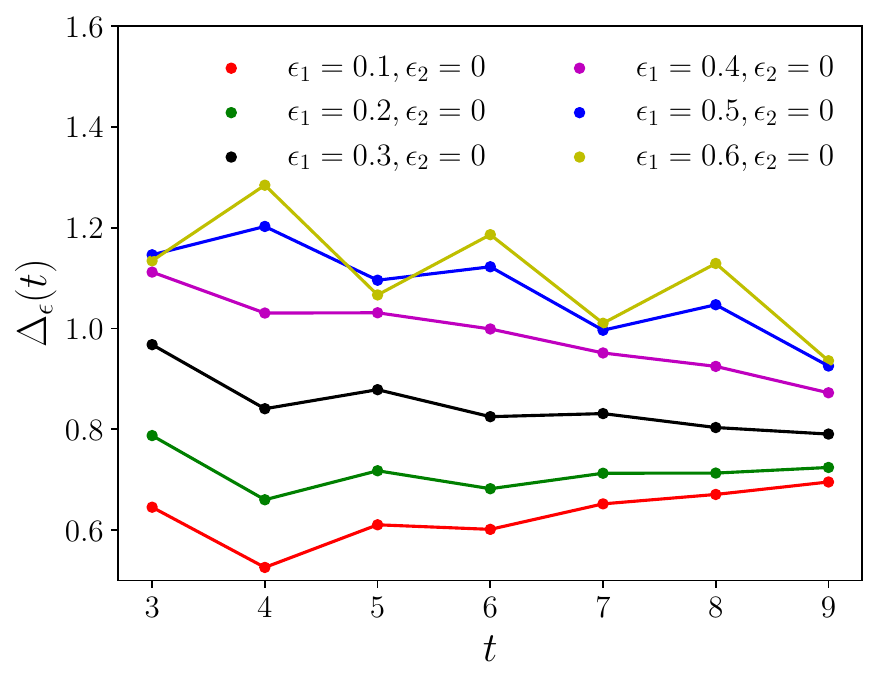}
\includegraphics[width=0.45\linewidth]{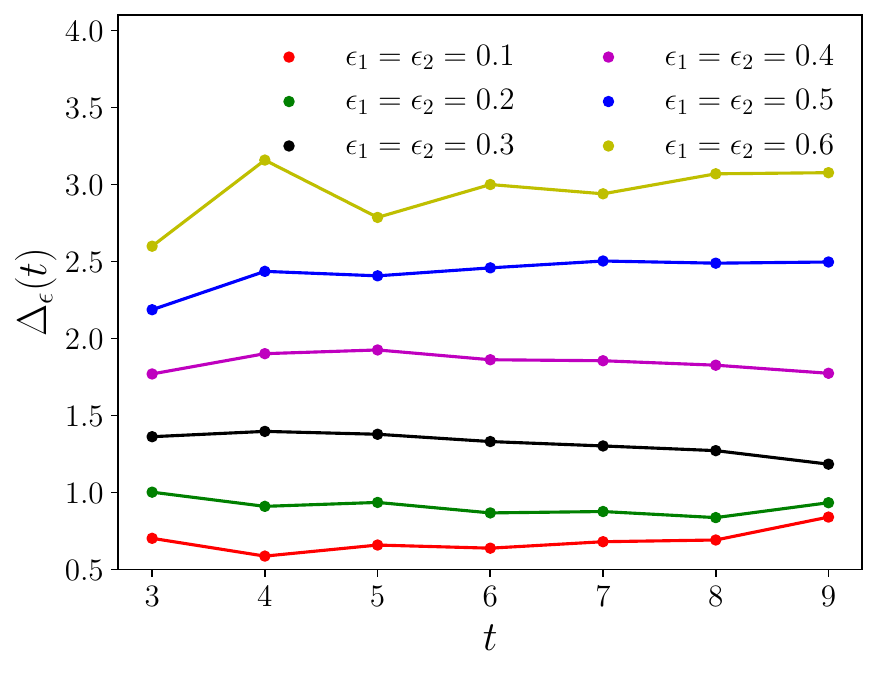}
\caption{The gap $\Delta = |\lambda_0| - |\lambda_1|$ versus time $t$ for various choices of perturbation. This data was retrieved using the Arnoldi method in the $(\nu,\nu') = (0,0)$ symmetry sector.}
\label{fig:epsgap}
\end{figure*}
\begin{figure*}[t]
\centering
\includegraphics[width=0.49\linewidth]{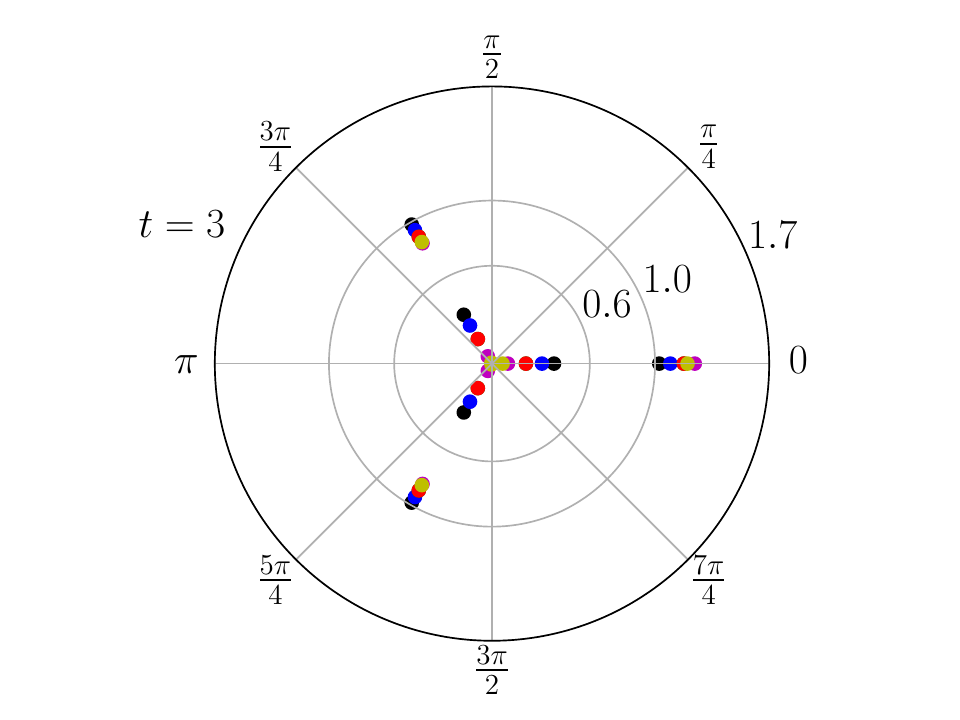}
\includegraphics[width=0.49\linewidth]{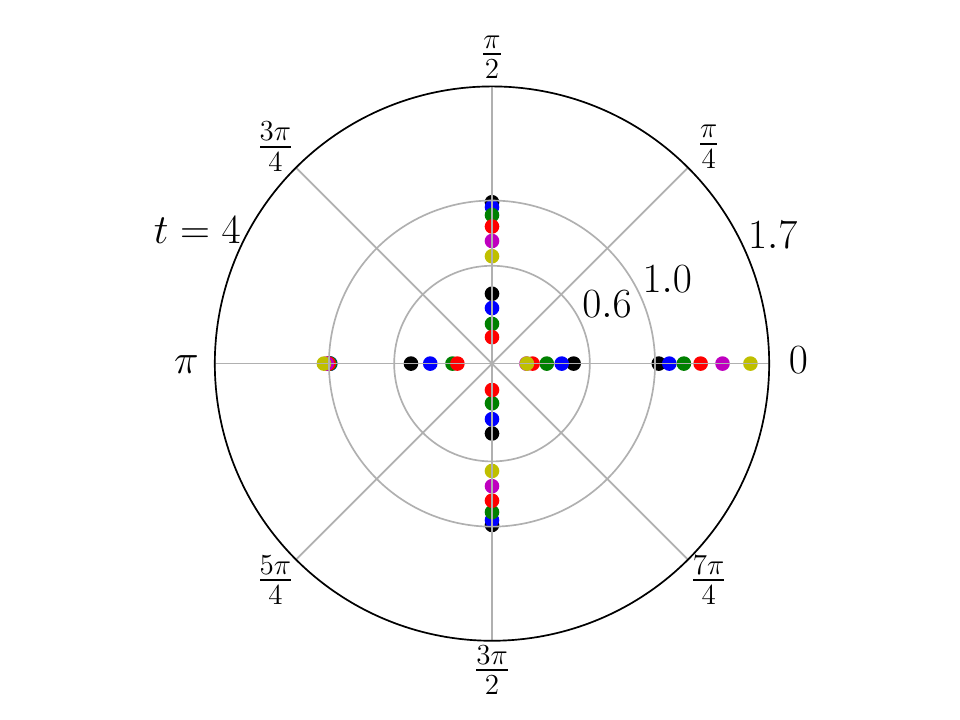}
\includegraphics[width=0.49\linewidth]{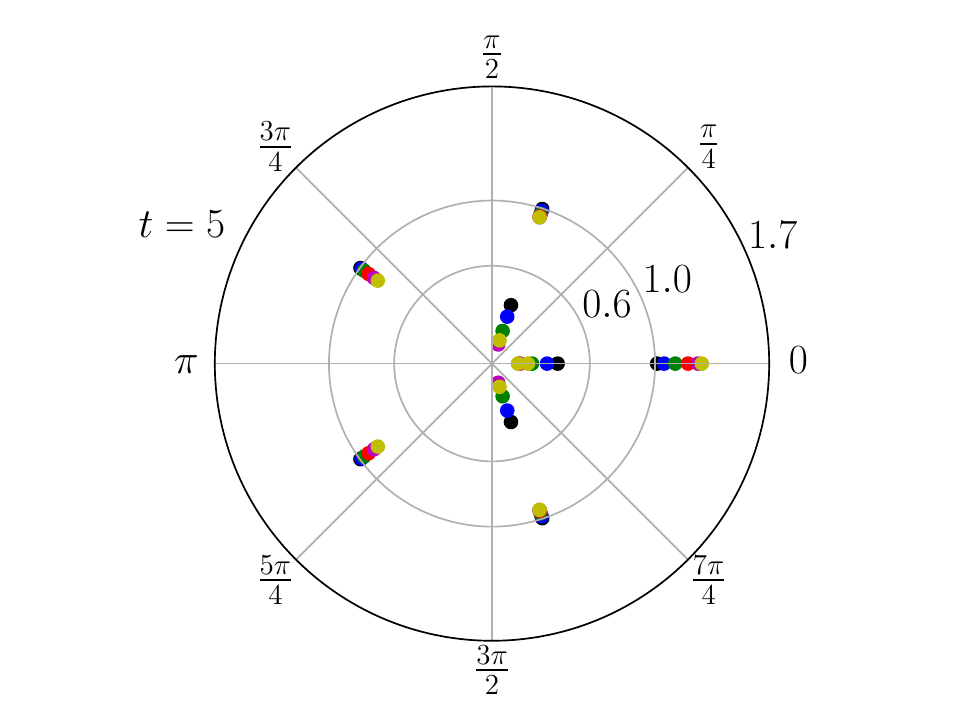}
\includegraphics[width=0.49\linewidth]{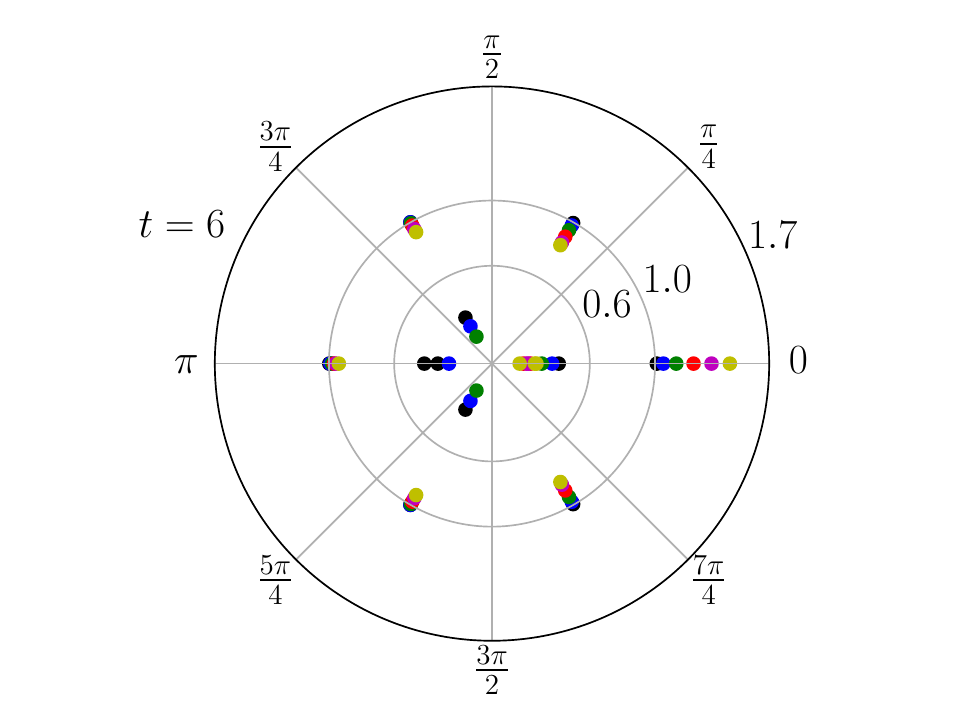}
\includegraphics[width=.6\linewidth]{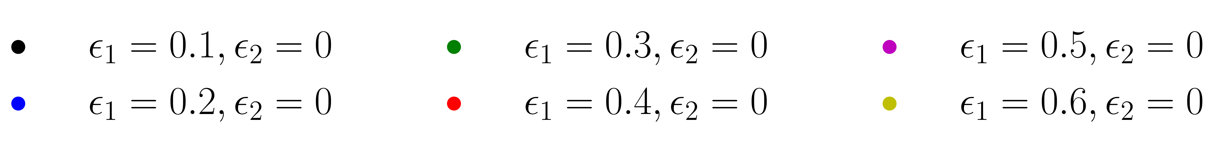}
\caption{Full spectrum $\lambda_{n,(\nu,\nu)}$ analysis (all diagonal double momentum sectors) of $\mathcal{T}$ for various $\epsilon$ and $t$. Points are plotted in polar coordinates, with the radius being the magnitude of the eigenvalue (this plot therefore covers up some degeneracy in the sub-leading eigenvalues). The polar angle is $2\pi \nu/t$, where $\nu$ labels the symmetry sector $(\nu,\nu)$. Results were obtained by an Arnoldi method converging $n=12$ eigenvalues at the edge of the spectrum.  }
\label{fig:fullspec}
\end{figure*}

Let us now proceed to substantiate these expectations using our exact numerical approaches. Even though maximal-magnitude eigenvalues exist in all the $(\nu,\nu)$ symmetry sectors, we begin by focussing our attention on the $(0,0)$ sector (and temporarily drop the sector label) as this gives us the ability to investigate longer times (larger time-lattice sizes). At the end of this section we will discuss all symmetry sectors simultaneously. We observe that in the $(0,0)$ sector, $\epsilon \neq 0 $ implies $\lambda_0(t) > 1$ for all the observable times $t$, that is, perturbing away from the dual unitary point increases the leading eigenvalue (Fig.~\ref{fig:fullspec}). Conversely,  in the other sectors we detect a decreasing in the size of the maximal eigenvalue, i.e., $|\lambda_{0,(\nu,\nu)}|<1$ for $\nu\neq 0$ and $\epsilon\neq 0$.

In Fig.~\ref{fig:lamexpdecay} we report $\ln(\lambda_0(t)-1)$ versus $t$ for both Cases I and II and variety of values of $\epsilon$, all relatively small. We find good agreement with the following scaling form  
\begin{equation}\label{eq:lambda0expdecay}
    \lambda_0(t) \approx 1 + c(\epsilon) e^{-\gamma(\epsilon) t},
\end{equation}
where $\gamma(\epsilon)>0$ and $c(\epsilon)$ are constants depending solely on $\epsilon$. In general, we observe that $\gamma(\epsilon)$ decreases monotonically with increasing $\epsilon$. In particular, we find that $\gamma(\epsilon)$ is substantially larger in Case I than in Case II (cf.\ Sec.~\ref{sec:minexample}). More specifically, we observe that taking $\epsilon$ twice as big in Case I compared to Case II produces similar $\gamma(\epsilon)$. For example, $\gamma(0.05)|_{\rm Case II} \approx 0.6191$ while $\gamma(0.1)|_{\rm Case I} \approx 0.5962$. This persists for all $\epsilon$ tested. We similarly observe $\gamma(0.3)_{\rm Case II} \approx 0.0687$ and $\gamma(0.6)_{\rm Case I} \approx 0.0805$.

For larger values of $\epsilon$ the exponent $\gamma(\epsilon)$ becomes too small to be reliably determined in the time window accessible by our exact methods. However, our numerics indicate that for Case I $\gamma\left(\epsilon \to \frac{\pi}{4}\right)$ is non-zero. This suggests that for Case I ergodicity is stable for the entire parameter regime. In principle one can explore larger times by using the time-evolution based approach introduced in Ref.~\cite{garratt2021local} (cf.\ Sec.~\ref{sec:numerics}), however, here our focus is chiefly on small $\epsilon$. We further note the apparent zig-zag pattern in the data of Fig.~\ref{fig:lamexpdecay} differentiating even and odd $t$. We associate this behaviour with the fact that for even $t$ there is an additional parity symmetric sector of $\mathcal T$ besides $(0,0)$, i.e., $(\pi,\pi)$. This changes the structure of the eigenspaces (and their dimensions) also affecting the value of $\lambda_{0,(0,0)}$.

Having discussed the behaviour of the leading eigenvalue in the $(0,0)$ sector we now move on and study the gap between the latter and the rest of the spectrum. In particular, in Fig.~\ref{fig:epsgap} we report 
\be
\Delta(t) := |\lambda_0(t)|-|\lambda_1(t)|, 
\ee
as a function of time and for different choices of $\epsilon$ (again for both Cases I and II). In all the cases explored we find that the gap satisfies $\Delta(t)>0$ in $(0,0)$ sector and, therefore, $\lambda_0(t)$ is sufficient to characterise the large $L$ behaviour of the SFF. This means that, for small $\epsilon$, the $\nu=0$ component of  Eq.~\eqref{eq:K_approx} is confirmed.  

To conclude our numerical test of Eqs.~\eqref{eq:K_approx} and \eqref{eq:eigenvalueapprox} it remains to check that this behaviour is mirrored in other symmetry sectors. We achieve this by using the Arnoldi method to check all sectors simultaneously. Our results are reported in Fig.~\ref{fig:fullspec}. Interestingly, the leading eigenvalue in the $(0,0)$ sector is always observed to be the dominant one. Along with this property, it is observed to converge to $1$ more slowly than the other leading eigenvalues. For example the $t=6$, $\epsilon=0.6$ data set for Case I has $\lambda_{0,(0,0)} = 1.45898$ while the second furthest from unity is $\lambda_{0,(1,1)} = \lambda_{0,(5,5)} = 0.838665$. Leading eigenvalues are always found to be real numbers, while sub-leading values in general are real or complex with magnitude smaller than unity. The leading eigenvalue in the $(0,0)$ sector is the only one to be consistently greater than unity, other leading eigenvalues can oscillate around $1$ as functions of time.

In summary, our numerical analysis is consistent with the expectation that for small $\epsilon$ the spectrum of $\cal T$ is a smooth deformation of that at the dual unitary point: We have $t$ real eigenvalues close to unity while the rest have significantly smaller magnitudes. We stress that this is the case despite the fact that our minimal model (\ref{eq:model}, \ref{eq:model2}) involves maximal disorder strength. Interestingly, in all our numerical experiments we also observed that the $(0,0)$ sector has the largest eigenvalue. This is consistent with the symmetry breaking picture of Ref.~\cite{garratt2021manybody}: as the system is perturbed away from the maximally ergodic point, i.e., the dual unitary point, a single symmetric eigenvalue becomes the dominant one. Further numerical surveys suggest that this phenomenology continues also when moving away from our minimal model (\ref{eq:model}, \ref{eq:model2}).

\section{Perturbation theory} 
\label{sec:pertubationtheory}

In this section we propose an analytical explanation for the main observation of Sec.~\ref{sec:independentnumerics}. Namely, that the dual-unitary eigenvalue structure is maintained for finite $\epsilon$ suggesting structural stability of the dual-unitary phase. The idea is to fix $\epsilon$ and write the maximal eigenvalue of $\mathcal{T}$ in each diagonal double-momentum sector $(\nu,\nu)$ --- we denote it by $\lambda_{(\nu,\nu)}\equiv\lambda_{0,(\nu,\nu)}$ --- as a perturbative series in an auxiliary parameter. Studying this series we then show that, if two assumptions are fulfilled,  then $|\lambda_{(\nu,\nu)}-1|$ is bounded by a term that is exponentially small in $t$, implying that our circuit models are ergodic at the considered value of $\epsilon$. Remarkably, this happens even for the minimal model with maximal disorder strength discussed in Sec.~\ref{sec:minexample}. 

Specifically, our two assumptions are
\begin{itemize}
\item[(i)] No ``maximal level" crossing occurs in the perturbative expansion, i.e., the evolution of the maximal eigenvalue in each sector can be followed by tracing the smooth deformation of the maximal eigenvalue at the dual-unitary point. 
\item[(ii)] The ($\epsilon$-dependent) coefficients of our perturbative series are bounded by an exponentially decaying function of $t$ and grow at most exponentially in $n$, where $n$ is the perturbative order. 
\end{itemize} 
A more precise formulation of these assumptions is given in the upcoming derivation. 

The first assumption is safe for generic enough perturbations: if the perturbation couples all the eigenvectors in a given sector, all adjacent level encounters are avoided crossings. Therefore, Assumption (i) can be rephrased by saying that we assume our perturbation to be sufficiently generic. This assumption is consistent with our numerical survey of Sec.~\ref{sec:independentnumerics} (cf. Fig.~\ref{fig:epsgap}).  The second assumption is our main one and, as we discuss in Sec.~\ref{sec:discussion}, can be partly justified by an analytical argument. In Sec.~\ref{sec:discussion} we also show that Assumption (ii) fails at the trivially localised point (i.e.\ for Case II and  $\epsilon=\pi/4$, cf.\ Sec.~\ref{sec:minexample}) while we give numerical evidence of it holding for small~$\epsilon$. 

We now proceed to show that when (i) and (ii) hold the quantum circuit \eqref{eq:evolutionoperator} is ergodic. We begin considering the eigenvalue equation for the maximal eigenvalue $\lambda$ of the transfer matrix resolved to the double momentum sector $(\nu,\nu)$, namely 
\be
\label{eq:maxeigenvalueeq}
\mathcal T \ket*{\lambda} = \lambda \ket*{\lambda}.
\ee
Here and in the following, we drop the dependence on the subscript $(\nu,\nu)$ whenever it is not ambiguous to do so. Since for $\epsilon=0$ only the diagonal sectors contain the maximal eigenvalues, and the latter are unique, Eq.~\eqref{eq:maxeigenvalueeq} contains all the necessary information to characterise the spectral form factor for small enough $\epsilon$. It is not obvious, however, that this will continue to hold also for finite $\epsilon$. Here we assume this to be the case, namely we make the following postulate
\begin{assumption}[No leading eigenvalue crossing]
\label{asmp:asmp1}
The leading eigenvalues of $\mathcal T$ for small enough $\epsilon$ are obtained by smooth deformation of those of $\mathcal T|_{\epsilon=0}$. Namely, $\lambda$ is a smooth function of $\epsilon$. 
\end{assumption}

By virtue of Assumption~\ref{asmp:asmp1} we can limit our treatment to diagonal double-momentum sectors and only consider Eq.~\eqref{eq:maxeigenvalueeq}, which we treat using a ``dressed'' perturbative approach. Specifically, we proceed as follows.  First we set 
\be
\mathcal T_0 \equiv  \mathcal T |_{\epsilon=0}, \qquad \bar{\cal T}_\epsilon \equiv  \frac{\mathcal  T -\mathcal  T_{0}}{\epsilon}, 
\ee 
and rewrite Eq.~\eqref{eq:maxeigenvalueeq} as  
\be
(\mathcal T_0 + \epsilon \bar{\cal T}_\epsilon)\ket*{\lambda} = \lambda \ket*{\lambda}\,. 
\ee
Next we solve 
\be
\label{eq:perturbativeeeq}
(\mathcal T_0 + x \bar{\cal T}_\epsilon)\ket*{\lambda} = \lambda \ket*{\lambda}, 
\ee
in perturbation theory in $x$ for fixed $\epsilon$. Finally we recover a perturbative solution of Eq.~\eqref{eq:maxeigenvalueeq} by setting $x=\epsilon$ in the end. The advantage of this approach is that it involves only a first order correction to the transfer matrix as it happens in standard time-independent perturbation theory in quantum mechanics, at the cost of making the perturbation manifestly $\epsilon$-dependent. 

Explicitly, we expand both $\ket{\lambda}$ and $\lambda$ in $x$
\begin{align}
\ket*{\lambda} &= \sum_{k=0}^{\infty} x^k \ket*{{\lambda}}^{(k)}, & \ket*{\lambda}^{(0)}&=\ket{1}, \label{eq:lambdax}\\
\lambda &= \sum_{k=0}^{\infty} x^k \lambda^{(k)}, & {\lambda^{(0)}}&={1},
\label{eq:lambdax} 
\end{align}
and impose \eqref{eq:perturbativeeeq} order by order in $x$. As shown explicitly in Appendix~\ref{app:PT}, this yields 
\be
\begin{aligned}
&\ket*{\lambda}^{(1)}  =
 \mathcal G \bar{\cal T}_\epsilon \ket{{1}},\\
&\ket*{\lambda}^{(n>1)} = \sum_{\ell=1}^{n-1} \!\!\sum_{\substack{k_1,\ldots,k_\ell=1 \\ k_1+\ldots + k_\ell=n-1}}^{n-1} \!\!\!\!\!\!\!\!\!\!\!\! \mathcal G  {\mathcal K}_{k_1} \mathcal G  {\mathcal K}_{k_2}  \cdots \mathcal G  {\mathcal K}_{k_\ell} \mathcal G \bar{\cal T}_\epsilon \ket{1},
\end{aligned}
\ee
and 
\be
\begin{aligned}
&\lambda^{(1)} = \mel{1}{\bar{\cal T}_\epsilon}{1}, \quad\lambda^{(2)} = \mel{1}{ \bar{\cal T}_\epsilon  \mathcal G  \bar{\cal T}_\epsilon }{1},\\
&\lambda^{(n>2)} = \sum_{\ell=1}^{n-2} \!\!\!\! \sum_{\substack{k_1,\ldots,k_\ell=1 \\ k_1+\ldots + k_\ell=n-2}}^{n-2}  \!\!\!\!\!\!\!\!\!\!\!\mel{1}{\bar{\cal T}_\epsilon  \mathcal G {\mathcal K}_{k_1} \mathcal G  \mathcal{K}_{k_2}  \cdots \mathcal{K}_{k_\ell} \mathcal G \bar{\cal T}_\epsilon}{1},
\end{aligned}
\label{eq:lambdaxn}
\ee
where we set
\begin{align}
&\mathcal Q= \1- \ketbra*{1_{(\nu,\nu)}},\\  
&\mathcal G = \sum_{n=0}^\infty \mathcal Q \mathcal{T}^n \mathcal Q = \mathcal Q (1-\mathcal T\mathcal Q)^{-1}, \\ &\mathcal{K}_{k} = 
\delta_{k,1}\mathcal{\bar{T}}_\epsilon - \lambda^{(k)} \1. 
\label{eq:Kktilde}
\end{align}
Note that $\mathcal G$ is the resolvent $\mathcal R(z)=(z \1-\mathcal T_0)^{-1}$ of $\mathcal T_0$ projected away from the leading-eigenvalue subspace and evaluated at $z=1$. The projection makes this operator well defined.

Using these expressions one can write $\lambda^{(n)}$ in terms of the following expectation values of products of the perturbation $\bar{\mathcal T}_\epsilon$ and the projected resolvent $\mathcal G$  
\be
\label{eq:coefficients}
\begin{aligned}
\!\!{[k_1,\ldots,k_m]} &= \mel*{1}{\bar{\cal T}_\epsilon  \mathcal G^{k_1} \bar{\cal T}_\epsilon \mathcal G^{k_2}\cdots \mathcal G^{k_m} \bar{\cal T}_\epsilon }{1},\,\, k_j\geq1,\\
\!\!{[\phantom{,}]} &=\mel*{1}{\bar{\cal T}_\epsilon}{1}.
\end{aligned}
\ee
Explicitly, the first few orders read as
\be
\begin{aligned}
\lambda^{(1)} =&  [\phantom{,}], 
\label{eq:lambda1}\\
\lambda^{(2)} =& [1],\\
\lambda^{(3)} =& [1,1]-[\phantom{,}][2],\\
\lambda^{(4)} =& [1,1,1]-[1][2]-[\phantom{,}][2,1]-[\phantom{,}][1,2]+[\phantom{,}]^2 [3]\,,\\
\lambda^{(5)} =& [1,1,1,1]\!-\![1,1][2]\!+\![\phantom{,}][2]^2\!-\![1,2][1]\!-\![2,1][1]\\
&+\!2[0][1][3]\!-\![\phantom{,}][2,1,1]\!-\![\phantom{,}][1,2,1]\!-\![\phantom{,}][1,1,2]\\
& +\![\phantom{,}]^2 [3,1] \!+\! [\phantom{,}]^2 [2,2] \!+\! [\phantom{,}]^2 [1,3]\!-\![\phantom{,}]^3 [4]\,.
\end{aligned}
\ee 
Continuing to arbitrary order we have 
 \begin{align}
\lambda^{(n>2)} = & \sum_{q=1}^\infty \sum_{p_1, \ldots, p_q =1}^{\infty} \sum_{ k_{11},\ldots, k_{q p_q} =1}^{n-1} \!\!\!\!\!\!\!\!(-1)^{q+1}  C[\{k_{ij}\}]  \notag\\
&\quad\times\delta\!\!\left[\sum_{i=1}^q \sum_{j=1}^{p_q} k_{ij} -(n-1)\right] \theta\!\!\left[n-\sum_{i=1}^q (p_i + 1)\right]\notag\\
&\quad\times [\phantom{,}]^{n-\sum_{i=1}^q (p_i + 1)} \prod_{m=1}^q {[k_{m1},\ldots,k_{m p_m}]},
\label{eq:perturbativesymbols}
 \end{align}
where $\delta[x]$ and $\theta[x]$ are equal to one when $x=0$ and $x\geq0$ respectively and to 0 otherwise, while $C[\{k_{ij}\}]$ counts the combinatorial multiplicity of a given term (we do not need its explicit expression). Here $q$ denotes the number of ${[k_1,\ldots,k_m]}$ symbols appearing in a given term, $p_m$ is the length of the $m$-th symbol. Each symbol contains $p_m+1$ factors of the perturbation $\mathrm{\bar{\cal T}_\epsilon}$, while each $[\phantom{,}]$ contains one factor. Therefore, the final line of  Eq.~\eqref{eq:perturbativesymbols} contains in total $n$ factors of the perturbation, consistent with it being an $n$-th order perturbative term.

To find the constraints we used that ${p_i} +1$ is the number of $\bar{\cal T}_\epsilon$ in ${[k_{i1},\ldots,k_{i p_i}]}$ and $\sum_{j=1}^{p_q} k_{ij}$ is the number of $\mathcal G$s. Since the total number of $\bar{\cal T}_\epsilon$ in each term at order $n$ equals $n$ we have  
\be
\sum_{i=1}^q  ({p_i}+1) = n-m \,, 
\ee
where $m$ is the number of $[\phantom{,}]$'s in the term. On the other hand from the last of \eqref{eq:lambdaxn} and the last of \eqref{eq:Kktilde} we have that each term contributing to the $n$-th order in perturbation theory contains $n-1$ occurrences of $\mathcal G$. Therefore we find  
\be
\sum_{i=1}^q \sum_{j=1}^{p_q} k_{ij} = n-1\,. 
\ee
Our goal is to bound from above the magnitude of the perturbative correction $\lambda^{(n)}$ to the maximal eigenvalue. To this end we make the following assumption on the scaling of the symbols ${[k_1,\ldots,k_m]}$
\begin{assumption}
The symbols ${[k_1,\ldots,k_m]}$ and $[\phantom{,}]$can be bounded as follows 
\label{asmp:asmp2}
\be
\begin{aligned}
|{[k_1,\ldots,k_m]}| &\leq e^{-\beta (t-t_0)}  e^{\alpha \sum_{j=1}^m k_j} e^{\gamma (m+1)},\\
|{[\phantom{,}]}| &\leq e^{-\beta (t-t_0)},
\end{aligned}
\label{eq:mainconjecturemod}
\ee
where $\beta > 0, \alpha, \gamma, t_0 $ are independent of $m,k_j$ and $t$.
\end{assumption}

Using Assumption~\ref{asmp:asmp2} in Eq.~\eqref{eq:perturbativesymbols} we obtain the following bound 
\be
|\lambda^{(n)}| \leq   e^{- \alpha-\beta (t-t_0) } e^{(\alpha + \gamma) n}  {\mathcal N}_n,
\label{eq:lambdanboundmod}
\ee
where $\mathcal N_n$ is defined as 
 \begin{align}
 {\mathcal N}_n \equiv &  \sum_{q=1}^\infty \sum_{p_1, \ldots, p_q =1}^{\infty} \sum_{ k_{11},\ldots, k_{q p_q} =1}^{n-1} C[\{k_{ij}\}]\notag\\ 
 &\times\delta\!\!\left[\sum_{i=1}^q \sum_{j=1}^{p_q} k_{ij} \!=\! n-1\right] \theta\!\!\left[n\!-\!\sum_{i=1}^q  (p_i+1)\right]\!.
 \end{align}
 This number can be computed with three simple observations. First we note that \eqref{eq:lambdaxn} implies ${\mathcal N}_1= {\mathcal N}_{2}=1$. Next, we observe that ${\mathcal N}_{n>2}$ can be alternatively written as 
\be
{\mathcal N}_{n>2} =  \sum_{\ell=1}^{n-2} \sum_{\substack{k_1,\ldots,k_\ell=1 \\ k_1+\ldots + k_\ell=n-2}}^{n-2}   {\#}_{k_1,\ldots,k_\ell},
 \label{eq:Nkalt}
\ee
where ${\#}_{k_1,\ldots,k_\ell}$ denotes the number of terms in the expansion of $\mel{1}{ \bar{\mathcal T}_{\epsilon}  \mathcal G  {\mathcal K}_{k_1} \mathbb G \mathcal {\mathcal K}_{k_2}  \cdots  {\mathcal K}_{k_\ell} \mathcal G  \bar{\mathcal T}_{\epsilon}}{{1}}$. Finally, noting 
\be
\#_{k_1,k_2\ldots,k_\ell} = \begin{cases}
 \#_{k_1,k_2,\ldots,k_{\ell-1}} ( {\mathcal N}_1+1), & k_\ell=1,\\
 \\
 \#_{k_1,k_2,\ldots,k_{\ell-1}}  {\mathcal N}_{k_\ell}, & k_\ell>1,
 \end{cases}
\ee
we immediately find ${\mathcal N}_3 =3$ and the following recursive relation
\be
\!\!{\mathcal N}_{n>3} =  {\mathcal N}_{n-2}+ \sum_{p=1}^{n-3} ( {\mathcal N}_{p}+\delta_{p,1})  {\mathcal N}_{n-p}\!=\! \sum_{p=1}^{n-1}  {\mathcal N}_{p}  {\mathcal N}_{n-p},
\ee
where we used ${\mathcal N}_{1}= {\mathcal N}_2 =1$. This means that ${\mathcal N}_{n+1}$ fulfils the recursive relation of the Catalan numbers with the same initial condition. Therefore
\be
{\mathcal N}_{n} = {\mathcal C}_{n-1} = \frac{1}{n} \binom{2n-2}{n-1} \simeq \frac{4^{n}}{4 n^{3/2} \sqrt{\pi}}\,.
\ee
Plugging back into \eqref{eq:lambdax} we find 
\be
\begin{aligned}
|\lambda-1| &\leq \sum_{n=1}^\infty x^n |\lambda^{(n)}| \\
&= e^{- \alpha-\beta (t-t_0) }  \sum_{n=1}^\infty x^n e^{(\alpha +\gamma) n}  {\mathcal N}_n\,.
\end{aligned}
\ee
To conclude we observe that the sum on the r.h.s.\ is always convergent for small enough $x$. Namely we have convergence whenever 
\be
x \leq \epsilon \leq e^{-(\alpha +\gamma)}/4. 
\ee
For all the values of $\epsilon$ fulfilling the above bound we then have 
\be
|\lambda-1| \leq A(\gamma, \alpha) e^{-\beta t}.
\ee
This expression recovers Eq.~\eqref{eq:eigenvalueapprox} and shows that whenever Assumptions~\ref{asmp:asmp1} and~\ref{asmp:asmp2} hold the ergodic phase is stable. 

\begin{figure*}[t!]
\centering
\includegraphics[width=0.45\linewidth]{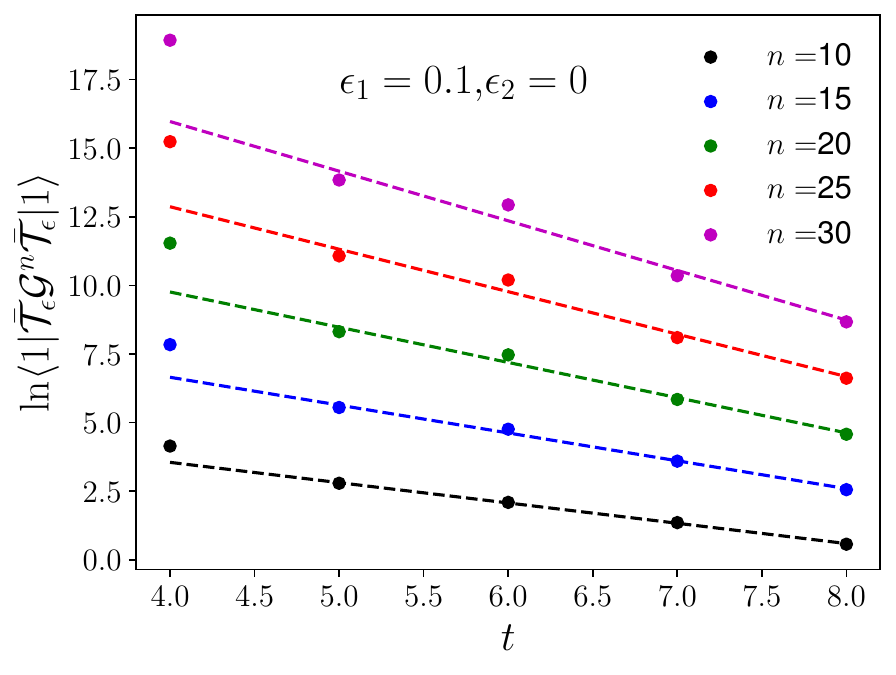}
\includegraphics[width=0.45\linewidth]{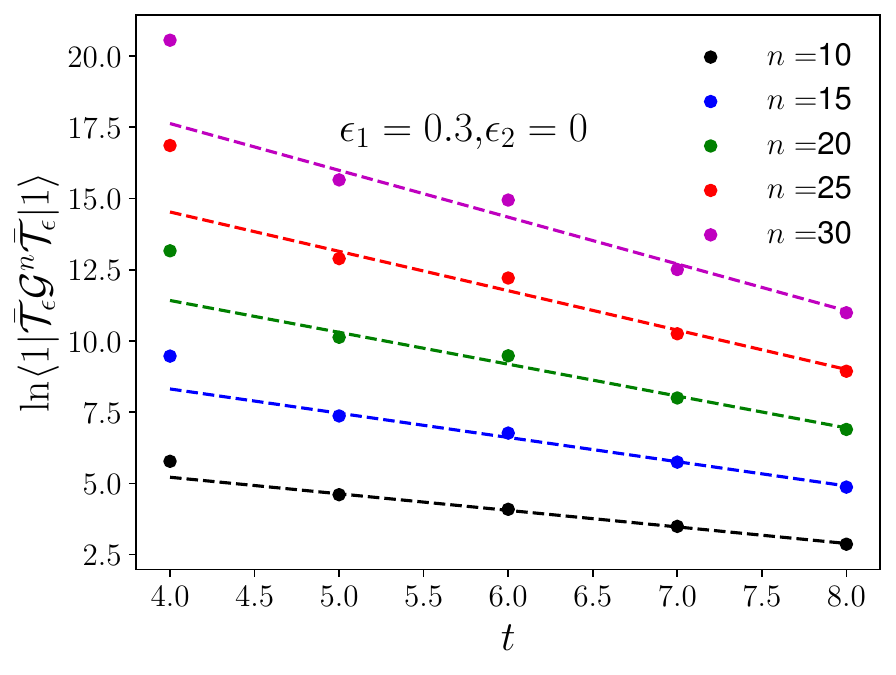}
\includegraphics[width=0.45\linewidth]{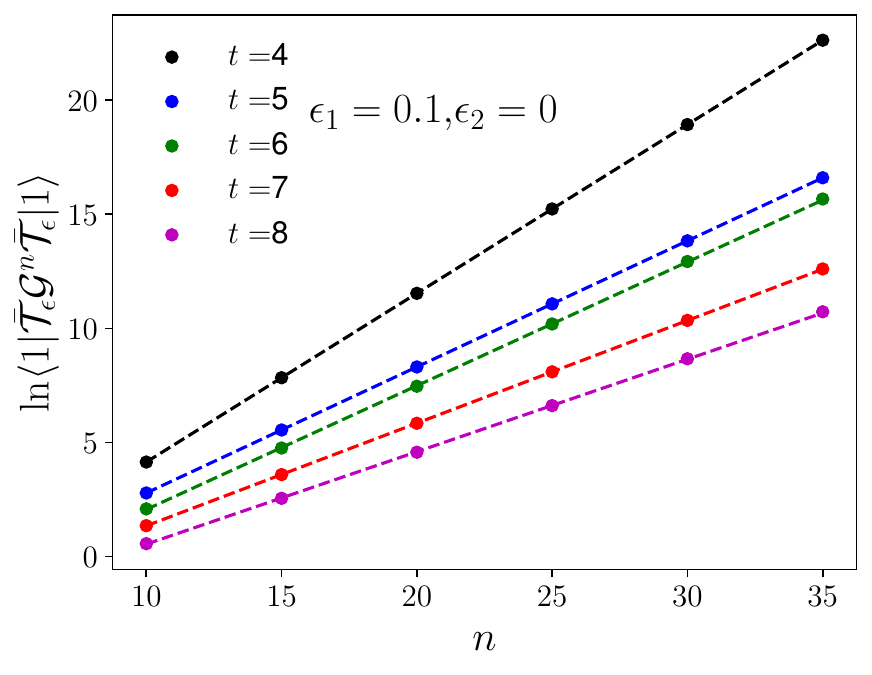}
\includegraphics[width=0.45\linewidth]{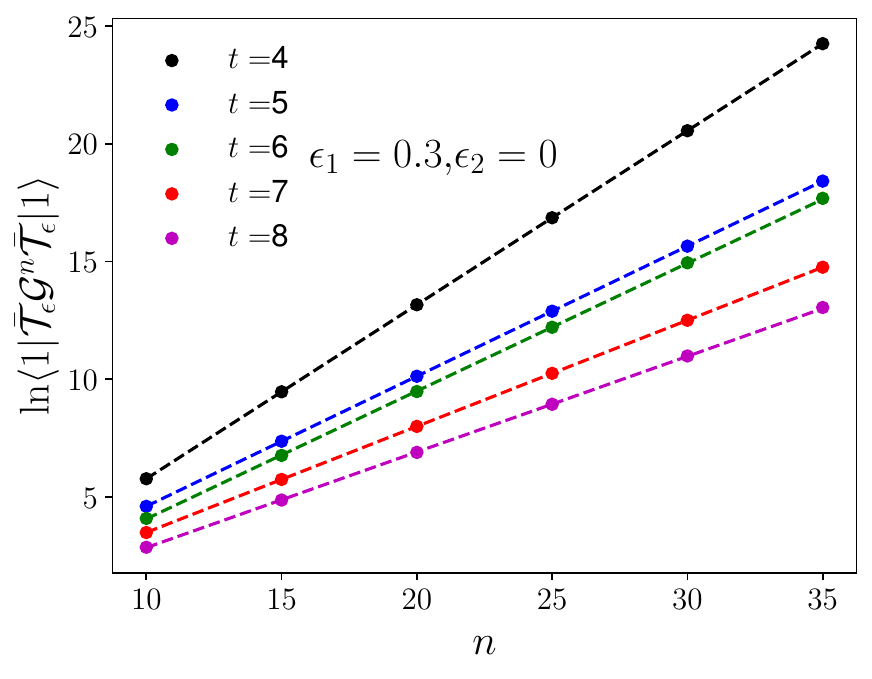}
\caption{ $[n]$ terms in perturbation theory as a function of $t$ and $n$. Solid dots represent the natural logarithm of data retrieved through exact evaluation of $[n]$ in the $(\nu,\nu') = (0,0)$ sector.  Dotted lines are numerical fits indicating exponential dependence on the independent variables $t,n$.  }
\label{fig:Rpeaked}
\end{figure*}
\begin{figure*}[t!]
\centering
\includegraphics[width=0.45\linewidth]{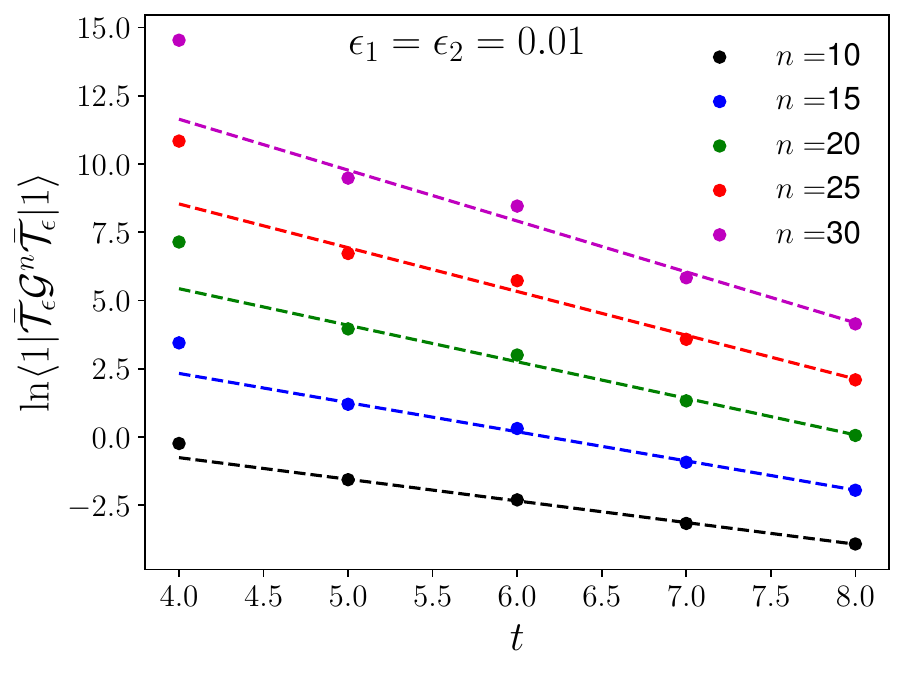}
\includegraphics[width=0.45\linewidth]{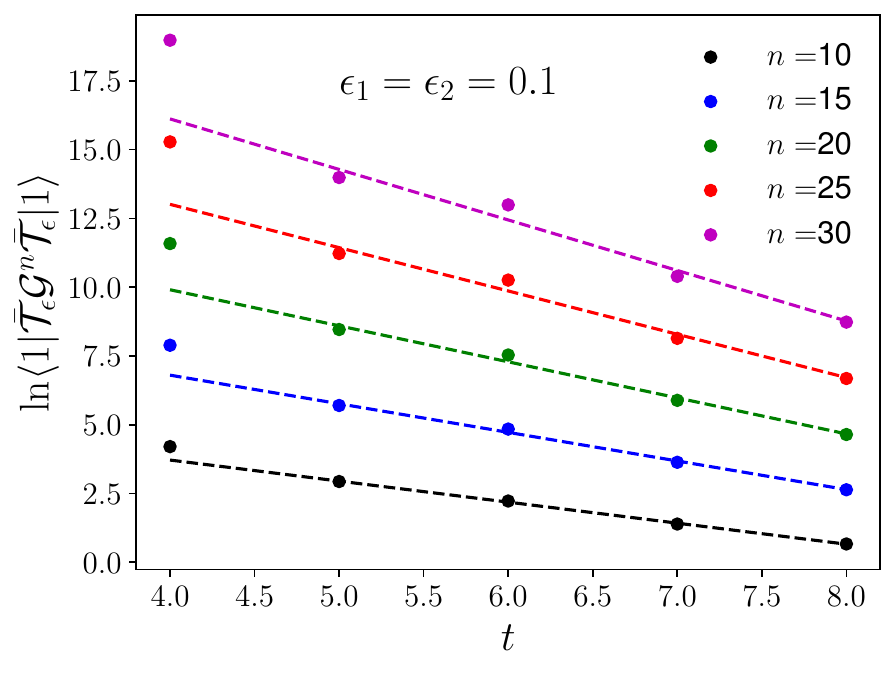}
\includegraphics[width=0.45\linewidth]{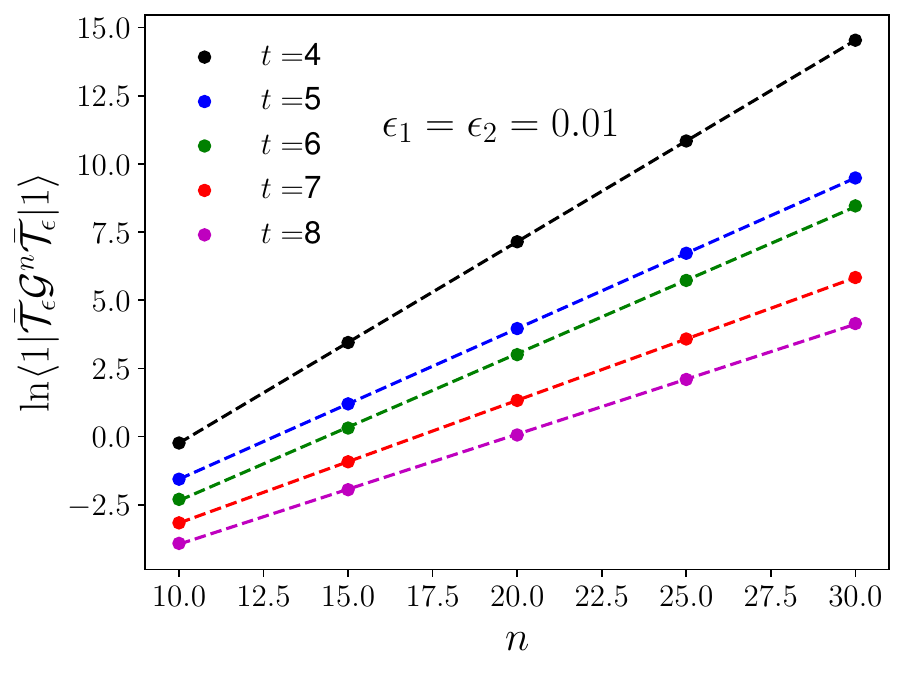}
\includegraphics[width=0.45\linewidth]{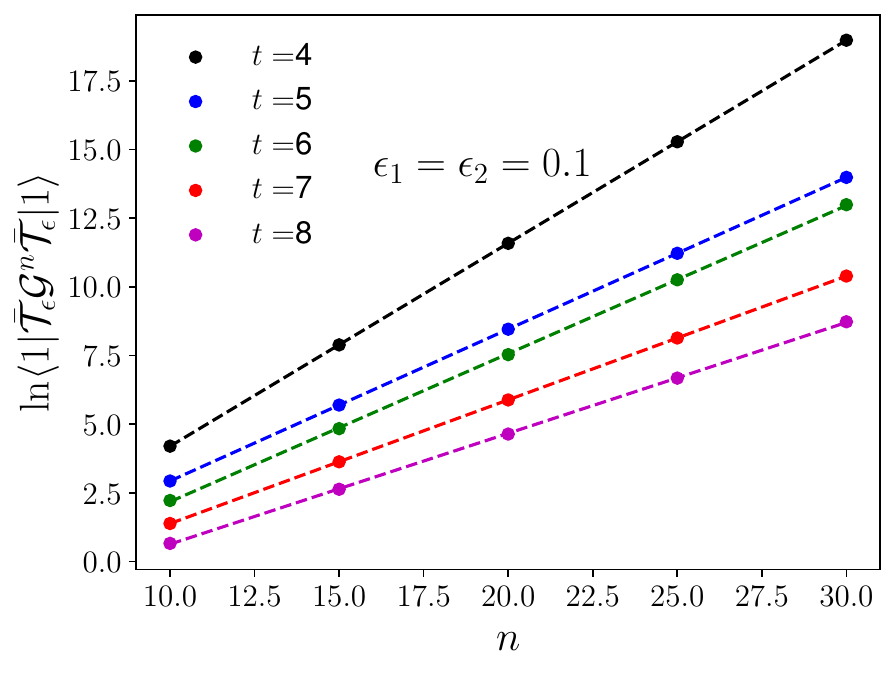}
\caption{ $[n]$ terms in perturbation theory as a function of $t$ and $n$. Solid dots represent the natural logarithm of data retrieved through exact evaluation of $[n]$ in the $(\nu,\nu') = (0,0)$ sector.  Dotted lines are numerical fits indicating exponential dependence on the independent variables $t,n$.  }
\label{fig:Rpeakedv2}
\end{figure*}

\section{Discussion of Assumption~\ref{asmp:asmp2}}
\label{sec:discussion}

Our Assumption~\ref{asmp:asmp2} on the behaviour of the perturbative coefficients  in Eq.~\eqref{eq:coefficients} can be justified by an analytical argument assisted by numerical observations. For definiteness we again focus on the sector $(\nu,\nu)=(0,0)$, although other double momentum sectors show similar behaviour. We begin by considering the simplest of the coefficients in Eq.~\eqref{eq:coefficients}, i.e., 
\begin{equation} 
\label{eq:peaked}
    [n] = \mel*{1}{\bar{\cal T}_\epsilon  \mathcal G^n  \bar{\cal T}_\epsilon }{1},
\end{equation}
and compute it numerically for $n=1,\ldots,30$ and $t= 3,\ldots,8$. Some representative examples of our results, for both Cases I and II, are reported in Figs.~\ref{fig:Rpeaked} and~\ref{fig:Rpeakedv2}. Overall we see that, in agreement with Eq.~\eqref{eq:mainconjecturemod}, the term increases exponentially as a function of $n$ and is exponentially suppressed as a function of $t$.

\begin{figure*}[t]
\centering
\includegraphics[width=0.45\linewidth]{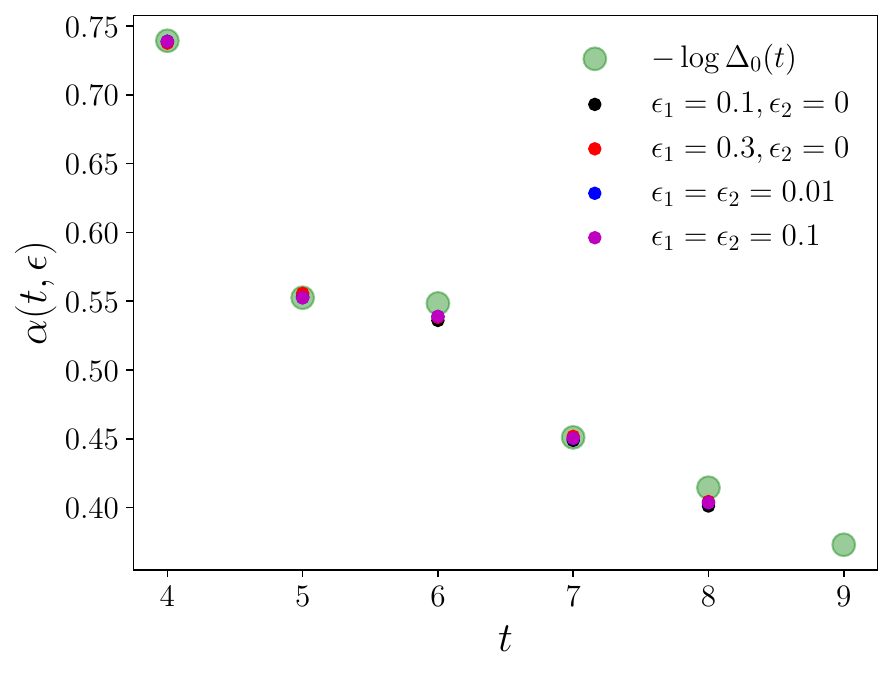}
\includegraphics[width=0.45\linewidth]{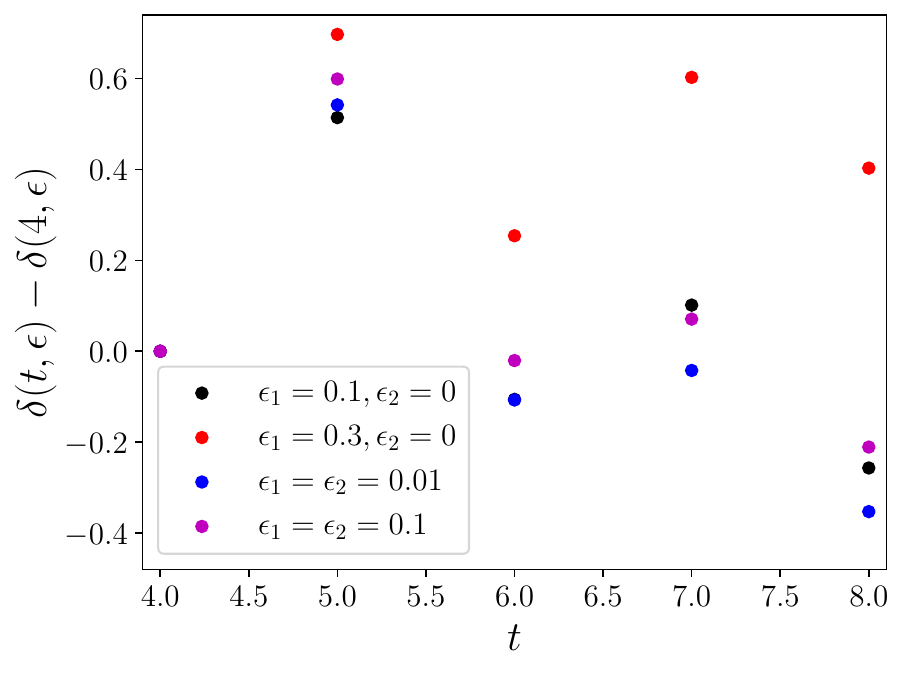}
\caption{ Data extracted from numerical fits in Fig. \ref{fig:Rpeaked} and \ref{fig:Rpeakedv2} for different choices of $\epsilon_1,\epsilon_2$. $\alpha(t,\epsilon),\delta(t,\epsilon)$ correspond to the quantities defined in Eq.~\eqref{eq:fit}. $\Delta_0(t)$ is the gap at the dual unitary point. Importantly we plot $\Delta_0(9)$ for reference, $\alpha(t,\epsilon), \delta(t,\epsilon)$ were not extracted for $t=9$.}
\label{fig:betalin}
\end{figure*}

 A more refined analysis is provided by fixing $t$, varying $n$, and performing a linear fit. Namely we set  
\be
\label{eq:fit}
\log |[n]| \approx \alpha(t, \epsilon) n + \delta(t, \epsilon), 
\ee
and find $\alpha (t, \epsilon)$ and $\delta(t, \epsilon)$ providing the best fit~\footnote{We called the slope $\alpha(t, \epsilon)$ as it plays the same role as $\alpha$ in Eq.~\eqref{eq:mainconjecturemod}.}. Studying these coefficients (cf.\ Fig.~\ref{fig:betalin}) we find that, very interestingly, the slope $\alpha(t, \epsilon)$ is roughly independent of $\epsilon$. Moreover --- and this is a key observation --- it matches remarkably well the logarithm of the inverse of the spectral gap calculated at the dual-unitary point, i.e., 
\be
\label{eq:numobs}
\alpha(t, \epsilon) \approx -\log \Delta_0(t).
\ee
In fact, since in this case the largest sub-leading eigenvalue is unique and real (cf. Tab.~\ref{table:dualgap}) we have $\Delta_0(t)= 1- \lambda_1(t)$. In fact, our Assumption~\ref{asmp:asmp2} requires $\delta(t, \epsilon)$ to decrease linearly in $t$. Here our data are less convincing and sensitive to the parity of $t$, but have the correct overall trend.

\begin{table}[b]
\centering 
\begin{tabular}{c | c | c} 
\hline
$t$ & $\lambda_1$ & $\lambda_2$ \\
\hline 
$3$ & $0.4081361$ & $-0.0982743\pm 0.1952433i$ \\
$4$ & $0.5225734$ & $0.3919454$  \\
$5$ & $0.4245154$ & $0.2848218$ \\
$6$ & $0.4221175$ & $0.3426469$ \\
$7$ & $0.3630244$ & $0.2755609$ \\ 
$8$ & $0.3392898$ & $0.3102323$ \\
$9$ & $0.3113076$ & $0.2756089$\\
\hline 
\end{tabular}
\caption{In this table we present the raw data used to calculate $\Delta_0(t)$ in Fig. \ref{fig:betalin}. We in general observe the value to be real for this choice of parameters. We also include $\lambda_2(t)$ which was extracted from the symmetry resolved Arnoldi method.} 
\label{table:dualgap} 
\end{table}

This result can be reproduced by making two assumptions on the structure of the dual-unitary transfer matrix $\mathcal T_0$. First, we assume that $\mathcal T_0$ is diagonalisable: this seems a reasonable assumption given that $\mathcal T_0$ is an average over matrices, and that defective (i.e.\ non-diagonalisable) matrices are non-generic. In fact, our upcoming reasoning continues to hold also when there are non-trivial Jordan blocks but only for the eigenvalue $0$. The latter requirement is easier to check numerically, and it is fulfilled in all our numerical observations~\footnote{Even though the Jordan decomposition is numerically unstable, one can exclude the presence of non-trivial Jordan blocks by verifying that the numerically-computed eigenvalues of $\mathcal T_0$ in a given sector are non-degenerate. In our numerical investigations we only saw degeneracy of the eigenvalue 0.}. Therefore, we write $\mathcal G$ as~\footnote{To lighten the notation from now on we drop the explicit dependence of $\lambda_j$ on $t$.} 
\be
\label{eq:spectraldecG}
\mathcal G = \sum_{j> 0} \frac{1}{(1-\lambda_j)} \frac{\ketbra{\lambda_j, {\rm r}}{\lambda_j, {\rm l}}}{\braket{\lambda_j, {\rm r}}{\lambda_j, {\rm l}}}\,,
\ee
where $\ket{\lambda_j, {\rm l}}$ and $\bra{\lambda_j, {\rm r}}$ are respectively the (orthogonal) right and left eigenvectors corresponding to the $j$-th sub-leading eigenvalue $\lambda_j$.
Next, assuming $1 - \lambda_1 \ll |1-\lambda_{j>1}|$ we truncate the spectral decomposition \eqref{eq:spectraldecG} to the leading eigenvalue 
\be
\label{eq:approxG}
\mathcal G \approx  \frac{1}{(1-\lambda_1)} \frac{\ketbra{\lambda_1, {\rm r}}{\lambda_1, {\rm l}}}{\braket{\lambda_1, {\rm r}}{\lambda_1, {\rm l}}}.
\ee
This gives 
\be
\label{eq:approxncoeff}
[n] \approx \frac{1}{(1-\lambda_1)^n} \frac{\mel{1}{\bar{\cal T}_\epsilon}{\lambda_1, {\rm r}}\mel{\lambda_1, {\rm l}}{\bar{\cal T}_\epsilon}{1}}{\braket{\lambda_1, {\rm r}}{\lambda_1, {\rm l}}},
\ee
which is consistent with the numerical observation \eqref{eq:numobs}. 

In fact, the approximate form \eqref{eq:approxG} of the resolvent can also be used to explain the exponential decay in time observed in the upper panels of Figs.~\ref{fig:Rpeaked} and~\ref{fig:Rpeakedv2}. We begin noting that for a large enough $\ell \in \mathbb N$ one can make the following approximation  
\be
\label{eq:approxproj}
\frac{\ketbra{\lambda_1, {\rm r}}{\lambda_1, {\rm l}}}{\braket{\lambda_1, {\rm r}}{\lambda_1, {\rm l}}} \approx \left(\frac{\mathcal T_0 \mathcal Q}{\lambda_1}\right)^\ell =  \left(\frac{\mathcal T_0}{\lambda_1}\right)^\ell \mathcal Q, 
\ee
where $\mathcal Q$ is the projector on the leading eigenspace of $\mathcal T_0$ (cf.\ Eq.~\eqref{eq:Kktilde}) and we neglected terms that are prima facile $O((\lambda_j/\lambda_1)^\ell)$. This approximation is in fact more subtle than it might appear because the terms $\ketbra{\lambda_j, {\rm r}}{\lambda_j, {\rm l}}/({\braket{\lambda_j, {\rm r}}{\lambda_j, {\rm l}}})$ can have large operator norm (possibly even exponentially large in $t$). This means that one might need to consider $\ell = O(t)$ to safely neglect higher order terms.  Here we assume that this is not the case and take $\ell$ to be $O(t^0)$. 

Using \eqref{eq:approxproj} we can rewrite Eq.~\eqref{eq:approxncoeff} as  
\be
\label{eq:appn2}
\begin{aligned}
[n] {(1-\lambda_1)^n} &\approx \frac{1}{\lambda_1}\mel*{1}{\bar{\cal T}_\epsilon  {\cal T}^\ell_0 \mathcal Q  \bar{\cal T}_\epsilon }{1}\\
&= \frac{1}{\lambda_1 \epsilon^2}(\mel*{1}{{\cal T}  {\cal T}^\ell_0 {\cal T} }{1}-\mel*{1}{{\cal T}}{1}^2). 
\end{aligned}
\ee
Next, we rewrite the last line in terms of the original time evolving gates, undoing the space-time duality transformation discussed in Sec.~\ref{sec:SFF}. The idea is to represent the expectation value of a product of $n$ transfer matrices on a state in terms of the time-evolution operator of a chain of $n$ qubits. The state translates into the boundary conditions imposed on the time-evolution operator. In our case the boundary conditions will pair forward and backward evolution, giving a non-unitary boundary term to the evolution operator. More concretely, considering for instance the first term we have  
\be
\label{eq:expansionterm}
\begin{aligned}
\mel*{1}{{\cal T}  {\cal T}^\ell_0 {\cal T} }{1} &= \frac{\sum_{\tau=0}^{t-1} \mel*{\1}{{\cal T} {\cal T}^\ell_0 {\cal T} }{\Pi_{2t}^{2\tau}}}{\sum_{\tau=0}^{t-1} \braket{\1}{\Pi_{2t}^{2\tau}}}\\
&\simeq \frac{1}{2^{2t}} \sum_{\tau=0}^{t-1} \mel*{\1}{{\cal T} {\cal T}^\ell_0 {\cal T} }{\Pi_{2t}^{2\tau}},
\end{aligned}
\ee
where $\ket{O}$ is the state corresponding to the operator $O$ under the mapping in Eq.~\eqref{eq:mapping}. In the second step we dropped terms that are at most $O(2^{-t})$ by using  
\be
\sum_{\tau=0}^{t-1} \braket{\1}{\Pi_{2t}^{2\tau}} = \sum_{\tau=0}^{t-1} 2^{2 {\rm gcd}(t,
\tau) }= 2^{2t} + \sum_{\tau=1}^{t-1} 2^{2 {\rm gcd}(t,\tau) },
\ee
and noting that the sum on the r.h.s.\ is bounded by $2^t$.

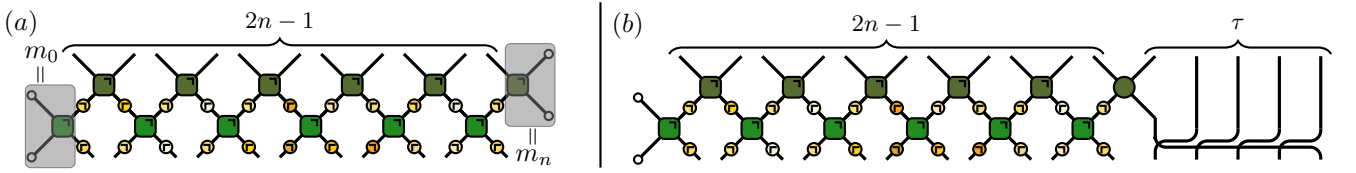
\begin{figure*}
\begin{tikzpicture}[baseline=(current  bounding  box.center), scale=0.55]
\Wgategreen{-10}{0}\Wgategreen{-8}{0}\Wgategreen{-6}{0}\Wgategreen{-4}{0}\Wgategreen{-2}{0}\Wgategreen{0}{0}
\Wgateolivegreen{-9}{1}\Wgateolivegreen{-7}{1}\Wgateolivegreen{-5}{1}\Wgateolivegreen{-3}{1}\Wgateolivegreen{-1}{1}\Wgateolivegreen{1}{1}
\draw [thick, black,decorate,decoration={brace,amplitude=5pt,mirror},xshift=0.0pt,yshift=-0.0pt](0.5,1.8) -- (-10,1.8) node[black,midway,yshift=0.4cm] {$2n-1$};
\draw[ thick, fill=myyellow1, rounded corners=2pt] (0.5,-0.5) circle (.15);
\draw[ thick, fill=myyellow2, rounded corners=2pt] (-0.5,-.5) circle (.15);
\draw[ thick, fill=myyellow3, rounded corners=2pt] (-1.5,-.5) circle (.15);
\draw[ thick, fill=myyellow4, rounded corners=2pt] (-2.5,-.5) circle (.15);
\draw[ thick, fill=myyellow5, rounded corners=2pt] (-3.5,-.5) circle (.15);
\draw[ thick, fill=myyellow6, rounded corners=2pt] (-4.5,-.5) circle (.15);
\draw[ thick, fill=myyellow7, rounded corners=2pt] (-5.5,-.5) circle (.15);
\draw[ thick, fill=myyellow8, rounded corners=2pt] (-6.5,-.5) circle (.15);
\draw[ thick, fill=myyellow9, rounded corners=2pt] (-7.5,-.5) circle (.15);
\draw[ thick, fill=myyellow10, rounded corners=2pt] (-8.5,-.5) circle (.15);
\draw[ thick, fill=myyellow1, rounded corners=2pt] (-9.5,-0.5) circle (.15);
\draw[ thick, fill=myyellow8, rounded corners=2pt] (0.5,0.5) circle (.15);
\draw[ thick, fill=myyellow9, rounded corners=2pt] (-0.5,0.5) circle (.15);
\draw[ thick, fill=myyellow1, rounded corners=2pt] (-1.5,0.5) circle (.15);
\draw[ thick, fill=myyellow3, rounded corners=2pt] (-2.5,0.5) circle (.15);
\draw[ thick, fill=myyellow2, rounded corners=2pt] (-3.5,0.5) circle (.15);
\draw[ thick, fill=myyellow4, rounded corners=2pt] (-4.5,0.5) circle (.15);
\draw[ thick, fill=myyellow10, rounded corners=2pt] (-5.5,0.5) circle (.15);
\draw[ thick, fill=myyellow9, rounded corners=2pt] (-6.5,0.5) circle (.15);
\draw[ thick, fill=myyellow8, rounded corners=2pt] (-7.5,0.5) circle (.15);
\draw[ thick, fill=myyellow7, rounded corners=2pt] (-8.5,0.5) circle (.15);
\draw[ thick, fill=myyellow1, rounded corners=2pt] (-9.5,0.5) circle (.15);
\draw[thick, fill=white] (-10.75,-0.75) circle (0.1cm); 
\draw[thick, fill=white] (-10.75,0.75) circle (0.1cm); 
\draw[thick, fill=white] (1.75,1.75) circle (0.1cm); 
\draw[thick, fill=white] (1.75,0.25) circle (0.1cm); 
\foreach \jj in {-0.5}{
\foreach \i in {1,...,5}{
\draw[ thick] (-2*\i+1.4,-.95-\jj) -- (-2*\i+1.55,-.95-\jj) -- (-2*\i+1.55,-1.1-\jj);}
\foreach \i in {1,...,6}{
\draw[ thick] (-2*\i+2.4,0.05-\jj) -- (-2*\i+2.55,0.05-\jj) -- (-2*\i+2.55,-0.1-\jj);}
\foreach \i in {1,...,5}{
\draw[ thick]  (-2*\i+1.45,-0.1-\jj) -- (-2*\i+1.45,0.05-\jj) -- (-2*\i+1.6,0.05-\jj);}
\foreach \i in {0,...,5}{
\draw[ thick]  (-2*\i+.45,-1.1-\jj) -- (-2*\i+.45,-0.95-\jj) -- (-2*\i+.6,-0.95-\jj);}
}
\Text[x=-11,y=2.5]{$(a)$}
\draw[ thick]  (3,-1) -- (3,3);
\draw[thin, rounded corners=2pt, fill= gray, opacity=0.5] (1-10.7,1) rectangle (-.2-10.7,-1);
\draw (-10.5,-.75+2) node[rotate=90] {$=$};
\Text[x=-10.5,y=-.25+2]{$m_0$}
\draw[thin, rounded corners=2pt, fill= gray, opacity=0.5] (.7,2) rectangle (1.9,0);
\draw (1.4,-.75+.5) node[rotate=90] {$=$};
\Text[x=1.4,y=-1.25+.5]{$m_n$}
\end{tikzpicture}
\hspace{-.15cm}
\begin{tikzpicture}[baseline=(current  bounding  box.center), scale=0.55]
\Wgategreen{-10}{0}\Wgategreen{-8}{0}\Wgategreen{-6}{0}\Wgategreen{-4}{0}\Wgategreen{-2}{0}\Wgategreen{0}{0}
\Wgateolivegreen{-9}{1}\Wgateolivegreen{-7}{1}\Wgateolivegreen{-5}{1}\Wgateolivegreen{-3}{1}\Wgateolivegreen{-1}{1}\CircularGate{1}{1}
\draw [thick, black,decorate,decoration={brace,amplitude=5pt,mirror},xshift=0.0pt,yshift=-0.0pt](0.5,1.8) -- (-10,1.8) node[black,midway,yshift=0.4cm] {$2n-1$};
\draw [thick, black,decorate,decoration={brace,amplitude=5pt,mirror},xshift=0.0pt,yshift=-0.0pt](6,1.8) -- (1.5,1.8) node[black,midway,yshift=0.4cm] {$\tau$};
\draw[ thick, fill=myyellow1, rounded corners=2pt] (0.5,-0.5) circle (.15);
\draw[ thick, fill=myyellow2, rounded corners=2pt] (-0.5,-.5) circle (.15);
\draw[ thick, fill=myyellow3, rounded corners=2pt] (-1.5,-.5) circle (.15);
\draw[ thick, fill=myyellow4, rounded corners=2pt] (-2.5,-.5) circle (.15);
\draw[ thick, fill=myyellow5, rounded corners=2pt] (-3.5,-.5) circle (.15);
\draw[ thick, fill=myyellow6, rounded corners=2pt] (-4.5,-.5) circle (.15);
\draw[ thick, fill=myyellow7, rounded corners=2pt] (-5.5,-.5) circle (.15);
\draw[ thick, fill=myyellow8, rounded corners=2pt] (-6.5,-.5) circle (.15);
\draw[ thick, fill=myyellow9, rounded corners=2pt] (-7.5,-.5) circle (.15);
\draw[ thick, fill=myyellow10, rounded corners=2pt] (-8.5,-.5) circle (.15);
\draw[ thick, fill=myyellow1, rounded corners=2pt] (-9.5,-0.5) circle (.15);
\draw[ thick, fill=myyellow8, rounded corners=2pt] (0.5,0.5) circle (.15);
\draw[ thick, fill=myyellow9, rounded corners=2pt] (-0.5,0.5) circle (.15);
\draw[ thick, fill=myyellow1, rounded corners=2pt] (-1.5,0.5) circle (.15);
\draw[ thick, fill=myyellow3, rounded corners=2pt] (-2.5,0.5) circle (.15);
\draw[ thick, fill=myyellow2, rounded corners=2pt] (-3.5,0.5) circle (.15);
\draw[ thick, fill=myyellow4, rounded corners=2pt] (-4.5,0.5) circle (.15);
\draw[ thick, fill=myyellow10, rounded corners=2pt] (-5.5,0.5) circle (.15);
\draw[ thick, fill=myyellow9, rounded corners=2pt] (-6.5,0.5) circle (.15);
\draw[ thick, fill=myyellow8, rounded corners=2pt] (-7.5,0.5) circle (.15);
\draw[ thick, fill=myyellow7, rounded corners=2pt] (-8.5,0.5) circle (.15);
\draw[ thick, fill=myyellow1, rounded corners=2pt] (-9.5,0.5) circle (.15);
\draw[thick, fill=white] (-10.75,-0.75) circle (0.1cm); 
\draw[thick, fill=white] (-10.75,0.75) circle (0.1cm); 
\foreach \jj in {-0.5}{
\foreach \i in {1,...,5}{
\draw[ thick] (-2*\i+1.4,-.95-\jj) -- (-2*\i+1.55,-.95-\jj) -- (-2*\i+1.55,-1.1-\jj);}
\foreach \i in {1,...,6}{
\draw[ thick] (-2*\i+2.4,0.05-\jj) -- (-2*\i+2.55,0.05-\jj) -- (-2*\i+2.55,-0.1-\jj);}
\foreach \i in {1,...,5}{
\draw[ thick]  (-2*\i+1.45,-0.1-\jj) -- (-2*\i+1.45,0.05-\jj) -- (-2*\i+1.6,0.05-\jj);}
\foreach \i in {0,...,5}{
\draw[ thick]  (-2*\i+.45,-1.1-\jj) -- (-2*\i+.45,-0.95-\jj) -- (-2*\i+.6,-0.95-\jj);}
}
\foreach \jj in {0,...,3}{
\draw[very thick, rounded corners]  (1.75+\jj,-0.75) -- (1.75+\jj,-0.3) -- (2.75+\jj,-0.3) -- (2.75+\jj,0);
}
\foreach \jj in {0,...,3}{
\draw[very thick, rounded corners]  (2.75+\jj,0) -- (2.75+\jj,1.75);
}
\draw[very thick, rounded corners]  (2.75+3,-0.75) -- (2.75+3,-0.45) -- (1.75,-0.45) -- (1.75,0) -- (1.75,.25);
\Text[x=-11,y=2.5]{$(b)$}
\end{tikzpicture}
\caption{Diagrammatic representation of $\mathcal B_{n,0}$ (a) and $\mathcal B_{n,\tau\neq0}$ (b).}
\label{fig:Bmap}
\end{figure*}

The terms on the r.h.s.\ of Eq.~\eqref{eq:expansionterm} can be easily translated in the time evolving picture because the states $\ket*{\Pi_{2t}^{2j}}$ implement simple pairings between backward and forward evolution. For instance, the term with $\tau=0$ is written as  
\be
\label{eq:TtoB}
\mel*{\1}{{\cal T} {\cal T}^\ell_0 {\cal T} }{\1} = \mathbb E\left[\tr[\mathcal B_{\ell+2,0}^{t}]\right],
\ee
where we introduced the $4^{2n-1} \times 4^{2n-1}$ matrix 
\be
\mathcal B_{n,0} = (( \mathbb U \otimes_r \mathbb U^*) \otimes m_{n}) (m_0\otimes (\mathbb W \otimes_r \mathbb W^*)).
\ee
The latter is written in terms of the time evolution operators for $2n-2$ qubits ($\in {\rm End}(\mathbb C^{2^{2n-2}})$)
\begin{align}
& \mathbb U = U_{0} \otimes \ldots \otimes U_{n-1}, \\
& \mathbb W = U_{1/2} \otimes \ldots \otimes U_{n-1/2}, 
\end{align}
and the boundary matrices ($\in {\rm End}(\mathbb C^{4})$)
\begin{align}
&[m_0]_{ij} =  \frac{1}{2} \sum_{r,s=1}^{2} [U_{-1/2}]_{s,j_1}^{r,i_1} ([U_{-1/2}]_{s,j_2}^{r,i_2})^* ,\\
&[m_{n}]_{ij} =  \frac{1}{2} \sum_{r,s=1}^{2} [U_{n}]_{j_1,s}^{i_1,r} ([U_{n}]_{j_2,s}^{i_2,r})^*. 
\end{align}
Introducing a convenient diagrammatic representation for objects acting on both the forward and backward time sheets 
\begin{align}
&\begin{tikzpicture}[baseline=(current  bounding  box.center), scale=.8]
\def\eps{0.5};
\draw[very thick] (-0.5, 0.5) -- (0.5,-0.5);
\draw[very thick] (-0.5,-0.5) -- (0.5,0.5);
\draw[ thick, fill=mygreen, rounded corners=2pt] (-0.25,+0.25) rectangle (0.25,-0.25);
\draw[thick] (0,0.15) -- (0.15,0.15) -- (0.15,0);
\Text[x=-0,y=-0.6]{}
\end{tikzpicture}
=
\begin{tikzpicture}[baseline=(current  bounding  box.center), scale=.8]
\draw[thick] (-1.65,0.65) -- (-0.65,-0.35);
\draw[thick] (-1.65,-0.35) -- (-0.65,0.65);
\draw[ thick, fill=myblue, rounded corners=2pt] (-1.4,0.4) rectangle (-.9,-0.1);
\draw[thick] (-1.15,0) -- (-1,0) -- (-1,0.15);
\draw[thick] (-2.25,0.5) -- (-1.25,-0.5);
\draw[thick] (-2.25,-0.5) -- (-1.25,0.5);
\draw[ thick, fill=myred, rounded corners=2pt] (-2,0.25) rectangle (-1.5,-0.25);
\draw[thick] (-1.75,0.15) -- (-1.6,0.15) -- (-1.6,0);
\Text[x=-2.25,y=-0.6]{}
\end{tikzpicture}
= U\otimes U^{*}\,,\notag\\ 
&\begin{tikzpicture}[baseline=(current  bounding  box.center), scale=.8]
\def\eps{0.5};
\draw[very thick] (-0.5, 0.5) -- (0.5,-0.5);
\draw[very thick] (-0.5,-0.5) -- (0.5,0.5);
\draw[ thick, fill=OliveGreen, rounded corners=2pt] (-0.25,+0.25) rectangle (0.25,-0.25);
\draw[thick] (0,0.15) -- (0.15,0.15) -- (0.15,0);
\Text[x=0,y=-0.6]{}
\end{tikzpicture}
=
\begin{tikzpicture}[baseline=(current  bounding  box.center), scale=.8]
\draw[thick] (-1.65,0.65) -- (-0.65,-0.35);
\draw[thick] (-1.65,-0.35) -- (-0.65,0.65);
\draw[ thick, fill=myblue4, rounded corners=2pt] (-1.4,0.4) rectangle (-.9,-0.1);
\draw[thick] (-1.15,0) -- (-1,0) -- (-1,0.15);
\draw[thick] (-2.25,0.5) -- (-1.25,-0.5);
\draw[thick] (-2.25,-0.5) -- (-1.25,0.5);
\draw[ thick, fill=myorange, rounded corners=2pt] (-2,0.25) rectangle (-1.5,-0.25);
\draw[thick] (-1.75,0.15) -- (-1.6,0.15) -- (-1.6,0);
\Text[x=-2.25,y=-0.6]{}
\end{tikzpicture}
= W\otimes W^{*}\,,\\
 \begin{tikzpicture}[baseline=(current  bounding  box.center), scale=.7]
\draw[very thick] (-4.25,0.5) -- (-4.25,-0.5);
\draw[ thick, fill=myYO, rounded corners=2pt] (-4.25,0) circle (.15);
\draw[thick, rotate around = {-45:(0.525-4.77,0.375-0.4)}]  (.45-4.77,0.3-0.4) -- (.45-4.77,0.45-0.4) -- (.6-4.77,0.45-0.4);
\Text[x=-4.25,y=-0.75]{}
\end{tikzpicture}
 =
\begin{tikzpicture}[baseline=(current  bounding  box.center), scale=.7]
\draw[ thick] (-4,0.5) -- (-4,-0.5);
\draw[ thick, fill=mygray4, rounded corners=2pt] (-4,0) circle (.15);
\draw[thick, rotate around = {135:(0.525-4.27-0.25,0.375-0.35)}]  (.45-4.27-.25,0.3-0.35) -- (.45-4.27-.25,0.45-0.35) -- (.6-4.27-.25,0.45-0.35);
\draw[ thick] (-4.25,0.5) -- (-4.25,-0.5);
\draw[ thick, fill=myblue10, rounded corners=2pt] (-4.25,0) circle (.15);
\draw[thick, rotate around = {-45:(0.525-4.77,0.375-0.4)}]  (.45-4.77,0.3-0.4) -- (.45-4.77,0.45-0.4) -- (.6-4.77,0.45-0.4);
\Text[x=-4.25,y=-0.75]{}
\end{tikzpicture}
& = u_{x}\otimes u_{x}^{*},\,\, w_{x}\otimes w_{x}^{*}\,,\qquad 
\begin{tikzpicture}[baseline=(current  bounding  box.center), scale=.7]
\draw[very thick] (-0.15,0.25) -- (-0.15,-0.251);
\draw[thick, fill=white] (-.15,-0.25) circle (0.1cm); 
\end{tikzpicture}
=
\frac{1}{\sqrt{2}}\,\,
\begin{tikzpicture}[baseline=(current  bounding  box.center), scale=.7]
\draw[thick] (-2,0.25) -- (-2,-0.251);
\draw[thick] (-1.5,0.4) -- (-1.5,-0.101);
\draw[thick] (-2,-0.25) to[out=-85,in=-80] ( (-1.505,-0.1);
\end{tikzpicture}\,,
\label{eq:circstate}
\end{align}
we can depict $\mathcal B_{n,0}$ as in Fig.~\ref{fig:Bmap}a. Analogously, a generic term with $\tau\neq 0$ is written as 
\be
\mel*{\1}{{\cal T} {\cal T}^\ell_0 {\cal T} }{\Pi_{2t}^{2\tau}} = \mathbb E\left[\tr[\mathcal B^{t}_{\ell+2,\tau}]\right],
\ee
where we introduced the matrices
\begin{align}
&\mathcal B_{n,\tau} = ((\mathbb U \!\otimes_r\! \mathbb U^*) \!\otimes\! b_{\tau} \otimes \1^{\otimes 2(\tau-1)}) \notag\\
&\qquad\quad \times (m_0\!\otimes\!((\mathbb W\otimes \Pi_{\tau}) \!\otimes_r\! (\mathbb W^*\otimes \Pi^*_{\tau}))),\\
&[b_\tau]_{i j} = \frac{1}{d} [U_{n}]_{j_1,j_3}^{i_1,j_4} ([U_{n}]_{j_2,i_3}^{i_2,i_4})^*.
\end{align}
Note that $\mathcal B_{n,\tau}\in{\rm End}(\mathbb C^{4^{2n+\tau-1}})$ and $b_\tau \in{\rm End}(\mathbb C^{4^{2}})$. Introducing the following diagrammatic representation for $b_\tau$,  
\be
b_\tau= 
\begin{tikzpicture}[baseline=(current  bounding  box.center), scale=.7]
\draw[very thick] (-0.5,  +0.5) -- (0.5,-0.5);
\draw[very thick] (-0.5,-0.5) -- (0.5,0.5);
\draw[thick, fill=OliveGreen, rounded corners=2pt] (0,0) circle (.35);
\Text[x=0,y=-0.6]{}
\end{tikzpicture},
\ee
we can depict $\mathcal B_{n,\tau}$ as in Fig.~\ref{fig:Bmap}b.

The traces of $\mathcal B_{n,\tau}$ can be treated following Ref.~\cite{kos2021thermalization}. In particular, using Theorem 1 of the aforementioned reference we have that if there are no $x\leq y$ such that 
\begin{align}
U_x (\1\otimes a) U^\dag_x = \1\otimes a', \label{eq:leftcondition}\\
U_y (b\otimes \1) U^\dag_y = b'\otimes \1, \label{eq:rightcondition}
\end{align}
for some local operators $a,a',b,b'$, then 
\be
\label{eq:Bt}
\tr[\mathcal B_{n,0}^{t}] = 1 + O(e^{-\beta t})\,, \quad \beta>0.
\ee
Note that, although $\beta>0$ for all values of $n$, Ref.~\cite{kos2021thermalization} gives no information on its $n$ dependence.  

With a similar reasoning we prove in Appendix~\ref{app:proofBtau} that if \eqref{eq:leftcondition} does not hold for any $x$, then $\rho(\mathcal B_{n,\tau})<1$, where we used $\rho(\cdot)$ to denote the spectral radius. This implies  
\be
\label{eq:Btaut}
\tr[\mathcal B_{n,\tau}^{t}] = O(e^{-\beta t}), \qquad \tau >0\,. 
\ee
Since for $(\epsilon_1,\epsilon_2)\neq (\pi/4,\pi/4)$ the gates fulfilling \eqref{eq:leftcondition} and \eqref{eq:rightcondition} have measure zero in the disorder average, we conclude that 
\be
\label{eq:TTonT}
\mel*{1}{{\cal T}  {\cal T}^\ell_0 {\cal T} }{1} \simeq 1+ A(\epsilon) e^{-\beta(\epsilon) t},
\ee
where the constant $A(\epsilon)$ vanishes for $\epsilon=0$ because the l.h.s.\ is trivially equal to one for $\epsilon=0$ while Fig.~\ref{fig:Rpeakedv2} suggests that $\beta(0)$ is finite. Note that, since we do not control the $\beta$ dependence on $\ell$, we cannot exclude that it approaches 0 in the limit of infinite $\ell$. This is why we had to assume $\ell=O(t^0)$ in Eq.~\eqref{eq:approxproj}. 

Proceeding analogously we find 
\be
\mel*{1}{{\cal T}}{1}  \simeq 1+ B(\epsilon) e^{-\beta(\epsilon) t}\,,
\ee
with $B(\epsilon)\simeq A(0)' \epsilon/2$ for small $\epsilon$. Putting all together in Eq.~\eqref{eq:appn2} we then have 
\be
\label{eq:ntermfinal}
[n]  \approx \frac{C_2(\epsilon) e^{-\beta(\epsilon) t}}{\lambda_1^2 (1-\lambda_1)^{n}}, 
\ee
where $C_2(\epsilon)$ is $O(1)$ for small $\epsilon$. If we compare Eq.~\eqref{eq:ntermfinal} with our two parameter fit for $[n]$ (Eq.~\eqref{eq:fit}), we predict that $-\beta$ is the slope of  $\delta$ with respect to $t$. The right panel of Fig.~\ref{fig:betalin} then suggests that $\beta(\epsilon)$ depends weakly on $\epsilon$ for small enough $\epsilon$.

\begin{figure*}[t]
\centering
\includegraphics[width=0.45\linewidth]{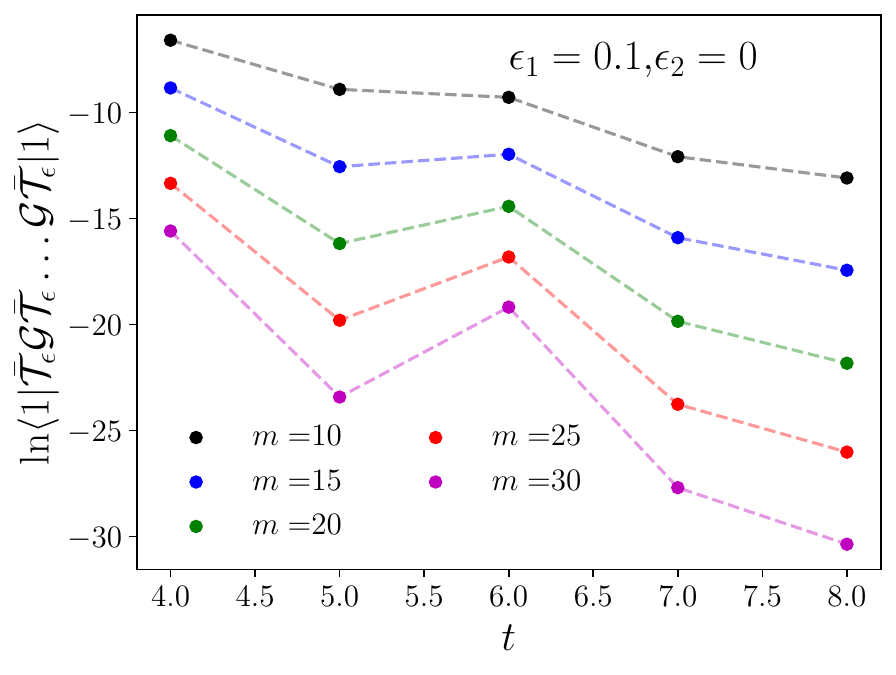}
\includegraphics[width=0.45\linewidth]{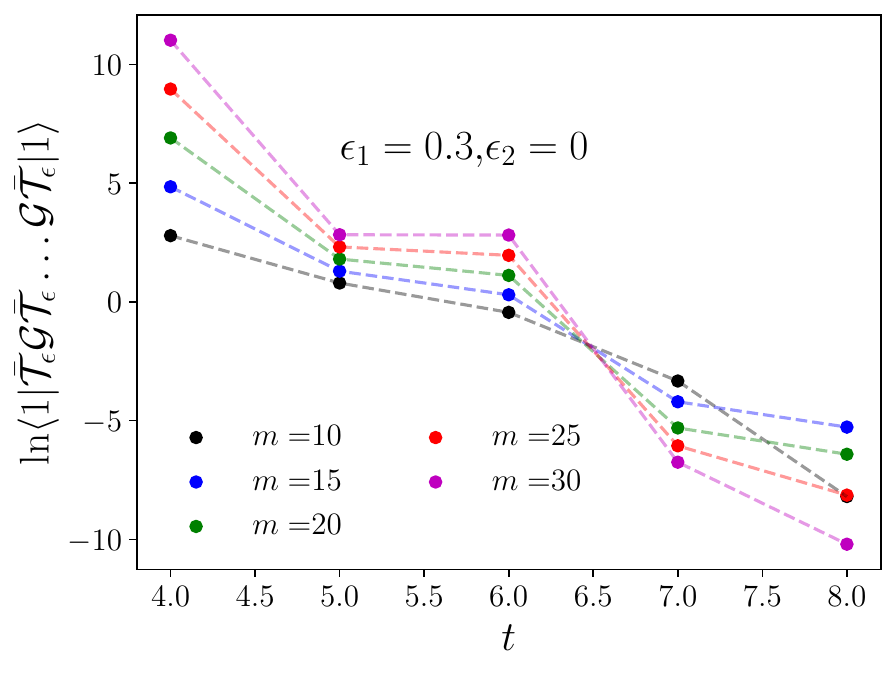}
\includegraphics[width=0.45\linewidth]{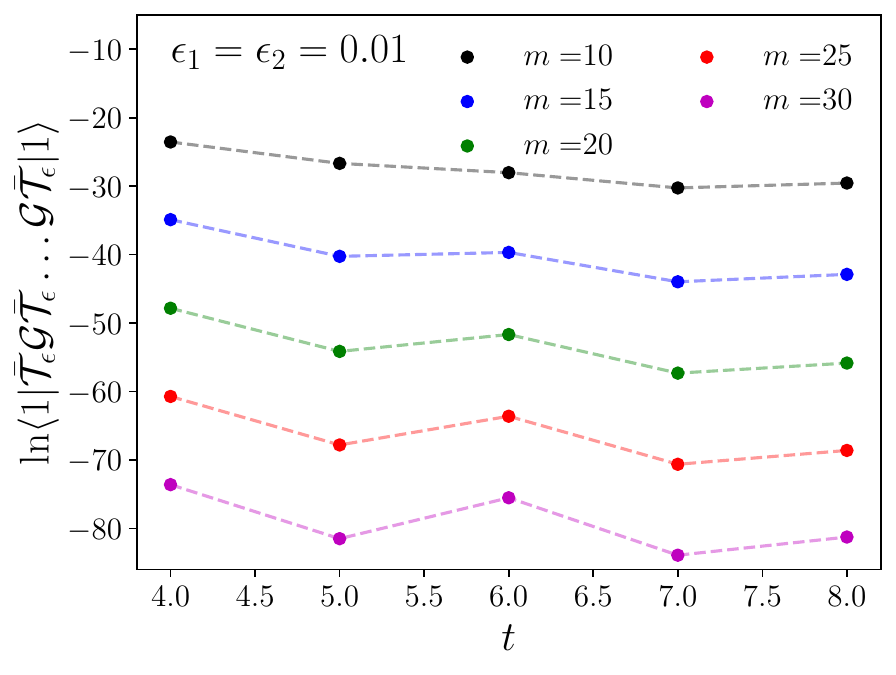}
\includegraphics[width=0.45\linewidth]{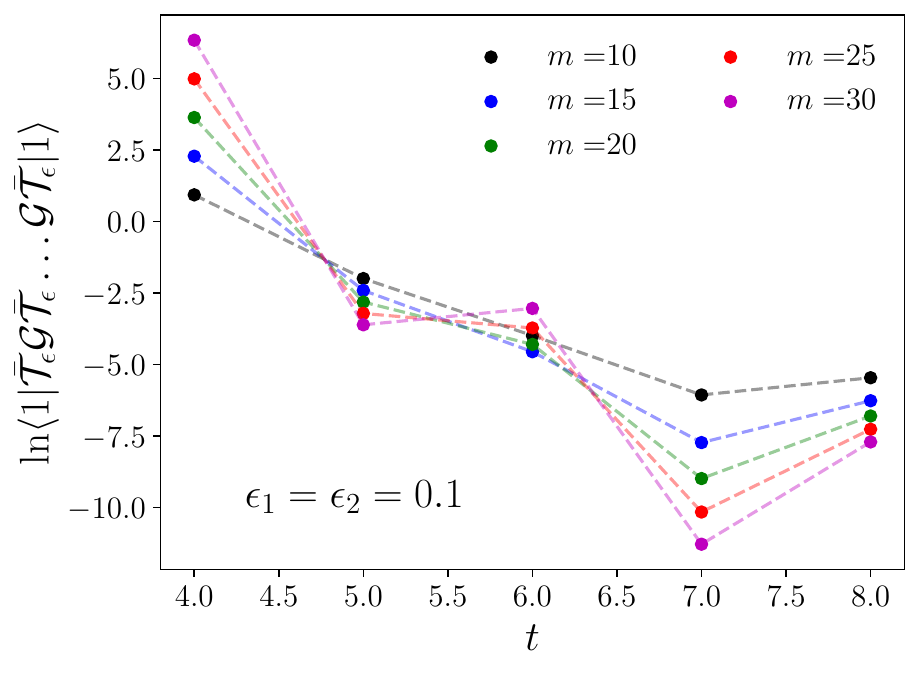}
\caption{Flat coefficients $[1,1,1\dots 1]$ as a function of $t$ for different values of $m$ and $\epsilon$. The top panels report two examples of Case I ($(\epsilon_1, \epsilon_2)=(0.1,0)$ and $(\epsilon_1,\epsilon_2)=(0.3,0)$) while the bottom ones report two examples of Case II  ($\epsilon_1 = \epsilon_2= 0.01$ and $\epsilon_1 = \epsilon_2= 0.1$).  }
\label{fig:flat1}
\end{figure*}

\begin{figure*}[t]
\centering
\includegraphics[width=0.45\linewidth]{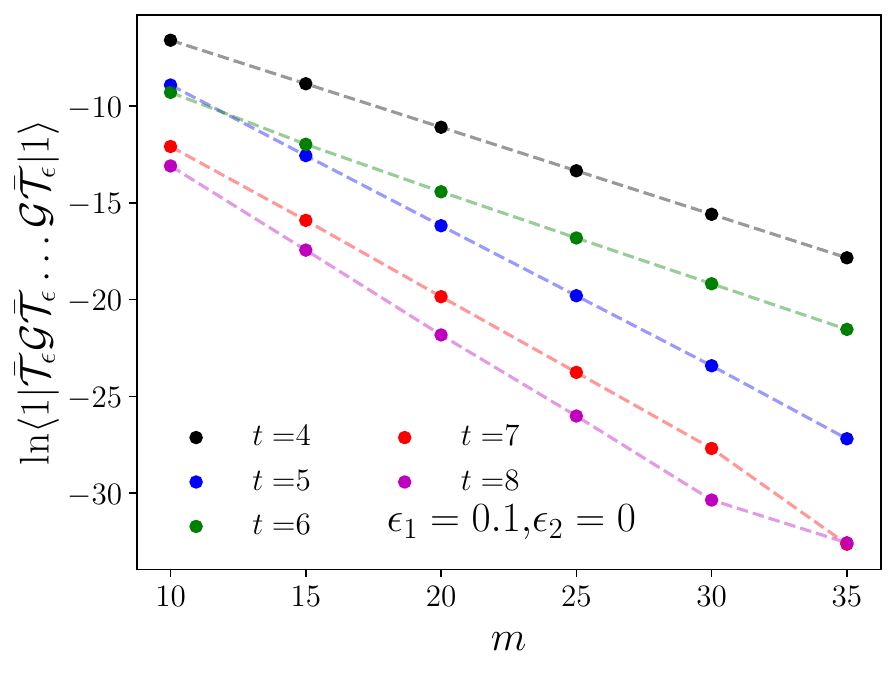}
\includegraphics[width=0.45\linewidth]{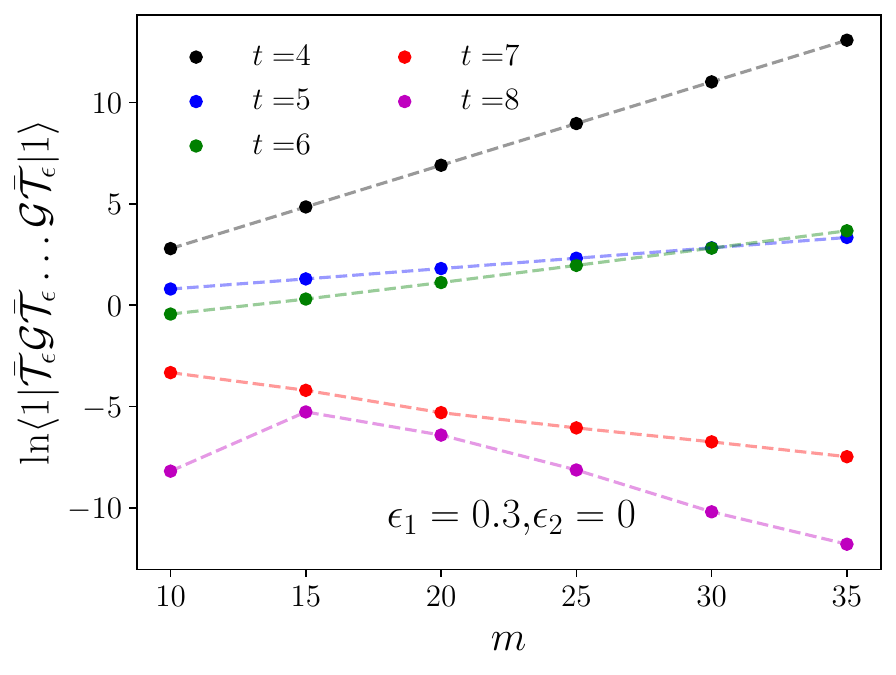}
\includegraphics[width=0.45\linewidth]{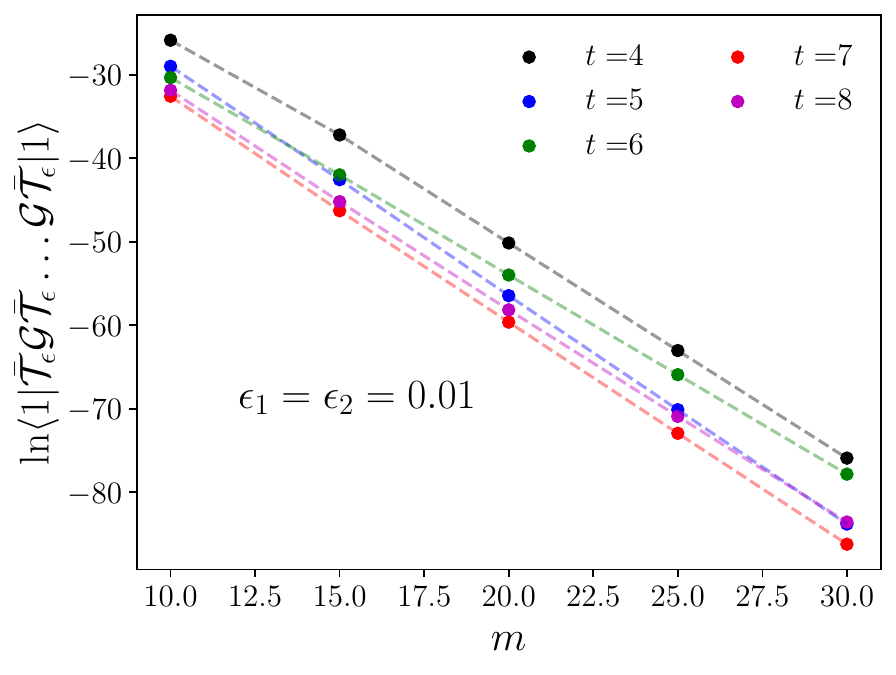}
\includegraphics[width=0.45\linewidth]{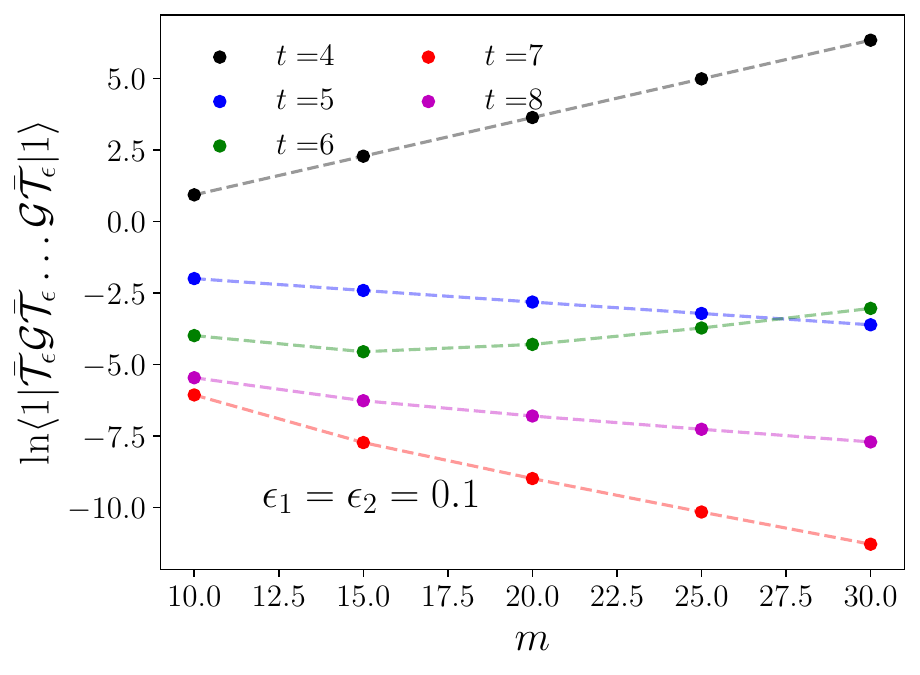}
\caption{Flat coefficients $[111\dots 1]$ as a function of $m$ for different values of $t$ and $\epsilon$. The top panels report two examples of Case I ($(\epsilon_1, \epsilon_2)=(0.1,0)$ and $(\epsilon_1,\epsilon_2)=(0.3,0)$) while the bottom ones report two examples of Case II  ($\epsilon_1 = \epsilon_2= 0.01$ and $\epsilon_1 = \epsilon_2= 0.1$). }
\label{fig:flat2}
\end{figure*}

Proceeding along similar lines we can estimate all the coefficients in Eq.~\eqref{eq:coefficients}. In particular, using the approximations \eqref{eq:approxG} and \eqref{eq:approxproj} we have that the generic coefficient ${[k_1,\ldots,k_m]} $ is written as 
\be
{[k_1,\ldots,k_m]}  \approx \frac{\mel*{1}{\bar{\cal T}_\epsilon  (\mathcal T_0 \mathcal Q \bar{\cal T}_\epsilon)^m}{1}}{\lambda_1^{m}(1-\lambda_1)^{k_1+\ldots+k_m}} .
\ee
While applying \eqref{eq:Bt} and \eqref{eq:Btaut} we have 
\be
\label{eq:estimates}
\begin{aligned}
{[k_1,\ldots,k_m]}  &\approx \frac{C_{m+1}(\epsilon) e^{-\beta t}}{\lambda_1^{m}(1-\lambda_1)^{k_1+\ldots+k_m}},\\   
{[\phantom{,}]}  &\approx {C_{1}(\epsilon) e^{-\beta t}},
\end{aligned}
\ee
with $C_m(\epsilon)= O(1)$ for small $\epsilon$. This provides an analytical justification to Assumption~\ref{asmp:asmp2}. In Figs.~\ref{fig:flat1} and \ref{fig:flat2} we provide an independent numerical test of this assumption by considering the behaviour of the flat coefficients
\begin{eqnarray} \label{eq:flat}
    [1,1\dots 1] = \mel*{1}{\bar{\cal T}_\epsilon  \mathcal G \bar{\cal T}_\epsilon \mathcal G\cdots \mathcal G \bar{\cal T}_\epsilon }{1}. 
\end{eqnarray}
These contributions are those for which the approximation \eqref{eq:approxG} is the least justified as the latter becomes exact only when the resolvent is taken to infinite power. Nevertheless, from Fig.~\ref{fig:flat1} we clearly see an exponential decay in time in agreement with Eq.~\eqref{eq:estimates} and hence with Assumption~\ref{asmp:asmp2}. This despite the relatively short times accessible in our numerical simulations. Instead, Fig.~\ref{fig:flat2} reports the behaviour of $[1,1\dots 1]$ as a function of $m$, i.e.\  the number of ones in the coefficient. We see that  the coefficient decays exponentially in $m$. This suggests that $C_{m+1}(\epsilon)$ in Eq.~\eqref{eq:estimates} is bounded by an exponentially decaying constant. In fact we note that, at least for Case I and small enough $\epsilon$, Fig.~\ref{fig:flat1} shows that the exponential decay in time becomes stronger when $m$ increases. This can be explained by our asymptotic form Eq.~\eqref{eq:estimates} if we admit that the factor $\lambda_1(1-\lambda_1)$ appearing in the denominator increases as a function of time. Note that this growth cannot be unbounded as $\lambda_1(1-\lambda_1)\leq 1/4$.

Finally we stress that Eq.~\eqref{eq:estimates} do not hold at the trivially localised point $\epsilon_1 = \epsilon_2= \pi/4$. Indeed, in that case Eqs.~\eqref{eq:Bt} and \eqref{eq:Btaut} do not apply as Eqs.~\eqref{eq:leftcondition} and \eqref{eq:rightcondition} are clearly satisfied for all $x$ and $y$. As a result, in this case there is no exponential decay in time of the coefficients~\eqref{eq:coefficients}. For instance, using the simple form of $\mathcal T$ at the localised point (see, e.g., Ref.~\cite{bertini2022exactspectral}) it is easy to show that 
\be
{[\phantom{,}]}|_{\epsilon_1 = \epsilon_2= \pi/4}  = \frac{4}{\pi} + O(t^{-\alpha}). 
\ee

\section{Conclusions}
\label{sec:conclusions}

In this work we laid down a general framework to investigate the structural stability of dual-unitary spectral correlations, which can be loosely thought of as the quantum many-body analogue of the theory of structural stability of hyperbolic flows in classical chaotic dynamical systems~\cite{robbin,robinson}. 

Our guiding principle has been that, contrary to integrable systems, dual-unitary systems should be robust under typical perturbations as they are quantum chaotic. Therefore, the spectral correlations of a perturbed dual-unitary system, or at least their universal part, should be accessible by devising an appropriate perturbation theory. Here we formulated such a perturbation theory and identified two key assumptions needed for a rigorous proof of its convergence. We then provided a compelling numerical evidence for the validity of these assumptions in a particular family of perturbed Floquet dual-unitary circuits, and corroborated them with a heuristic analytical argument (supported by numerical evidence) involving the spectral decomposition of the resolvent of the unperturbed transfer matrix.

Besides their consequences on the structural stability of dual-unitary correlations, our findings have an important consequence on the interplay between ergodicity and disorder. Indeed, since spatial disorder does not affect dual-unitarity breaking (only two-body couplings can break dual-unitarity), our results imply that whenever a quantum many-body system is close to an interacting (non-SWAP) dual unitary point its spectral correlations are always random-matrix-like, irrespective of the disorder strength --- for instance, the particular family used in our numerical analysis has maximal disorder strength. This rules out the possibility of Floquet MBL in the thermodynamic limit for systems close enough to the dual-unitary point.   

Although these findings are remarkable, they are in many ways only a stepping stone to the development of a comprehensive theory of structural stability for Floquet dual-unitary circuits, and many key questions remain open. In particular, it would be important to explore the degree of generality of the structural stability we found in dual unitary circuits by investigating more general perturbations. Moreover, further research is required to corroborate and expand our perturbative approach. To this end, we identify two main directions.

The first is to find quantitative estimates or bounds for the radius of convergence of our perturbative expansion. Indeed, at the moment we have merely shown that, under our two assumptions, the radius is finite. However, we gave no information on its value. A quantitative estimate of the radius of convergence could potentially lead to the identification of the point of transition to the non-ergodic (i.e.\ localised) regime, which might be occur for a finite value of two-body coupling or only at the trivially localised point where the qubits are disconnected. A related question is whether one can identify the transition point expanding around the trivially localised point.

The second direction is, of course, to provide rigorous mathematical proofs of our assumptions, in particular to the second one that appears the more substantial. We believe that the heuristic analytical argument we provided in support of that assumption can be used as blueprint for such a proof.

\begin{acknowledgments}
We thank Pavel Kos for collaboration on related topics and Juan Garrahan for useful discussions. We acknowledge financial support from the Royal Society through the University Research Fellowship No.\ 201101 (J.\ R.\ and B.\ B.), a UKRI Future Leaders Fellowship MR/T040947/1 (C.\ K.), and Grants P1-0402, N1-0334, N1-0219 of Slovenian Research and Innovation Agency (T. P.). J.\ R., T.\ P.\ and  B.\ B.\ warmly acknowledge the hospitality of the Simons Center for Geometry and Physics during the program ``Fluctuations, Entanglements, and Chaos: Exact Results'' where part of the project has been carried out.
\end{acknowledgments}

\appendix

\section{Perturbation Theory to an Arbitrary Order}
\label{app:PT}

Writing explicitly the coefficient of $x^n$ in Eq.~\eqref{eq:perturbativeeeq} we have 
\be
\label{eq:pertordern}
(1-\mathcal T_0)\ket*{\lambda}^{(n)}  = \bar{\cal T}_\epsilon\ket*{\lambda}^{(n-1)}  - \sum_{k=1}^{n} \lambda^{(k)} \ket*{\lambda}^{(n-k)} ,  
\ee
where we assumed $n>0$ as the equation is trivially satisfied for $n=0$. Next, we note that fixing the arbitrary phase and normalisation of $\ket{\lambda}$ as 
\be
\braket{\lambda}{\lambda}^{(0)}=1,
\ee
gives $|\lambda\rangle^{(n)}\perp |\lambda\rangle^{(0)}$ for $n>0$. Therefore, we can rewrite Eq.~\eqref{eq:pertordern} as 
\begin{align}
\!\!\!\lambda^{(n)} &=  \langle \lambda|^{(0)}\bar{\cal T}_\epsilon \ket{\lambda}^{(n-1)}, \label{eq:recursiveeval}\\
\!\!\!\!\!\!\!\ \ket*{\lambda}^{(n)} \!\!  &= \mathcal G \bar{\cal T}_\epsilon \ket{\lambda}^{(n-1)}\! -\! \sum_{k=1}^{n-1} \lambda^{(k)} \mathcal G  \!\ket*{\lambda}^{(n-k)} \notag\\
&=\!\!  \sum_{k=1}^{n-1} \mathcal G \mathcal{K}_k \!\ket*{\lambda}^{(n-k)} \!\!.
\label{eq:recursivelambdavec}
\end{align}
To obtain the first equation we took the scalar product of Eq.~\eqref{eq:pertordern} with $\ket*{\lambda}^{(0)}$ and to obtain the second we multiplied it by $\mathcal Q$, used $\ket{\lambda}^{(n>0)}=\mathcal Q \ket{\lambda}^{(n>0)}$, and noted $\mathcal G = (\mathcal Q (1-\mathcal T_0) \mathcal Q)^{-1}$. Finally, we recalled the definition of $\mathcal{K}_k$ from Eq.~\eqref{eq:Kktilde}.

Using \eqref{eq:recursivelambdavec} we find 
\be
\ket*{\lambda}^{(1)}  = \mathcal G \bar{\cal T}_\epsilon \ket*{\lambda}^{(0)} , 
\ee
recovering the first of \eqref{eq:lambdaxn}. Moreover, for $n\geq 2$ we obtain  
\be
\ket*{\lambda}^{(n)}   = \mathcal G \mathcal{K}_{n-1} \mathcal G \bar{\cal T}_\epsilon \ket*{\lambda}^{(0)}  + \sum_{k=1}^{n-2} \mathcal G \mathcal{K}_k \ket*{\lambda}^{(n-k)} \,.
\ee  
\begin{widetext}
Using Eq.~\eqref{eq:recursivelambdavec} we then have 
\begin{align}
\ket*{\lambda}^{(n)}   &= \mathcal G \mathcal{K}_{n-1} \mathcal G \bar{\cal T}_\epsilon \ket*{\lambda}^{(0)}  + \sum_{k_1=1}^{n-2}\sum_{k_2=1}^{n-k_1-1} \mathcal G \mathcal{K}_{k_1} \mathcal G \mathcal{K}_{k_2} \ket*{\lambda}^{(n-k_1-k_2)} \notag\\
&=  \mathcal G \mathcal{K}_{n-1} \mathcal G \bar{\cal T}_\epsilon \ket*{\lambda}^{(0)}  + \sum_{k_1=1}^{n-2} \mathcal G \mathcal{K}_{k_1} \mathcal G \mathcal{K}_{n-k_1-1} \mathcal G \bar{\cal T}_\epsilon \ket*{\lambda}^{(0)}  + \sum_{k_1=1}^{n-2}\sum_{k_2=1}^{n-k_1-2} \mathcal G \mathcal{K}_{k_1} \mathcal G \mathcal{K}_{k_2} \ket*{\lambda}^{(n-k_1-k_2)} \notag\\
&=  \mathcal G \mathcal{K}_{n-1} \mathcal G \bar{\cal T}_\epsilon \ket*{\lambda}^{(0)}  + \sum_{\substack{k_1,k_2=1 \\ k_1+k_2=n-1}}^{n-1} \mathcal G \mathcal{K}_{k_1} \mathcal G \mathcal{K}_{k_2} \mathcal G \bar{\cal T}_\epsilon \ket*{\lambda}^{(0)}  + \sum_{k_1=1}^{n-2}\sum_{k_2=1}^{n-k_1-2} \mathcal G \mathcal{K}_{k_1} \mathcal G \mathcal{K}_{k_2} \ket*{\lambda}^{(n-k_1-k_2)} \,.
\end{align}
We iterate this procedure $n-1$ times and note that
\be
 \sum_{k_1=1}^{n-2}\sum_{k_2=1}^{n-k_1-2}\cdots \sum_{k_n=1}^{n-k_1\cdots -k_{n-1}-1} \mathcal G \mathcal{K}_{k_1} \mathcal G \mathcal{K}_{k_2}\cdots G \mathcal{K}_{k_n}\ket*{\lambda}^{(n-k_1 \cdots -k_n)} =0, 
\ee
to obtain Eq.~\eqref{eq:lambdaxn}. Using that expression in Eq.~\eqref{eq:recursiveeval} gives Eq.~\eqref{eq:lambdaxn}. 
\end{widetext}

\section{Spectral Radius of $\mathcal B_{n,\tau}$}
\label{app:proofBtau}

In this appendix we show that if Eq.~\eqref{eq:leftcondition} is not satisfied for any $x$, then $\rho(\mathcal B_{n,\tau})<1$. We begin by noting that $\|\mathcal B_{n,\tau}\|_\infty=1$. This can be easily seen writing 
\begin{align}
&\mathcal B_{n,\tau}\mathcal B_{n,\tau}^\dag =\notag\\
&\!\!\!= U_{1/2}(m_0 m^\dag_0 \otimes \1) U^\dag_{1/2}\otimes \1^{\otimes (2n-2)} \otimes b_\tau b_\tau^\dag \otimes \1^{\tau -1}
\end{align}
and observing $\|m_0\|=\|b_\tau\|=1$. Since 
\be
\rho(\mathcal B_{n,\tau}) \leq \|\mathcal B_{n,\tau}\|_{\infty}=1,
\ee
our goal is then to show that Eq.~\eqref{eq:leftcondition} is not satisfied for any $x$, $\mathcal B_{n,\tau}$ does not have a unit-magnitude eigenvalue. Let us proceed by contradiction and assume that there exists a $\ket{\Lambda}$(normalised) such that 
\be
\label{eq:eigeneqLambda}
\mathcal B_{n,\tau} \ket{\Lambda}= e^{i \theta} \ket{\Lambda}. 
\ee
Using $\|\mathcal B_{n,\tau}\|_\infty=1$ we then have 
\begin{align}
&\mathcal B_{n,\tau}^\dag \mathcal B_{n,\tau} \ket{\Lambda}= \ket{\Lambda},\\
&\mathcal B_{n,\tau} \mathcal B_{n,\tau}^\dag \ket{\Lambda} = \ket{\Lambda},\label{eq:BBdag}
\end{align}
which also give 
\be
\mathcal B^\dag_{n,\tau} \ket{\Lambda}= e^{-i \theta} \ket{\Lambda}\,. 
\ee
We now proceed along the lines of Ref.~\cite{kos2021thermalization} and note that \eqref{eq:BBdag} and the fact that Eq.~\eqref{eq:leftcondition} is not satisfied for any $x\leq y$ imply 
\be
\ket{\Lambda}= \ket{\mcirc}\otimes \ket{\Lambda'}, 
\ee
where $\ket{\mcirc} = \sum_{k=1}^2 \ket{k}\otimes_r\ket{k}/\sqrt{2}$ (cf.\ Eq.~\eqref{eq:circstate}). Contracting \eqref{eq:eigeneqLambda} with this state we find 
\be
\mathcal B'_{n,\tau} \ket{\Lambda'}= e^{i \theta} \ket{\Lambda'}. 
\ee
where we introduced 
\begin{align}
&\mathcal B'_{n,\tau} = (m_0 \otimes (\mathbb U' \!\otimes_r\! \mathbb U^{\prime*}) \!\otimes\! b_{\tau} \otimes \1^{\otimes 2(\tau-1)})\notag\\
&\qquad\qquad\times (((\mathbb W\otimes \Pi_{\tau}) \!\otimes_r\! (\mathbb W^*\otimes \Pi^*_{\tau})))
\end{align}
where $\mathbb U'$ is obtained from $\mathbb U$ by removing the leftmost gate. Proceeding as before we find 
\be
\ket{\Lambda'} = \ket{\mcirc}\otimes\ket{\Lambda''}. 
\ee
This procedure can be iterated $2n-2$ times and gives 
\be
\ket{\Lambda} = \ket{\mcirc}^{\otimes 2(n-1)}\otimes \ket{\nu},
\ee
where $\ket{\nu}$ fulfils 
\be
(b_{\tau} \otimes \1^{\otimes 2(\tau-1)}) (m_0\otimes (\Pi_{\tau} \otimes_r\! \Pi^*_{\tau})) \ket{\nu} = e^{i \theta} \ket{\nu}\,. 
\ee
This implies that 
\be
\|b_{\tau} (m_0\otimes \1)\|_\infty =1. 
\ee
Namely that there exists a $\ket{\lambda}$ such that 
\be
\expval{(m_0^\dag\otimes \1) b_{\tau}^\dag b_{\tau} (m_0\otimes \1)}{\lambda}=1. 
\ee
Recalling that $\|m_0\|=\|b_\tau\|=1$ this also means 
\begin{align}
& (m_0^\dag m_0\otimes \1)\ket{\lambda} = \ket{\lambda}, \\
& b_{\tau}^\dag b_{\tau} (m_0\otimes \1)\ket{\lambda} = (m_0\otimes \1)\ket{\lambda}. 
\end{align}
The first of these equations implies $\ket{\lambda}=\ket{\mcirc}\otimes\ket{a}$ so that we finally have 
\be
b_{\tau}^\dag b_{\tau} \ket{\mcirc}\otimes\ket{a} = \ket{\mcirc}\otimes\ket{a}. 
\ee
Rewriting this equation in terms of the local gate it reads as 
\be
U_n (\1\otimes a) = b\otimes \1\,, \qquad b= \frac{1}{2}  {\rm tr}_{2}[U_n (\1\otimes a)]. 
\ee
This equation is solved only by $a$ unitary and $U_n=\1\otimes a^\dag$. Since such a gate fulfils Eq.~\eqref{eq:leftcondition}, we have a contradiction.

\section{Working in a Fixed Double-Momentum Sector} 
\label{sec:doublemoneta}

In this section we briefly describe how to reduce the numerical analysis to a given double-momentum sector. As mentioned in the main text the transfer matrix $\mathcal{T}$ is invariant under two-site translations in the forward and backward lattices
\begin{equation}
    [\Pi^{2\tau_1}_{2t} \otimes \Pi^{2\tau_2}_{2t} ,\mathcal{T}] = 0,\qquad \tau_1,\tau_2=0,\ldots,t-1.
\end{equation}
For the forward and backward lattice the most natural basis to work in is therefore the eigenbasis of the two-site shift operator. Considering the forward time lattice, we have $2t$ qubits, and we know we have $\Pi_{2t}^{2t} = 1$, giving eigenvalues $e^{{2\pi} i \nu/t}$ with $ \nu = 0, \dots t - 1$. To generate the eigenbasis we select a set of reference states $|f\rangle$ (taken to be product states in the computational basis) and write 
\begin{equation}
    |f_\nu\rangle = \frac{1}{\sqrt{L_f}} \sum_{r = 0}^{L_f-1} e^{{2\pi} i \nu r /t}\Pi_{2t}^{2r} |f\rangle,
\end{equation}
where $L_f$ is the period of the reference state $\Pi_{2t}^{2L_f}|f\rangle = |f\rangle$. Typically $L_f = t$ however some special states will have periods that are integer multiples of $t$. Computationally this representation reduces our overall storage from $2^{2t} \to {2^{2t}}/{t}$ approximately, with the largest sector being the $\nu =0$ sector (see, e.g., Ref.~\cite{Sandvik_2010} for more details). For the purposes of discussing complexity later in this section we will call $D = 2^{2t}$ and $D_\nu = {2^{2t}}/{t}$.  

An important observation about $\mathcal{T}$ is that it is made up of a product of operators which independently are translation invariant under two shifts. This is evident from Eq.~\eqref{eq:SFFTM2} and implies that we can update vectors in distinct steps. First let us treat the case where we do not couple the forward and backward lattice, i.e., we consider an operator of the form
\begin{equation}
   \tilde{\mathbb{U}}_o  = \tilde{U}_0 \otimes\cdots\otimes \tilde{U}_{2t-2}, 
\end{equation}
where $\tilde{U}$ is a two-local operator (it acts non-trivially only on a pair of nearest neighbours). We will use the above as an example to illustrate working in the momenta basis for operators with this general structure. Since 
\begin{equation}
    [\Pi_{2t}^{2r}, \tilde{\mathbb{U}}_o ] = 0,
\end{equation}
we can write 
\begin{equation}
    \tilde{\mathbb{U}}_o|\psi_\nu \rangle = \frac{1}{\sqrt{L_f}} \sum_f c_f \sum_{r = 0}^{L_f-1}e^{{2\pi} i \nu r /t} \Pi_{2t}^{2r} \tilde{\mathbb{U}}_o|f\rangle.
    \label{eq:Uopsi}
\end{equation}
Where the sum over $f$ is taken over the set of representation basis states.
One memory inefficient way to evaluate this expression is to simply evaluate $\tilde{\mathbb{U}}_o|f\rangle$ in the full Hilbert space, and then compress the state back into the translation invariant representation. This approach is computationally costly and likely memory bound, severely slowing down the code. It also removes the advantage of working in the symmetry resolved basis by increasing storage requirements to $D$, which, once we couple the forward and backward lattice, will eliminate our advantage with this approach. In fact, an open question we were not able to answer is how to evaluate  $\tilde{\mathbb{U}}_o|\psi_\nu\rangle$ faster than $O(D_\nu^2)$. The operations similarity to a discrete Fourier transform indicates this may be possible.

Instead of updating the vector directly from the expression \eqref{eq:Uopsi} we focus our efforts on computing matrix elements of the operator
\begin{equation} \label{eq:resolveU}
    \!\!\langle m_\nu | \tilde{\mathbb{U}}_o | f_\nu \rangle \!=\!\!  \sum_{p = 0}^{L_m-1}\sum_{r = 0}^{L_f-1}  \frac{e^{{2\pi} i \nu (r-p) /t}}{\sqrt{L_m L_f }}\langle m |\Pi_{2t}^{2(r-p)} \tilde{\mathbb{U}}_o|f\rangle.
\end{equation}
The above equation is simpler to evaluate. To see this without loss of generality take $r=p=0$. Because $\tilde{\mathbb{U}}_o$ is made up of a product of commuting terms the expression factorises 
\begin{equation}
    \langle m|\tilde{\mathbb{U}}_o|f\rangle = \prod_{j=0}^{t}\langle m_{2j} m_{2j+1} | \tilde{U}_o | f_{2j} f_{2j+1} \rangle,
\end{equation}
where we have broken the representative state into its computational basis form for individual time lattice qubits. Note that Eq.~\eqref{eq:resolveU} can be computed in $O(t)$ steps due to repeated computations. For convenience we will call this new symmetry resolved operator
\begin{equation}
  \tilde{\mathbb{U}}_{0,(m,f)}^{(\nu)}\equiv \langle m_\nu | \tilde{\mathbb{U}}_o | f_\nu \rangle.
\end{equation}
This allows us to store it in a $D_\nu \times D_\nu$ dimensional matrix, the same size as we will see, as the many body vectors once we combine the forward and backward lattice.
A vector on the full space can be represented by
\begin{equation}
    |\psi_{(\nu,\nu')}\rangle = \sum_{f,b} C_{f,b} |f_\nu \rangle |b_{\nu'}\rangle,
\end{equation}
where the sum is taken over the set of representation basis states. Updating the full forward and backward lattice state  with the symmetry resolved operator is now given by 
\begin{equation}
   \left( \tilde{\mathbb{U}}_o \otimes \tilde{\mathbb{U}}_o^* \right) |\psi_{(\nu,\nu')}\rangle \to  \tilde{\mathbb{U}}_o^{(\nu)} C \tilde{\mathbb{U}}_o^{(l) \dagger}.
\end{equation}

The final piece of the puzzle is to understand the operation
\begin{equation}
    \mathcal{O}_{0}^{(3)} |\psi_{(\nu,\nu')}\rangle = \sum_{f,b} C_{f,b} \mathcal{O}_{0}^{(3)} |f_\nu \rangle |b_{\nu'}\rangle.
\end{equation}
We focus here on $\mathcal{O}_{0}^{(3)}$ as it is diagonal in the computational basis. Other non-diagonal terms can be evaluated with a simple basis rotation, and then following the steps we will outline. The action of this operator on the translation invariant basis is trivial, we have, 

\begin{equation}
    \!\!\!\!\mathcal{O}_{0}^{(3)} \!|f_\nu \rangle |b_{\nu'}\rangle \!=\!\! \begin{cases}
        |f_\nu \rangle |b_{\nu'}\rangle & \sum_{m=0}^{t-1} f_{2m} = \sum_{m=0}^{t-1} b_{2m} \\
        0 & \text{ otherwise}
    \end{cases}\!\!.
\end{equation}
Where we again used the computational basis representation of our reference states. This concludes all necessary steps to reduce the overall storage required by the problem by a factor of $t^2$, along with working with the symmetry resolved transfer matrix.

\bibliography{bibliography.bib}  

\end{document}